\documentclass[reprint,aps,titlepage,nofootinbib]{revtex4-2}
\usepackage{graphicx}
\usepackage{amsmath,amssymb}
\usepackage{bbold}
\usepackage{bm}
\usepackage{comment}
\usepackage[T1]{fontenc}
\usepackage{textgreek}
\usepackage{lmodern}
\usepackage{hyperref}
\usepackage{xcolor}
\usepackage{soul}

\begin{document}

\title{Microscopic theory of nonlinear phase space filling in polaritonic lattices}

\author{Kok Wee Song}
\affiliation{Department of Physics and Astronomy, University of Exeter, Stocker Road, Exeter EX4 4QL, United Kingdom}

\author{Salvatore Chiavazzo}
\affiliation{Department of Physics and Astronomy, University of Exeter, Stocker Road, Exeter EX4 4QL, United Kingdom}

\author{Oleksandr Kyriienko}
\affiliation{Department of Physics and Astronomy, University of Exeter, Stocker Road, Exeter EX4 4QL, United Kingdom}

\begin{abstract}
We develop a full microscopic theory for a nonlinear phase space filling (NPSF) in strongly coupled two-dimensional polaritonic lattices. Ubiquitous in polaritonic experiments, the theoretical description of NPSF, also known as nonlinear optical saturation, remains limited to perturbative treatment and homogeneous samples. In this study, we go beyond the existing theoretical description and discover the broad scope of regimes where NPSF crucially modifies the optical response. Studying the quantum effects of non-bosonicity, cooperative light-matter coupling, and Coulomb blockade, we reveal several regimes for observing the nonlinear Rabi splitting quench due to the phase space filling. Unlike prior studies, we derive nonlinear Rabi frequency scaling all the way to the saturation limit and show that the presence of a lattice potential leads to qualitatively distinct nonlinearity.
We concentrate on the three regimes of NPSF: 1) planar; 2) fractured; and 3) ultralocalized. For the planar saturation, the Rabi frequency decreases exponentially as a function of exciton density. For the fractured case, where excitons form a lattice with sites exceeding the exciton size, we discover fast NPSF at low occupations. This is followed by slower NPSF as the medium becomes fully saturated. This behavior is particularly pronounced in the presence of Coulomb (or Rydberg) blockade, where regions of fast and slow NPSF depend on the strength of repulsion. For the ultralocalized NPSF, we observe the square-root saturation typical to the collection of two-level systems. Our findings shed light on recent observations of strong nonlinearity in heterobilayers of transition metal dichalcogenides where moir{é} lattices emerge naturally [Nature \textbf{591}, 61 (2021)]. Finally, the developed theory opens the prospects for engineering strongly nonlinear responses of polaritonic lattices with patterned samples, driving polaritonics into the quantum regime.
\end{abstract}

\maketitle

\section{Introduction}

Strong light-matter coupling (SC) hybridizes photons and matter excitations, leading to an emergence of polaritons \cite{BasovAsenjo2021,CarusottoCiuti2013,Deng2010,Liew2011}. Systems where SC can be achieved includes atomic vapors and lattices~\cite{Hammerer2010,Chang2018}, collection of color centers~\cite{Fehler2019,Radulaski2019,Eisenach2021}, quantum dots~\cite{Diniz2011,Trivedi2019}, microwave circuits~\cite{Blais2004}, molecular complexes~\cite{Sanchez-Barquilla2022,Wang2021,Yuen-Zhou2022}, semiconductor quantum wells~\cite{Brodbeck2017,BallariniDeLiberato2019}, and two-dimensional (2D) materials~\cite{liu2014,lundt2016,Sidler:NatPhys13(2016),Dufferwiel:NatPhoto11(2017),Schneider2018,Wang:RMP90(2018),Emmanuele2020,Lackner2021,Anton-Solanas2021,Zhumagulov2022} (see Ref.~\cite{BasovAsenjo2021} for the full panorama). In semiconductor nanostructures, the prominent example is strong coupling to excitonic modes, leading to exciton-polaritons~\cite{CarusottoCiuti2013}. The very essence of polaritonic response is in acquiring a nonlinearity for light~\cite{Chang2014}, ultimately being visible even at the few-photon occupation~\cite{Munoz-Matutano2019,Delteil2019,Zasedatelev2021,Kuriakose2022}. The utility of nonlinearity ranges from generating solitons~\cite{Sich2012,Walker2017,Maitre2020} to quantum information processing and gates at the single-photon level~\cite{Sanvitto2016,KyriienkoLiew2016,Ghosh2020,Kavokin2022}. In semiconducting microcavities, nonlinearity leads to emergent fluids of light in planar geometries~\cite{Amo2009,Amo2011,Liew2008,YangKim2022} and highly nontrivial dynamics in lattice-based polaritonic systems~\cite{Kim2013,Jacqmin2014,Ohadi2017,Whittaker2018,Cerda-Mendez2018,Scafirimuto2021,Alyatkin2021,Cookson2021,Topfer2021,Harder2021,Zvyagintseva2022}.

Polaritonic nonlinearity can originate from several sources, which lead to different types of nonlinear processes. First, quasiparticles in an underlying medium can interact via Coulomb repulsion or attraction~\cite{Ciuti1998,Tassone1999,Combescot2007,Glazov2009,Brichkin2011,Shahnazaryan2016,Shahnazaryan2017,Barachati2018,Bleu2020}, leading to energy shifts from hybridized polaritonic modes, and thus power-dependent response~\cite{Estrecho2019}. This type of nonlinearity prevails for Wannier-type excitons, manifested in exchange-dominated scattering for 2D systems~\cite{Ciuti1998,Tassone1999}, and dipole-dipole interaction for dipolar excitons~\cite{Butov2001,Kyriienko2012,Kyriienko2014a,Togan:PRL2018,Hubert2019}. Similar interactions are observed for Rydberg states~\cite{Browaeys2016}, enabling a plethora of highly nonlinear effects for Rydberg atom gas~\cite{Lukin2001,Sevincli2011,Gorshkov2011}, atomic arrays~\cite{Labuhn2016,Bernien2017,Henriet2020}, and recently Rydberg excitons~\cite{Kazimierczuk2014,Versteegh2021,Gallagher2022,Orfanakis2022}. The ultimate limit of Coulomb-driven nonlinearity comes from a Coulomb blockade~\cite{Alhassid2000} --- inability of creating more excitation at specific sites due to large energy penalties, and subsequent nonlinear impact on coupled photonic modes~\cite{Imamoglu1997,Verger2006}.

The second type of nonlinearity comes from statistical properties of matter excitations and corresponds to the nonlinear phase space filling (NPSF), which can be referred to interchangeably as a nonlinear optical saturation~\cite{Schmitt-Rink1985,Tassone1999,Brichkin2011}. Namely, creating two excitons in exactly the same state of electrons (e) and holes (h) is forbidden by their fermionic statistics. Similarly, a collection of two-level systems (TLS) can only be excited until further excitations are prevented by Pauli statistics~\cite{Hammerer2010}. Thus, increased number of excitations leads to filling the available phase space (no room for creating new quasiparticles)~\cite{Combescot2008,Combescot:EPJB68(2009)}, and effectively reduced light-matter coupling in remaining sites or area. At high powers NPSF can lead to the power-dependent quench (or collapse) of the associated Rabi splitting~\cite{Emmanuele2020}. For Wannier excitons (delocalized e-h pairs) this nonlinear mechanism was discussed already in the seminal paper of Tassone\,\&\,Yamamoto~\cite{Tassone1999}, where first-order power-dependent contribution to the reduction of Rabi splitting was described. However, in III-V semiconductors this nonlinearity was not considered to be dominant~\cite{Brichkin2011}. For Frenkel-type excitons (quasiparticles based on localized e-h pairs in molecular lattices) NPSF plays a major role~\cite{Betzold2020,Arnardottir2020}, and is ubiquitous in various experiments~\cite{Daskalakis2014,Keeling2020,Yagafarov2020,Betzold2020}. Finally, recent results in transition metal dichalcogenides (TMDC) show the importance of both contributions from Coulomb and nonlinear phase space filling~\cite{Emmanuele2020,Tan2020,Kravtsov2020,Kyriienko:PRL125(2020),Zhang:Nature591(2021),Stepanov:PRL126(2021),BastarracheaMagnani:PRL126(2021),Denning:PRB105(2022),Denning:PhysRevResearch(2022),Datta2022,Louca2022}.

In this work, we develop a unified treatment of nonlinear quantum optical effects based on phase space filling. Our theory is applicable to a wide range of exciton-polariton lattices, also in the presence of both Pauli and Coulomb blockade. We describe three distinct NPSF regimes being the planar, fractured, and ultralocalized regimes. For each case, we present an analysis and show that in the fractured case a sharp decrease of Rabi frequency in the low-density regime can be facilitated by the Coulomb blockade. Our theory can shed light onto recent experiments in moir\'{e} heterobilayers, and open a way for enhancing the nonlinear response.

\section{Theory of excitons coupled to a cavity mode}

\begin{figure}
    \centering
    \includegraphics[width=3.35in]{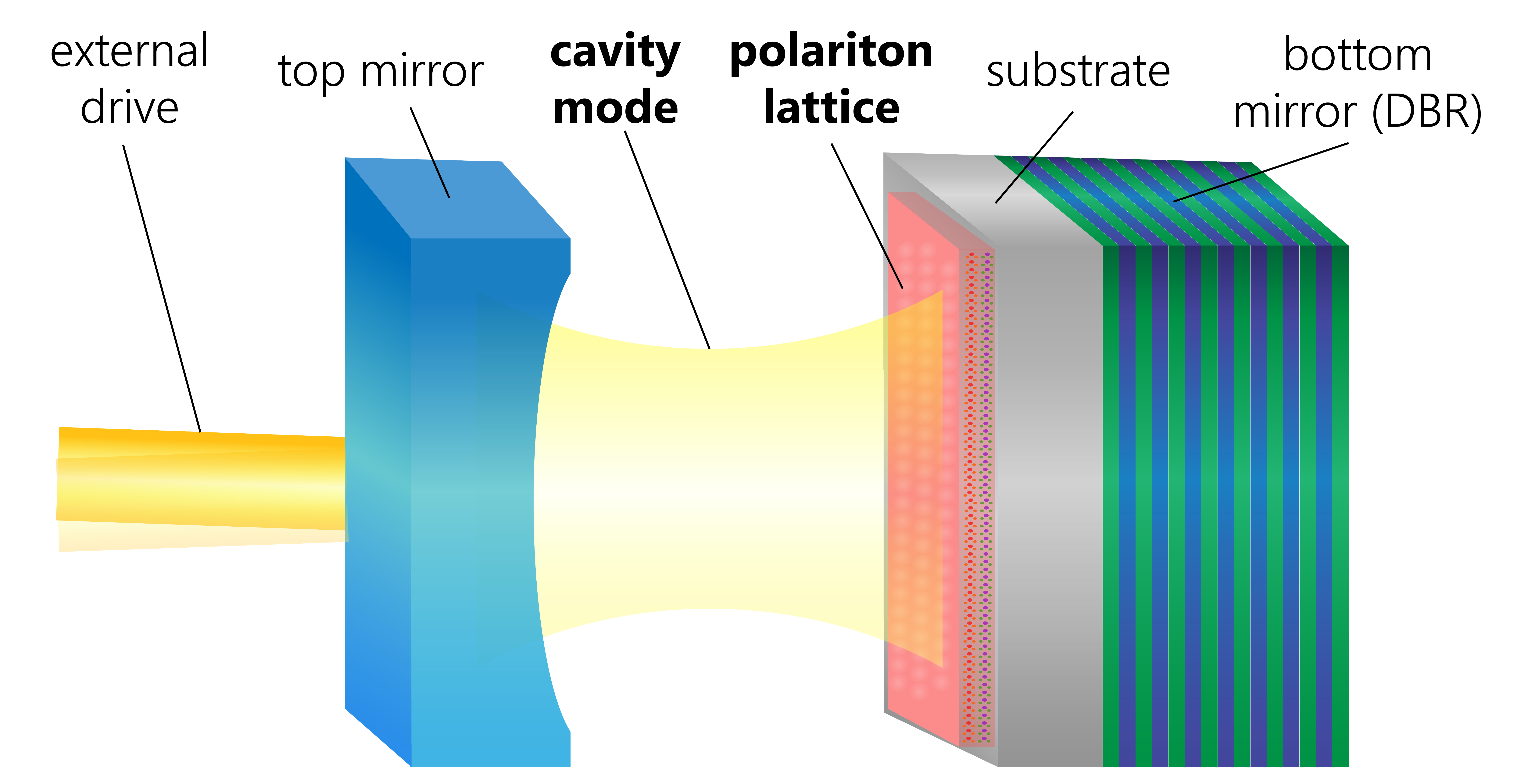}
    \caption{Sketch of an optical microcavity with polaritonic lattice as an active medium. The cavity is formed by a concave top mirror (e.g. fiber-based) and a bottom mirror as a distributed Bragg reflector (DBR) of high reflectivity. Polaritonic lattice is arranged as a patterned semiconductor with excitonic potential, as typically realized in heterobilayers of transition metal dichalcogenides (TMDC) with moir\'{e} potential.}
    \label{fig:cavity}
\end{figure}
\begin{figure}
    \centering
    \includegraphics[width=3.3in]{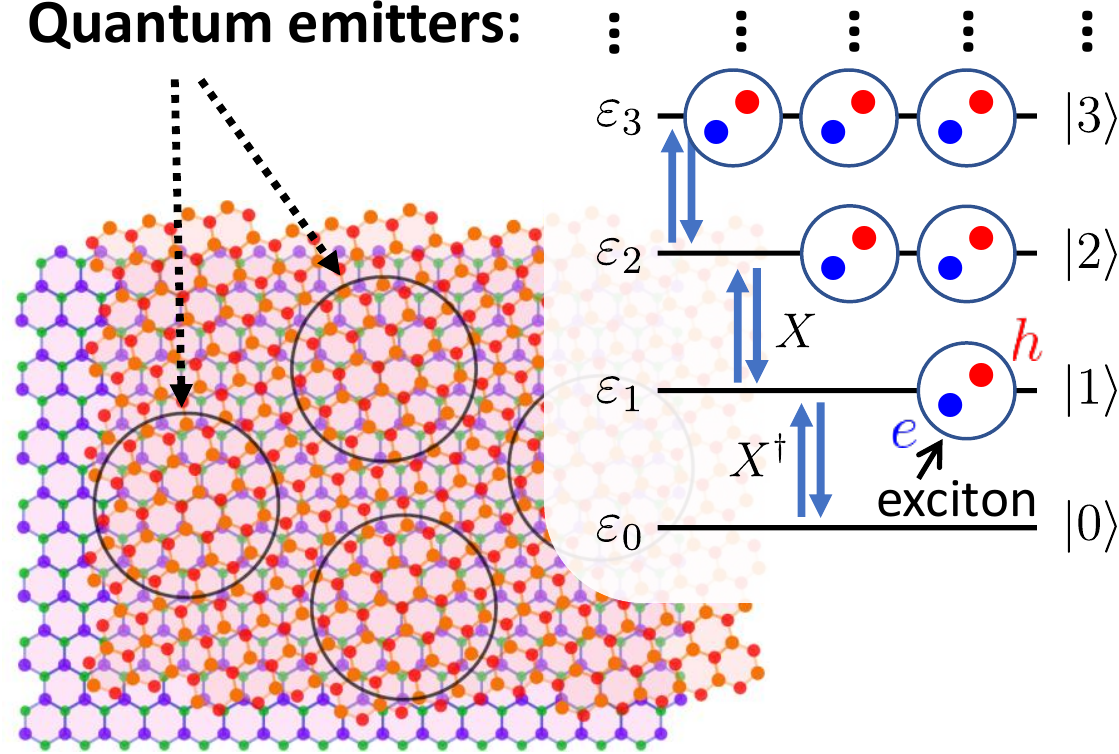}
    \caption{Quantum emitter. Each moir\'e cell of a bilayer can be viewed as a quantum emitter that hosts excitons in its excited state. The quantum states of the emitter with energy $\varepsilon_0,\dots,\varepsilon_n$ are denoted by $|0\rangle,\dots,|n\rangle$ in the diagram.} 
    \label{fig:level}
\end{figure}

To develop the microscopic model of nonlinear phase space filling, we start by considering excitons (electron-hole pairs) strongly coupled to a cavity mode. This is typically formed by distributed Bragg reflectors or fiber-based mirrors (see sketch in Fig.~\ref{fig:cavity}). 
\begin{figure*}
    \centering
    \includegraphics[width=0.95\linewidth]{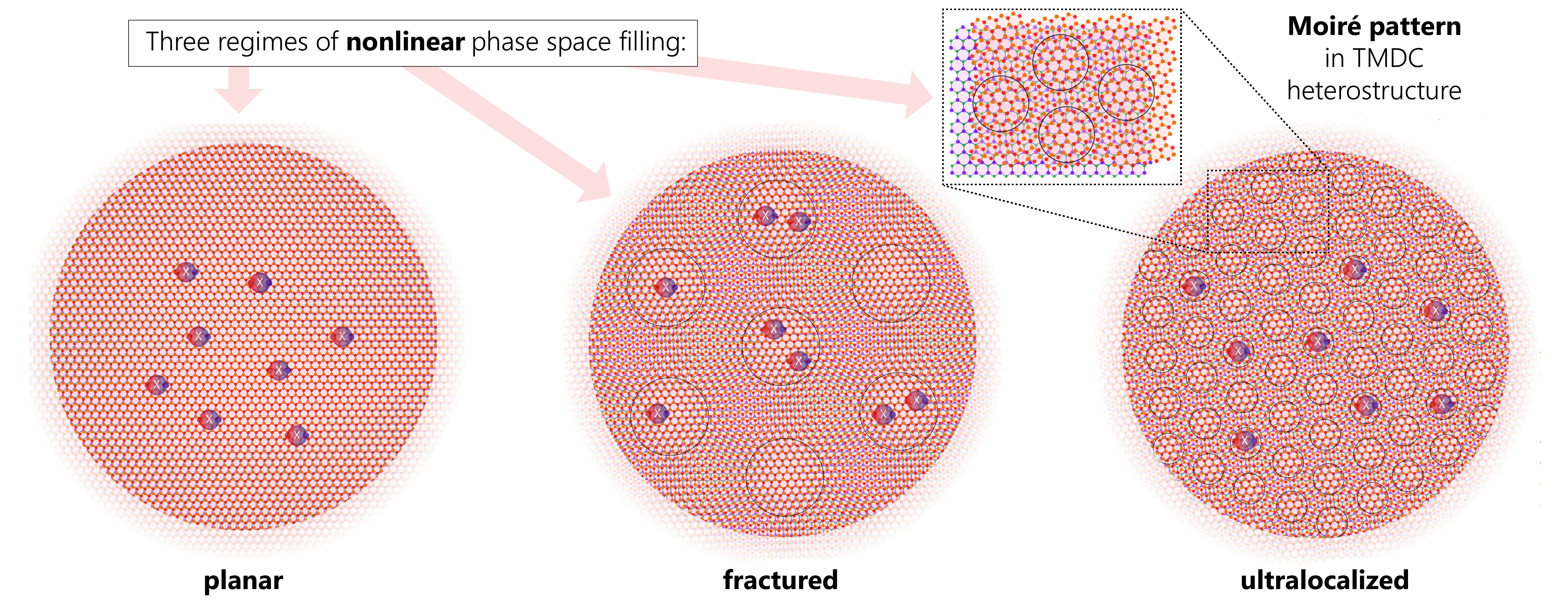}
    \caption{Regimes of NPSF. Here we sketch
different regimes of nonlinear phase space filling (saturation) in polaritonic lattices. As a prominent example, we present moir\'{e} heterostructures with different twist angles, demonstrating the planar, fractured, and ultralocalized NPSF at progressively increased twist angle.}
    \label{fig:moire}
\end{figure*}
Specifically, we consider bound excitons being confined to minima of potential, thus forming a polaritonic lattice (Fig. \ref{fig:moire}). A system Hamiltonian with $N_s$ lattice sites reads
\begin{equation}\label{eqn:model}
    \hat{H} = \omega_0 \hat{c}^\dagger \hat{c} + \sum^{N_{\rm s}}_{i=1} \hat{H}_{i}^{\mathrm{X}} + \sum^{N_{\rm s}}_{i=1} \Omega_0\Big( \hat{X}_i^\dagger \hat{c} + \hat{c}^\dagger\hat{X}_i \Big),
\end{equation}
where $\hat{c}^\dagger$ ($\hat{c}$) is a field operator creating (annihilating) a cavity photon with energy $\omega_0$ (we set $\hbar = 1$). The first term in Eq.~\eqref{eqn:model} thus describes the energy of the cavity mode at normal incidence (small photon momenta). The second term $\sum^{N_{\rm s}}_{i=1} \hat{H}_{i}^{\mathrm{X}}$ describes the available quantum states on each lattice sites, labeled by index $i$. This term may also be viewed as a lattice of \emph{identical} quantum emitters which can host up to $\ell$ excitons, depending on the confinement potential and interaction between excitons. Therefore, the Hamiltonian of each separate sites can be written as
\begin{equation} 
    \hat{H}^X_i=\mathbb{1}^{i-1} \otimes \left(\sum_{n=0}^{\ell}\varepsilon_{n}|n\rangle\langle n| \right)\otimes  \mathbb{1}^{N_s-i},
\end{equation}
where $\mathbb{1}$ is the identity operator in the Hilbert space of a single site, and $\mathbb{1}^{ n}=\bigotimes_{i=0}^n\mathbb{1}$. In the above, $|n\rangle$ denotes the $n$-exciton eigenstate with energies $\varepsilon_n$ in an emitter. The ground state of a emitter is $|0\rangle$ (no exciton) with energy $\varepsilon_0=0$, see Fig. \ref{fig:level}.

Intuitively, this model is analogous to a lattice of decoupled harmonic oscillators. However, we note that we assume a general form of the energy-level structure of $\hat{H}_{i}^{\mathrm{X}}$ to account for exciton-exciton interaction and arbitrary confinement potential, introducing some degree of unharmonicity. With this, we make sure that the microscopic model can describe excitons of different types (Wannier and Frenkel), as well as remaining valid for generic quantum emitters with $\ell+1$ levels. 

Finally, the third term in the system Hamiltonian \eqref{eqn:model} describes the strong light-matter coupling, where $\Omega_0$ is the bare exciton-photon coupling strength. This term defined the transition between the emitter's energies levels by absorbing or emitting a photon. These transitions at site $i$ is described by the \emph{excitonic ladder operator} as
\begin{equation}\label{eqn:X}
    \hat{X}^\dagger_i= \mathbb{1}^{i-1} \otimes \left( \sum_{n=0}^{\ell-1}\sqrt{r_{n}}|n+1\rangle\langle n| \right) \otimes  \mathbb{1}^{N_s-i},
\end{equation}
where $r_{n}\equiv|\langle n+1|x^{\dagger}|n\rangle|^2$ with $x^\dagger$ being the field operator for creating an exciton. This exciton \emph{transition matrix element} $r_n$ describes the effective rate for creating an additional exciton in a $n$-exciton state, $|n\rangle$. It is determined by the microscopic details of a $n$-exciton state.
We note that, although our polaritonic lattice model Eq.~\eqref{eqn:model} is similar to an atomic system, the optical transition of each emitter [Eq.~\eqref{eqn:X}] only takes place between the adjacent energy levels, $|n\rangle$ and $|n+1\rangle$. Namely, the optical multiexciton processes are forbidden. In contrast to atoms, the transitions between levels do not have such a constraint and they are determined by optical selection rules. Importantly, excitons are composite bosons with a nontrivial $n$-dependence of $r_n$. In contrast to atomic level transition, each atomic level cannot host more than one electron. This corresponds to $r_{n}=\delta_{n,0}$ in Eq. \eqref{eqn:X}, where the only allowable transition is between states $|0\rangle$ and $|1\rangle$.
This is the key difference between the two systems which leads to distinct optical saturation effects qualitatively. The transition matrix element $r_n$ depends on the underlying physics of the quantum emitter (exciton). The quantitative description for $r_n$ is presented in Section~\ref{sec:trnsRts}.

In the lattice of identical emitters, the exciton created by a uniform cavity mode must preserve the translational symmetry in the lattice. This implies that the photon can only couple to the lattice collective mode (bright state) in the form~\cite{Hammerer2010,Combescot:EPJB68(2009)}
\begin{equation}\label{eqn:B}
    \hat{B}^\dagger=\frac{1}{\sqrt{N_{\rm s}}}\sum_{i=1}^{N_{\rm s}}\hat{X}^\dagger_i.
\end{equation}
This operator creates a coherent lattice excitation where the probability of finding an exciton is uniformly distributed across the lattice instead of sitting on a particular site. It corresponds to a \emph{collective exciton} mode that participates in SC. This allows us to rewrite the coupling Hamiltonian,
\begin{equation}\label{eq:H_SC}
\sum_{i = 1}^{N_{\rm s}} \Omega_0 (\hat X^\dagger_i \hat c + \hat X_i \hat c^\dagger) = \sqrt{N_{\rm s}} \Omega_0 (\hat B^\dagger \hat c + \hat B \hat c^\dagger),
\end{equation}
meaning the coupling only acts in the subspace of bright states. The collective excitonic quasiparticle $\hat{B}^\dagger$ is exactly the mode that couples to light in optically active materials, with the corresponding $\sqrt{N_s}$ enhancement for Frenkel excitons \cite{BasovAsenjo2021,Betzold2020} and exciton area-based enhancement of light-matter coupling for Wannier excitons \cite{Deng2010}. Importantly, the collective exciton mode $\hat{B}^\dagger$ also acts like a quasiparticle with well-defined particle numbers, but it has a peculiar statistical property. In the low ratio of total number of excitons $N$ to $N_s$ in the lattice ($N/N_{\mathrm{s}}$, exciton number per site), the $B^\dagger$ quasiparticle excitations obey
statistics similar to the bosonic one. However, in a large $N/N_s$ regime, this quasiparticle excitation may strongly deviate from the Bose statistics due to the composite nature of exciton and blockade effects arising from the Coulomb interaction between excitons. This non-bosonicity determines how the phase space is depleted by $B^\dagger$-excitations. This will eventually appear as a nonlinear correction to the light-matter coupling of the lattice.

To see the relation between the phase space-filling effect and nonlinear light-matter coupling, we consider in the SC regime that the two states $\{(\hat{B}^\dagger)^{N+1}|\varnothing\rangle,\hat{c}^\dagger(\hat{B}^\dagger)^N|\varnothing\rangle\}$ hybridize and form a quasiparticle --- polariton. Here, we defined the global ground state $|\varnothing\rangle=|0\rangle_{\mathrm{c}}\bigotimes_{i=1}^{N_{\rm s}}|0\rangle$ as the product of emitter ground state $(\vert 0 \rangle)$ and cavity $(\vert 0 \rangle_{\mathrm{c}})$ ground states. To investigate the nonlinearity of the polariton in the lattice of emitters, we construct the polaritonic Hamiltonian as block diagonal matrices
\begin{equation}\label{eqn:H_N}
    H_N=\begin{bmatrix}
       E_{N+1}&\Omega_N\\
       \Omega_N&E_{N}+\omega_0
    \end{bmatrix},
\end{equation}
that couple states with $N+1$ excitons, containing $B^{\dagger}$-excitations in total. Here $E_{N}=\langle\varnothing |\hat{B}^{N}\hat{H}(\hat{B}^\dagger)^{N}|\varnothing\rangle/\mathcal{F}^{(N_{\mathrm{s}})}_N$ is the total energy of $N$ collective excitons and we have introduced the normalization factor of the $N$-exciton states~\cite{Combescot:EurPhysJB(2003),Combescot2008}
\begin{equation}
\label{eq:FN-def}
    \mathcal{F}^{(N_{\mathrm{s}})}_N=\langle\varnothing |\hat{B}^{N}(\hat{B}^\dagger)^{N}|\varnothing\rangle.
\end{equation}
This ensures the states are properly normalized such that the matrix elements in Eq.~\eqref{eqn:H_N} retain the correct physical meaning at increasing densities (as phase space filling increases). The off-diagonal elements then can be written as~\cite{Laussy:PRB73(2006)}
\begin{equation}\label{eqn:OmegaN}
\Omega_N=\Omega_0\sqrt{\mathcal{F}^{(N_{\mathrm{s}})}_{N+1}/\mathcal{F}^{(N_{\mathrm{s}})}_{N}},
\end{equation}
meaning that the light-matter coupling $\Omega_N$ (effective Rabi frequency) depends on the number of excitons. This introduces the nonlinearity in the system in the form of nonlinear phase space filling.

To study the effect of NPSF in full generality, we develop a strategy for calculating the normalization factor Eq.~\eqref{eq:FN-def} for a generic structure of onsite excitations and arbitrary $N_{\mathrm{s}}$. We do this by using a multinomial expansion. The compact expression of the normalization factor reads
\begin{align}\label{eqn:FN}
    \mathcal{F}^{(N_{\mathrm{s}})}_N\!
    =&\sum_{n_1+\dots+n_{N_{\rm s}}=N}\left(\frac{N!/N_{\rm s}^{N/2}}{n_1!\dots n_{N_{\rm s}}!}\right)^2\prod_{i=1}^{N_{\rm s}}P(n_i),
\end{align}
where $P(n_i)=r_0\dots r_{n_i-1}$ with $P(0)\equiv1$ (see Appendix \ref{app:FN} for the derivation). The renormalization factor in Eq.~\eqref{eq:FN-def} completely determines the NPSF of the Hamiltonian \eqref{eqn:model}. We note that while $\mathcal{F}_N^{(N_{\mathrm{s}})}$ was evaluated before in some limiting cases, such as Wannier exciton~\cite{Combescot:EurPhysJB(2003),Combescot2008} and Frenkel exciton (two-level emitter)~\cite{Combescot:EPJB68(2009)}, here we present the generic non-perturbative treatment. This is imperative for accessing the NPSF in the deep saturation regime, as demonstrated in the next sections.

To summarize, we have developed a general theoretical framework for the exciton-polariton on a lattice which is described by Eqs.\eqref{eqn:H_N}-\eqref{eqn:FN}. This reduces the study of the NPSF problem to specifying $r_n$ and $|n\rangle$, which are ultimately determined by the properties of the $n$-excitons state of an emitter. As we will discuss later, $r_n$ is the key parameter that gives rise to a different $\mathcal{F}_N^{(N_{\mathrm{s}})}$, $N$-dependence of the Rabi frequency, and thus enable various qualitative modes of NPSF. Before carrying out the full analysis, it is instructive to consider two special cases and recover known results. First, let us consider $|n\rangle$ being the state of purely bosonic excitations. In this case, the transition matrix element is $r_n=n+1$ (bosonic enhancement factor) and the number of available states is $\ell \rightarrow \infty$. Using this assumptions in Eq.~\eqref{eqn:FN}, we recover the usual bosonic normalization factor $\mathcal{F}^{(N_{\mathrm{s}})}_N=N!$ for $N$-boson state.
Next, let us then consider the opposite limit corresponding to two-level systems (fermionic limit). This is modeled by a quantum emitter with $r_0=1, \, r_{n \geq 1} = 0$. With this, Eq.~\eqref{eqn:FN} reduces to the quantum limit~\cite{Combescot:EPJB68(2009)} with $\mathcal{F}^{(N_{\mathrm{s}})}_N = N!N_{\rm s}!/(N_{\rm s}-N)!N_{\rm s}^{-N}$. In the rest of the paper, we demonstrate that the polariton Hamiltonian in Eqs. \eqref{eqn:H_N}-\eqref{eqn:FN} allows us to go beyond these two limits. We consider transition matrix elements that are neither purely bosonic nor purely fermionic. This is controlled by the confinement, composite particle properties, and interactions. Our theory reveals the behavior of $\mathcal{F}_{N}^{(N_{\mathrm{s}})}$ for a polaritonic lattice with strong NPSF (deep saturation regime). This regime of SC has not been explored before. In the following, we investigate this regime and discuss the relevant system in moir\'{e} structure of 2D materials.

\section{Introducing localization of excitons and polaritonic lattices}\label{sec:trnsRts}

In the previous section, we presented the general theory of excitons in a cavity and introduced the transition matrix elements $r_n$ that characterize the polaritonic system [Eq.~\eqref{eqn:X}]. As we mentioned previously, the form of $r_n$ has important implications for the nonlinear phase space filling of the lattice. As described in Eq.~\eqref{eqn:X}, the transition matrix element $r_n$ is a measure of the rate for creating an additional exciton-like quasiparticle in a $n$-excitons system. This quantity depends on the specific correlation effects between excitons. Calculating it for general cases is a highly non-trivial many-body problem. Therefore, in this section, we consider two microscopic mechanisms that induce NPSF. One is due to the correlation of the fermionic statistics (\emph{Pauli} blockade) where the electron and hole from excitons cannot occupy the same state in phase space. The other mechanism is due to the strong exciton-exciton interactions (\emph{Coulomb} blockade) where the high-energy states are effectively projected out in the phase space. In the Pauli blockade, we can evaluate $r_n$ exactly, and for the Coulomb blockade we use the phenomenological approach that allows to keep the system tractable in the presence of nonlinearity.

\subsection{Pauli blockade}

To investigate the Pauli blockade of a polaritonic lattice, we begin with the simplest case of $N_{\mathrm{s}}=1$ and extend the consideration to excitons with a spatial shape. For a single site (or emitter), we construct the $n$-exciton ground state as
\begin{equation}\label{eqn:n-states}
    |n\rangle=\frac{1}{\sqrt{\mathcal{F}^{(1)}_n}}(x^\dagger_0)^n|0\rangle, \quad
    \mathcal{F}^{(1)}_n=\langle0|x_{0}^n
    (x^\dagger_{0})^n|0\rangle,
\end{equation}
where the exciton creation field operator is
\begin{equation}\label{eqn:x-composite}
    x^\dagger_{\nu}=\int d\mathbf{r}_ed\mathbf{r}_h\Psi_{\nu}(\mathbf{r}_e,\mathbf{r}_h)a^\dagger_{\mathbf{r}_{e}}b_{\mathbf{r}_h}.
\end{equation}
Here, $\Psi_{\nu}(\mathbf{r}_e,\mathbf{r}_h)$ is the exciton wavefunction, with $\nu$ being the state index (quantum number), $\mathbf{r}_e$ and $\mathbf{r}_h$ are the electron and hole position vectors in real space, and $a^\dagger_{\mathbf{r}_e}$ and $b_{\mathbf{r}_h}$ are the electron and hole field operators, respectively.

Due to the finite sample size, there is a limited amount of quantum numbers to assign to each electron and hole composing the exciton. Moreover, two different fermions cannot be labeled by the same quantum numbers (Pauli exclusion principle), meaning that the single site cannot host an unlimited number of excitons. The site will saturate and exhibit a nonlinear optical response as the exciton density becomes higher and higher. In Eq.~\eqref{eqn:n-states} the effect of Pauli blockade appears in the normalization factor $\mathcal{F}^{(1)}_n$. We write the latter in the recursive form,
\begin{equation}\label{eqn:fn-emitter}
    \mathcal{F}^{(1)}_n=\sum_{m=1}^{n}\frac{(-1)^{m-1}}{n}\Big[\frac{n!}{(n-m)!}\Big]^2\sigma_m\mathcal{F}^{(1)}_{n-m},
\end{equation}
where we introduced the Pauli scattering terms $\sigma_m$ (see Appendix ~\ref{app:Pauli} and Refs.~\cite{Combescot2008,Laussy:PRB73(2006)}). The Pauli scattering is defined as an overlap between excitonic wavefunctions,
\begin{align}
    \sigma_m=&\prod_{i=1}^{m-1}\int d \mathbf{r}_{e_1}d\mathbf{r}_{h_1}\Psi^\ast_0(\mathbf{r}_{e_i},\mathbf{r}_{h_{i}})\Psi_0(\mathbf{r}_{e_i},\mathbf{r}_{h_{i+1}})\notag\\
    &\int d \mathbf{r}_{e_m}d\mathbf{r}_{h_m}\Psi^\ast_0(\mathbf{r}_{e_m},\mathbf{r}_{h_m})\Psi_0(\mathbf{r}_{e_m},\mathbf{r}_{h_1}). \label{eqn:sigma_n}
\end{align}
Intuitively, this term may be understood as the degree of bosonicity of the exciton. For the pure boson limit we have $\sigma_1=1$ and $\sigma_{m>1}=0$, leading to $\mathcal{F}_{n}^{(1)}=n!$ as a standard bosonic state normalization factor. In the Frenkel limit, we have $\sigma_m=1$ for all $m$, which yields $\mathcal{F}^{(1)}_{1}=1$ and $\mathcal{F}^{(1)}_{n>1}=0$. In between these two limits $\sigma_m$ is a monotonically decreasing function of $m$ with $\sigma_m\leq 1$. Away from the bosonic limit, the effect of nonlinear phase space filling is always present, thus lowering the effective probability for creating more excitons in confined regions. That is, $\mathcal{F}^{(1)}_n/n!$ is a monotonically decreasing function of the integer index $n$. The effect of the Pauli blockade can be seen as a modification of the transition matrix element $r_n$ defined through
\begin{equation}\label{eqn:rn-Pauli}
    r_n=|\langle n|x^\dagger_0|n-1\rangle|^2=\mathcal{F}^{(1)}_n/\mathcal{F}^{(1)}_{n-1}.
\end{equation}
Once $r_n$ is set for a single site, we can construct the $N_{\mathrm{s}}$-site of the polaritonic lattice by using Eq.~\eqref{eqn:FN} with $r_n$ specified in Eq.~\eqref{eqn:rn-Pauli}. In this case, calculating matrix elements in Eq.~\eqref{eqn:H_N} by using Eqs.~\eqref{eqn:FN} and \eqref{eqn:rn-Pauli} is not convenient, since the number of available excited states ($\ell$) may be large. Instead of using multinomial expansion, we derive the recursive formula for $\mathcal{F}^{(N_{\mathrm{s}})}_N$ (see Appendix~\ref{app:Pauli} for full derivation):
\begin{equation}\label{eqn:FN_recur}
    \mathcal{F}^{(N_{\mathrm{s}})}_N\!=\!\sum_{m=1}^{N}\frac{(-1)^{m-1}}{N}\left[\frac{N!}{(N-m)!}\right]^2 \frac{\sigma_m}{N_{\rm s}^{m-1}}\mathcal{F}^{(N_{\mathrm{s}})}_{N-m}.
\end{equation}
However, we note that when using Eq.~\eqref{eqn:FN_recur} for numerical calculation, one needs to ensure high accuracy for each terms in the sum due to large cancellations (we achieve this by developing a semi-symbolic treatment of expressions when only the final evaluation is numerical). As compared to the single site case in Eq.~\eqref{eqn:fn-emitter}, the multi-site result has a reduced Pauli scattering term $\sigma_m/N_{\rm s}^{m-1}$ which takes into account two contributing effects. One is the Pauli blockade that we have already described, relating the effect to the statistical properties of the particles. Second, we have the real space enhancement factor, meaning there are more sites to place our emitters.

\subsection{Coulomb blockade}

Next, we note that placing two excitons together may be impossible due to the fact that their constituents interact with each other via Coulomb potential. The enhancement of onsite exciton-exciton (X-X) Coulomb repulsion due to weak screening in two-dimensional structures translates into a strong Kerr shift for the excitonic mode. As the exciton energy is shifted out of resonance, this effectively reduces the transition matrix element between excited states, resulting in similar implications as the Pauli blockade. The main difference with the Pauli blockade is that the number of available excitations per site is sharply depleted at finite $\ell$. The cut-off value $\ell$ depends on the depth of localization potential and the cost of Coulomb energy by creating additional excitons at the same site. However, we note that the Coulomb blockade is a complex correlated problem since the onsite X-X Coulomb repulsion and the maximum number of excitons per site $\ell$ implicitly depend on the $n$-exciton rather than one-exciton wave function as in the Pauli blockade problem. Calculating the transition matrix element $r_n$ microscopically is beyond the scope of this paper, as it reduces to exclusively numerical modeling, while here we concentrate on predominantly analytical treatment. Nevertheless, our general theory in Eq.~\eqref{eqn:model} is still applicable, and we introduce the Coulomb blockade effect as the phenomenological reduction of the transition matrix element at increasing $n$. This may be modeled by
\begin{equation}\label{eqn:r_Coulomb}
    r_n\approx\begin{cases}n/(1+\delta^2_n),& n\leq\ell\\
    0,&\ell>n\end{cases},
\end{equation}
corresponding to the rate for driving the transition between energy levels with detuning $\delta_n = (n-1)U/\Omega_0$ (dimensionless) due to the Kerr shift. Here, $U$ is the strength of X-X interaction and $n$ is the number of excitons. In this case the Pauli statistics plays no role, where the exciton is considered as an elementary boson. However, we will see later that different blockade mechanisms eventually lead to similar qualitative results.  

Although the form of Coulomb blockade in Eq. \eqref{eqn:r_Coulomb} is considered for this paper, we remark that the physics of saturation effects due to X-X interaction is very rich. Particularly, the interaction depends on both the exciton’s wavefunction and the screened Coulomb potential which have important consequences for forming the $n$-exciton state $|n\rangle$. The details of $|n\rangle$ ultimately impact the scaling of the saturation effect. For instance, as we compare s-wave and p-wave exciton (Rydberg state), in the case of dipole-dipole interactions, the long-range part of the effective X-X interaction can set a blockade radius, thus effectively enhancing the saturation (we have now seen this in Cu$_2$O polaritons~\cite{Makhonin:LightSciAppl13(2024)}. For s-excitons, the exchange interaction is dominant and generally short-range. However, it depends heavily on the details of wavefunctions and this leads to a qualitatively different scaling from Rydberg state.


\section{Different regimes of Nonlinear Phase space filling}

Previously, we have derived a microscopic expression for light-matter coupling and nonlinear phase space filling in the general form, accounting for multiple localized excitonic sites, the spatial structure of excitons, and X-X interactions. Now, let us analyze the behavior of NPSF for some qualitatively distinct cases, in particular driven by a sample geometry. The size of excitons is described by the Bohr radius $a_{\mathrm{X}}$. This has to be compared with the exciton localization length $L$, defined by the lattice potential. Depending on the $L/a_{\mathrm{X}}$ ratio as well the number of sites $N_\mathrm{s}$, the NPSF contribution has different scaling, both in high and low occupation limits. We suggest the three regimes corresponding to: 1) planar; 2) fractured; and 3) ultralocalized NPSF.

We visualize the three sample geometries in Fig.~\ref{fig:moire}. The \textbf{planar} case corresponds to an exciton that is delocalized over the entire sample ($N_\mathrm{s}=1$; Fig.~\ref{fig:moire}, left). The saturation effect in this regime was studied before for Wannier excitons in III-V semiconductors~\cite{Tassone1999,Brichkin2011} and TMDC monolayers~\cite{Emmanuele2020,Shahnazaryan:PRB102(2020)}. In this case, the prior analysis was performed in the perturbative limit, where only the first nonlinear correction to $\Omega_N$ is derived as $\Omega_N \approx \Omega_0 \sqrt{N} (1 - \beta_1 N + \mathcal{O}[N^2])$, where the beta-factor $\beta_1=-1/(\Omega_0\sqrt{N})d\Omega_N/dN|_{N=0}$ depends on the ratio of exciton area to the total area~\cite{Tassone1999,Brichkin2011}, and the second-order correction was derived in Ref.~\cite{Emmanuele2020}.

Another regime corresponds to the opposite limit, where excitons are \textbf{ultralocalized} on the lattice of many sites, and the localization length is comparable to the size of the exciton (Fig.~\ref{fig:moire}, right). In this case, the exciton behaves like a Frenkel exciton~\cite{Combescot:EPJB68(2009)}, and we note that similar behavior can be attributed to trion polaritons~\cite{Kyriienko:PRL125(2020),Emmanuele2020}. The Rabi frequency of the ultralocalized case scales as a square root of the deviation from single occupation case.

Finally, we reveal that for the intermediate localized length, the \textbf{fractured} regime can be realized. In this case the analytical form of Rabi frequency is not known, and we will show that the NPSF scaling behavior of this regime is rather nontrivial. Namely, it is not a simple interpolating result between planar and ultralocalized cases. This may shed light on various polaritonic experiments where lattice sites can host multiple but finite number of excitons.

Let us now proceed with calculating the scaling of NPSF for the three regimes. To account for the effect of Pauli blockade we consider an excitonic wavefunction in the Gaussian form
\begin{equation}\label{eqn:Gaussian0}
    \Psi_0(\mathbf{r}_e,\mathbf{r}_h)\approx\frac{1}{\pi L a_{\mathrm{X}}}\mathrm{e}^{-\frac{1}{2}[(R/L)^2-(r/a_{\mathrm{X}})^2]} ,
\end{equation}
where $\bm{R}=\frac{m_{\mathrm{c}}}{M}\mathbf{r}_e+\frac{m_{\mathrm{v}}}{M}\mathbf{r}_h$ is the center of mass, and $\mathbf{r}=\mathbf{r}_e-\mathbf{r}_h$ is the relative position of an electron-hole pair. Here, $M=m_{\mathrm{c}} + m_{\mathrm{v}}$ is the mass of exciton, and $m_{\mathrm{c}}$ and $m_{\mathrm{v}}$ are the masses of electron and hole, respectively. We note that the similar ansatz was chosen for studying phase space filling of a quantum dot~\cite{Laussy:PRB73(2006)}, and that the Gaussian exciton shape is particularly suitable for TMDC heterostructures~\cite{Alexeev:Nature567(2019),Zhang:Nature591(2021),Louca2022} due to the quadratic scaling of an interlayer potential. We remind that $L$ is the localization length of the lattice potential and $a_{\mathrm{X}}$ defines the size of the exciton.  The exciton localization length can be characterized by the wavefunction spread in the exciton's center-of-mass motion which depends on the confinement potential. This can be dictated by an excitonic disorder, or induced by twisting in a TMDC bilayer forming the moir\'{e} potential~\cite{Tran:Nature567(2019),Tran:2DMat8(2020)} (visualized in Fig.~\ref{fig:moire}). The exciton Bohr radius $a_{\mathrm{X}}$ depends on the electron-hole screened interacting potential (defined as intralayer and interlayer potentials for TMDC bilayers). While other choices of wavefunction are possible, Gaussian shape allows evaluating many scattering terms efficiently, and we do not expect the shape modification to induce qualitative changes. 

\begin{figure}
    \centering
    \includegraphics[width=3.3in]{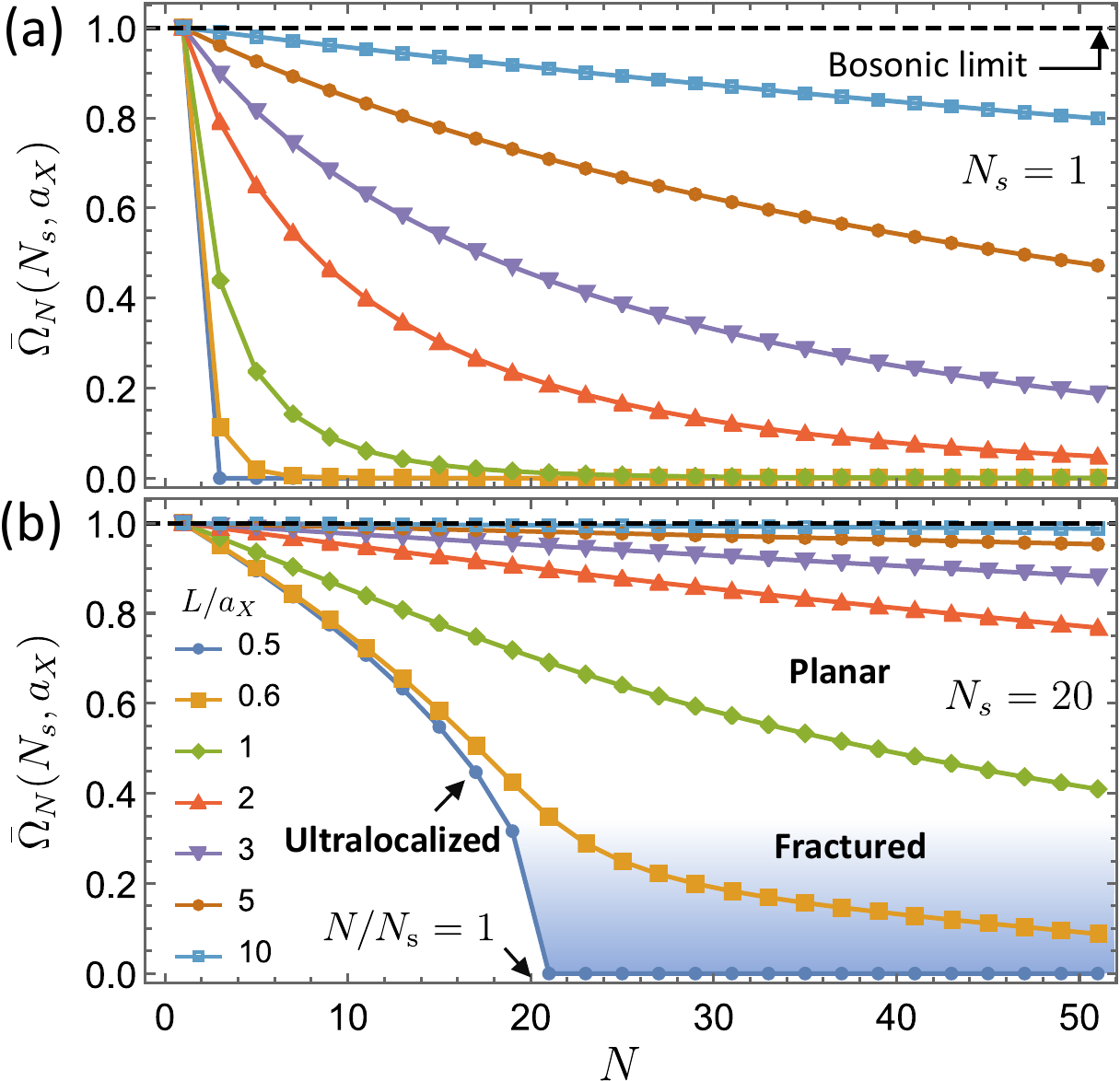}
    \caption{Pauli blockade and the
dependence. The dependence of the (renormalized) Rabi frequency $\Bar{\Omega}_N(N_{\rm s},a_{\mathrm{X}})= \Omega_N/\sqrt{N}$ on number of excitons $N$. (a) NPSF effect of a single localization site ($N_{\rm s}=1$) for different ratios of $L/a_\mathrm{X}$. $L$ is the exciton localized length, and the values are shown in the legend in (b). The black dashed line is the bosonic limit, where the exciton is treated as an elementary boson. (b) NPSF for $N_{\rm s}=20$ sites. We can see the transition from Frenkel ($L/a_{\mathrm{X}}=0.5$) to bosonic limit by changing the localized length $L$. Three qualitative regimes of the NPSF are labeled.  The planar regime smoothly transitions over to the ultralocalized regime via the fractured regime (blue-shaded region).}
    \label{fig:OmegaN}
\end{figure}

We proceed by evaluating Pauli scatterings for defining $r_n$. Substituting the wavefunction from Eq.~\eqref{eqn:Gaussian0} in Eq.~\eqref{eqn:sigma_n}, we obtain the analytic expression of $\sigma_n$. The compact expression reads (see details in Appendix \ref{app:sigma_n})
\begin{equation}\label{eqn:sn}
    \sigma_n=\prod_{k=0}^{n-1}\frac{(a_{\mathrm{X}}/L)^2}{(1\!+\!\gamma_c^2)(1\!+\!\gamma_v^2)-(1\!-\!\gamma_c\gamma_v)^2\cos^2(k\pi/n)},
\end{equation}
where $\gamma_{c,v}=(m_{c,v}/M)( a_{\mathrm{X}}/L)$~\cite{Laussy:PRB73(2006)}. We report the effects of the Pauli blockade in Fig.~\ref{fig:OmegaN}. We run the numerical simulation with various $L/a_{\mathrm{X}}$ ratios and $m_c=m_v$ by solving Eq.~\eqref{eqn:OmegaN}. In Fig.~\ref{fig:OmegaN}(a), we demonstrate the effects of saturation in the single site limit. This case corresponds to the single parabolic potential well with excitons that couple strongly to the cavity mode. In the limit of delocalized particles (light blue line), it approaches the ideal bosonic limit (gray dashed line), where NPSF is not present due to Bose statistics. The different curves and bullet points are summarized in the legend in Fig.~\ref{fig:OmegaN}(b). In the other limit of particles localized over the full sample size (dark blue curve with $L=0.5a_{\mathrm{X}}$, closest to the bottom), particles exhibit ``close-to-fermionic'' behavior,\footnote{Herein, the term fermionic is used in a loose sense, as the quasiparticles (excitons) do not exhibit anticommutation. Technically correct definition, albeit unusual, is a spin-1/2 or paulino statistics.} where the sample is saturated with more than one excitation. In the intermediate region, $0.5<L/a_{\mathrm{X}}<10$, one can see a smooth transition from bosonic-to-fermionic behavior. Also, the renormalized Rabi frequency in this intermediate region may be well approximated by the exponential decay with the exciton packing fraction in the sample $\eta=N(a_X/L)^2$,
\begin{equation}\label{eq:OmegaN-exp}
    \Omega_N\approx \Omega_0 \exp[-\tfrac{1}{2}(\eta v+\tfrac{1}{2}\eta^2v^2)+\mathcal{O}(\eta^3)
    ].
\end{equation} 
where we have used $\sigma_n\approx(\eta v)^{n-1}$ with $v=(1+\gamma_c^2)(1+\gamma_v^2)$ in Eq. \eqref{eqn:sn}.
Note that this result follows from the definition \eqref{eqn:OmegaN} and the exponential form of the normalization factors~\cite{Combescot2008,Combescot:EurPhysJB(2003)}, which is valid for very high orders in $\eta$. The exponential scaling in Eq.~\eqref{eq:OmegaN-exp} can be also recovered as a continuation of the diagrammatic expansion~\cite{Emmanuele2020}.

\begin{figure}
    \centering
    \includegraphics[width=3.3in]{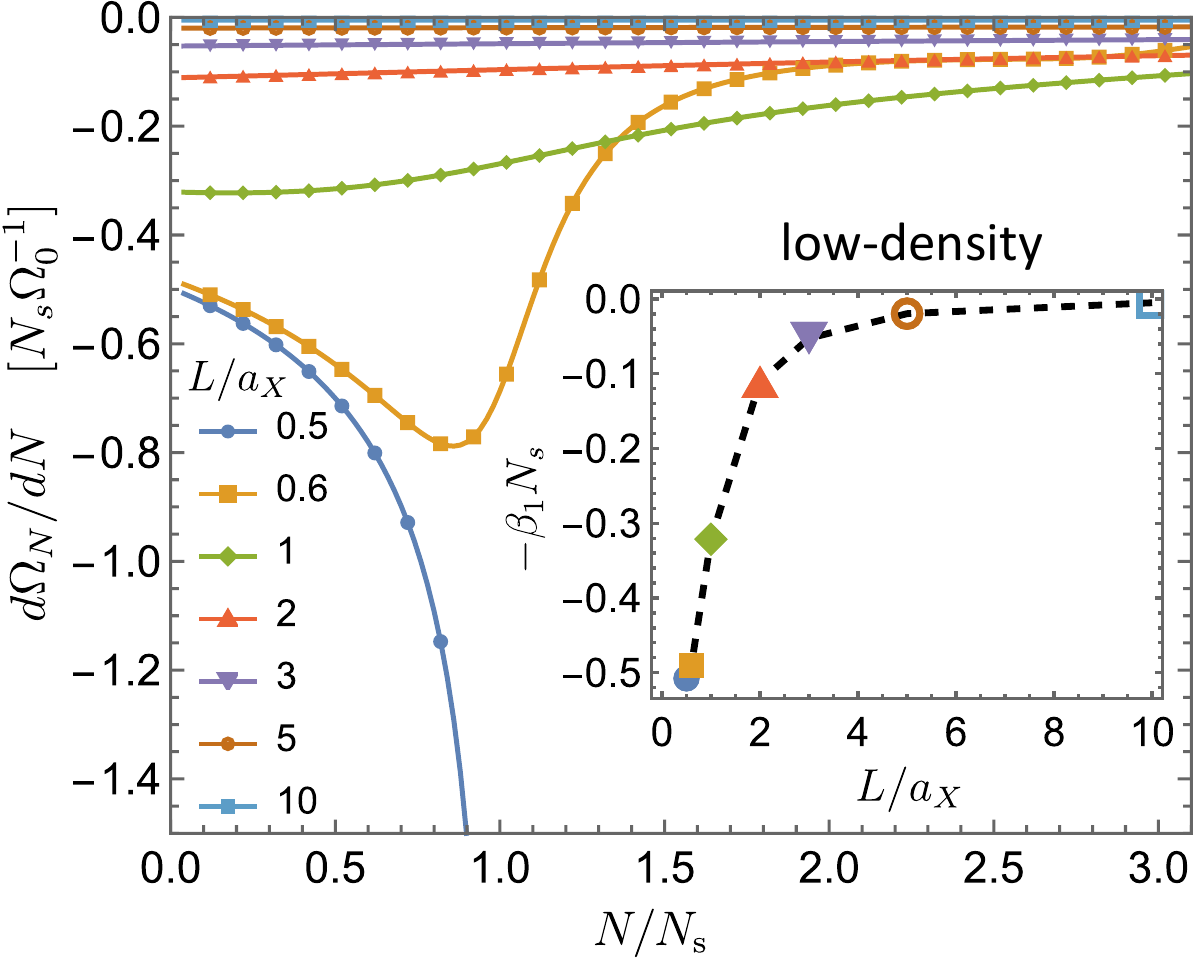}
    \caption{Saturation rate. Here we plot a rate of the Rabi splitting saturation, $d\Omega_N/d N$, as a function of exciton filling $N/N_{\mathrm{s}}$. (Inset) Saturation rate in a low-density regime ($\beta_1=-(\Omega_0\sqrt{N})^{-1}d\Omega_N/dN|_{N=0}$) with various ratios of localization length-to-exciton length, $L/a_X$.} 
    \label{fig:betaN}
\end{figure}

Next, we consider a lattice with many sites. As light couples to the lattice with $N_{\rm s} = 20$, the saturation effect exhibits a rather nontrivial behavior shown in Fig.~\ref{fig:OmegaN}(b). In this case, we observe similar behavior for large localization length with $L=10a_{\mathrm{X}}$ and recover the expected bosonic limit for point-like particles. As the exciton size increases ($L\simeq0.6a_{\mathrm{X}}$), we observe a kink in the proximity of $N \approx N_{\rm s}$ due to the transition from the low occupation regime with a fast saturation rate to the high occupation regime with a slower rate. This behavior may be understood as follows. As $N < N_{\rm s}$, the blockade of available phase-space is very effective which resembles Frenkel exciton (see the blue curve with $L=0.5a_{\mathrm{X}}$). At the point where $N = N_{\rm s}$, the $N$ excitons almost occupies the entire phase space with less available space. Therefore, the saturation due to the newly created excitations in $N>N_{\rm s}$ regime is less effective.  We can intuitively describe it as ``fill all sites by at least one exciton first, and only then fill the rest''.

To see the nontrivial changes in the saturation rate, in Fig.~\ref{fig:betaN} we plot a derivative of the Rabi frequency with respect to the number of particles, $d\Omega_N/dN$ (assuming $\Omega_N$ is a continuous function in $N$). We observe that in the planar regime (large $L/a_X$) the saturation rate is relatively small and constant in $N$. Once the exciton becomes localized with $L\leq a_X$, the saturation rate exhibits strong $N$-dependent behavior. In particular, in the fractured regime ($L/a_X\approx0.6$, yellow curve in Fig.~\ref{fig:betaN}), we can see clearly the transition of fast-to-slow saturation rate from low-occupation ($N<N_{\mathrm{s}}$) to high-occupation ($N>N_{\mathrm{s}}$) regime. In the dilute limit ($N\ll N_{\mathrm{s}}$), the saturation rate can be characterized by a constant $\beta_1$. As demonstrated in the inset of Fig.~\ref{fig:betaN}, this $\beta_1$ constant grows exponentially as the exciton localization length decreases.

As the localization length becomes smaller ($L=0.5a_{\mathrm{X}}$), the lattice localization eventually reaches the Frenkel limit~\cite{Combescot:EPJB68(2009)}. In this limit, the Rabi frequency can be calculated as
\begin{equation}
    \Omega_N=\Omega_0\sqrt{N_{\mathrm{s}}N} \sqrt{1- \frac{N-1}{N_{\mathrm{s}}}},
\end{equation}
which exhibits the square root scaling in $N$. We note that this expression follows from the normalization prefactor that is valid for all $N$ and $N_\mathrm{s}$, and includes nonperturbative corrections. We stress that the exponential and square root scalings are qualitatively distinct from the conjectured saturation dependence $\Omega_N = \Omega_0 / \sqrt{1 + N/N_\mathrm{s}}$ used in some studies~\cite{Datta2022,Zhang:Nature591(2021)}.

\begin{figure}
    \centering
    \includegraphics[width=3.3in]{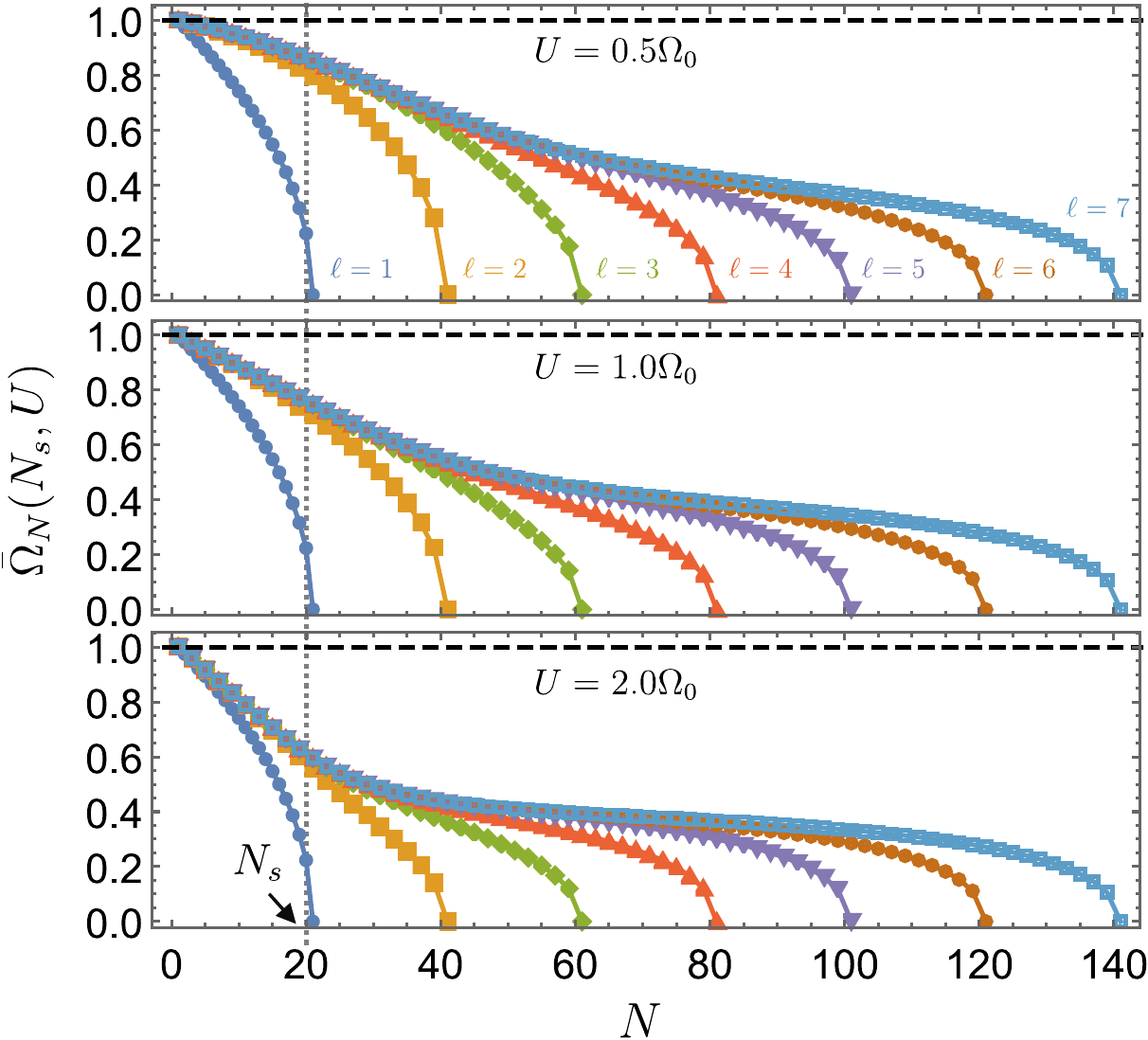}
    \caption{Coulomb blockade. The NPSF under the Coulomb blockade shown as a function of exciton occupation, for varying number of exciton states $\ell$ and X-X interaction strength $U$. We model the transition matrix element using Eq.~\eqref{eqn:r_Coulomb} with $\ell\leq 7$. Upper, middle and lower panels are plotted by using the different onsite X-X repulsion as $U/\Omega_0=0.5$, $U/\Omega_0=1$, and $U/\Omega_0=2$, respectively.}
    \label{fig:Coulomb}
\end{figure}

Now, let us add the Coulomb blockade. In this case, the number of excitons at each site is limited due to the finite confinement potential and Coulomb repulsion. Also, once many excitons occupy a single site, it becomes more difficult to drive the transition from $|n\rangle$ to $|n+1\rangle$ due to the Kerr shift of the excitonic mode. Using the blockade model in Eq.~\eqref{eqn:r_Coulomb}, we study the nonlinear phase space filling effect on the Rabi frequency. The results are shown in Fig.~\ref{fig:Coulomb}. Similar to the case of Pauli blockade (Fig.~\ref{fig:OmegaN}), the saturation curves exhibit similar behavior in each distinct regime. In this case, the fractured saturation may be realized as $\ell=2$, depending on the strength of Coulomb repulsion. The kink-like feature may not appear near $N\approx N_s$ if the Coulomb blockade is weak such that the $r_n$ does not decrease fast enough in large $n$, see Fig.~\ref{fig:Coulomb}. Moreover, the quench of Rabi frequency nearing the full saturation (Rabi collapse) behaves closer to the ultralocalized regime.

Summarizing the discussion in this section, the nonlinear phase space filling due to excitons localization can be divided qualitatively into three different regimes~(Fig. \ref{fig:moire}). Particularly, as the excitons started to localize at a different region of the lattice, NPSF enters the fractured regime. In this regime, the Rabi frequency of the polaritonic lattice is strongly renormalized. The phase space is saturated at a higher rate in the low ($N<N_{\rm s}$) density regime rather than in the high ($N>N_{\rm s}$) density regime, exhibiting an exponential tail.

\section{Saturation in TMDC moir\'{e} lattice}

In this section, we concentrate on drawing a connection between the proposed microscopic modeling of polariton lattice saturation with the recent exciting results for bilayers of TMDC. In the case of TMDC bilayers, the lattice ordering appears naturally, and we consider several possible contributing effects. Specifically, by twisting a TMDC bilayer one can drastically modify its optical properties~\cite{Wu:PRB97(2018),Tran:Nature567(2019),Tran:2DMat8(2020)}. One of the salient features of such a bilayer is the formation of modulating optical absorption with a moir\'{e} period. Additionally, the inhomogeneity due to the moir\'{e} pattern can also generate a potential landscape that results in the localization of excitons in a moir\'{e} cell~\cite{Tran:2DMat8(2020),RuizTijerina:PRB102(2020)}. Moreover, the variation of twisting angles and domain relaxation can lead to the formation of a lattice structure that is composed of smaller moir\'{e} patches. Therefore, twisted TMDC bilayers form an ideal system to realize near-perfect exciton-polaritonic lattices.
\begin{figure}
    \centering
    \includegraphics[width=3.3in]{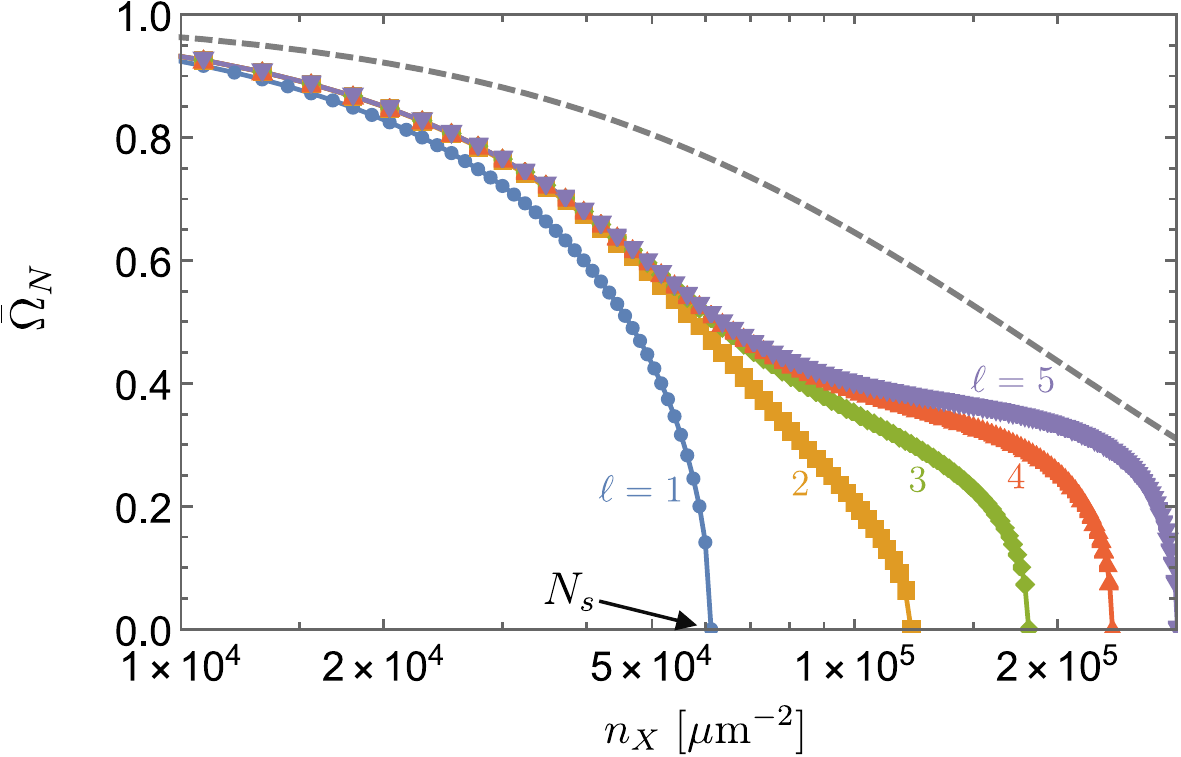}
    \caption{Moir\'{e} exciton saturation. The Rabi frequency renormalization is modeled by Pauli's blockade (gray dashed curve) and Coulomb blockade (colored solid curves) in the TMDC moir\'{e} lattice. We shown results as a function of exciton density $n_X$, and consider a varying number of excited levels $\ell = 1, 2, ..., 5$.}
    \label{fig:TMDC}
\end{figure}

moir\'{e} excitons in a twisted heterobilayer TMDC is the subject of current intense research~\cite{Alexeev:Nature567(2019),Tran:2DMat8(2020),Baek2020,Kremser2020,Andersen2021,Campbell2022}. In particular, the recent experiment~\cite{Zhang:Nature591(2021)} for the TMDC heterobilayers has found strong nonlinear saturation effects in moir\'{e} exciton-polariton. Experimental results suggest that there are several regimes of saturation where the saturation rate changes qualitatively. This was conjectured to originate from the Coulomb blockade due to the strong dipole-dipole interactions of moir\'{e} excitons, but a full explanation is missing.

We approach this system with the developed theory of NPSF. To investigate the optical nonlinearity of this twisted bilayer we use the lattice model in Eq.~\eqref{eqn:model}. Assuming $m_{\mathrm{c}}=m_{\mathrm{v}}$ and using $M=0.71~m_0$ from Ref.~\cite{Zhang:Nature591(2021)}, we plot the saturation curve for Rabi frequency due to Pauli blockade and compare it to Coulomb blockade with $\Omega_0=10$~meV and $U=30$~meV. Using the estimation of the moir\'{e} cells density in Ref.~\cite{Zhang:Nature591(2021)}, being $6\times 10^4 \mu\text{m}^{-2}$, this gives the estimated exciton density in the horizontal axis in Fig.~\ref{fig:TMDC}. As we can see, the Pauli blockade shows a rather smooth saturation in this bilayer system. In contrast to the Coulomb blockade, we can see a clear kink near $N\approx N_{\rm s}$. We see that this nontrivial Rabi frequency renormalization qualitatively resembles the saturation behavior observed in Ref.~\cite{Zhang:Nature591(2021)}. Namely, the saturation rate is fast in the low-density exciton regime where $N<N_{\rm s}$, and the rate reduces in the $N>N_{\rm s}$ case.

We note the kink-like feature in the saturation curve appears in the higher density as compared to the density estimated in the experimental study. One possible reason may be that our calculation assumes a perfect correlation of exciton density with the laser power. In reality, the relation between them may be complicated due to the presence of disorder such as residue strain and can be impacted by the lattice reconstruction~\cite{Tran:2DMat8(2020)}. This results in a smaller effective area of light-matter interaction. Furthermore, these effects also lead to the light interacting with the dark collective modes. Therefore, deducing the actual excitonic density in the sample may not be an easy task, since the number of excitons cannot be counted directly in the experiments. Nevertheless, keeping this as a hypothesis, we suggest that the experimentally observed nontrivial renormalization of Rabi frequency may be the manifestation of fractured NPSF which is facilitated by the Coulomb blockade.

\section{Conclusions}

In summary, we developed a non-perturbative microscopic theory for describing nonlinear optical effects arising from the phase space filling. While the developed quantum theory can be applied for many systems at strong coupling, we concentrate on the lattice geometries in the limit of strong NPSF (deep saturation). We unified different phase space filling mechanisms such as Pauli and Coulomb blockade by writing them in term of transition matrix elements between the excitations of an emitter, $r_n$. First, we described the effect of Pauli blockade arising from the finite system size and accounting for the exciton shape. This allows describing the change of NPSF from the planar quasi-bosonic case all the way to the case of two-level system saturation. We generalized the considerations to multiple sites, and studied the NPSF behavior between planar and ultralocalized (Frenkel) limit. Intriguingly, we find a distinct regime of fractured NPSF, where the Rabi frequency decrease has a kink-like feature at the number of sites, while the full saturation at high occupation reveals an exponential tail. This behavior is particularly pronounced in the presence of Coulomb blockade effects. Looking into specific examples, we analyzed the NPSF effects for moir\'{e} lattices of TMDC heterobilayers, suggesting that fractured NPSF may be relevant to the recent experimental observations of strong nonlinearity~\cite{Zhang:Nature591(2021)}.

Looking into the future, we note that the understanding of nonlinear phase space filling can help engineering moir\'{e} structures, or rather patterned samples, such that the nonlinearity is maximized. This will allow to push further the limits of quantum polaritonics~\cite{Munoz-Matutano2019,Kyriienko:PRL125(2020),Ghosh2020,Kavokin2022}. As for potential open questions, in our polaritonic lattice model we considered only the NPSF of a uniform collective excitation (bright state) [Eq.~\eqref{eqn:B}] coupled to cavity photons. However, cavity photons can couple to other non-uniform collective excitations (dark states) in the presence of lattice inhomogeneity arising from disorder, strain~\cite{Feierabend:PRB96(2017)}, lattice reconstruction~\cite{Weston:NatNano15(2020)}, and non-uniform cavity modes. This shall further enrich the NPSF behavior leading to unique nonlinear effects. Also, in this paper we did not take into account the tunneling between sites.  As indicated in Ref.~\cite{Zhang:Nature591(2021)}, the moir\'{e} band dispersion of the exciton center-of-mass motion is not very flat. This implies that excitons tunneling between localized regions in the moir\'{e} lattice may not be negligible and can be tunable. Another consideration in Ref. \cite{Estrecho2019} is that the light can induce changes in the exciton radius. This effect may add an extra contribution to the nonlinear saturation, and affect the Pauli and Coulomb blockade that we consider in this paper. We believe it is an interesting topic for future studies. It is also interesting to generalize our theoretical approach to investigate the saturation effect of moir\'{e} trions~\cite{Wang:NatNano16(2021),Liu:Nat594(2021),Brotons-GisbertPRX11(2021)}. Studying the trion-polariton NPSF in the lattice geometry and accounting for particle attraction~\cite{Song2022} is an important next step.

\begin{acknowledgments}
The authors are grateful to Prof. H. Deng for sharing the data and discussions on moiré heterostructures. We also thank I. Shelykh, D. Krizhanovskii, M. Richard, and A. Tartakovskii for valuable scientific discussions on the subject. We acknowl- edge the support from UK EPSRC Awards No. EP/X017222/1 and No. EP/V00171X/1, as well as NATO SPS Project No. MYP.G5860.  For the purpose of open access, the authors have applied a ‘Creative Commons Attribution (CC BY) licence to any Author Accepted Manuscript version arising from this submission, following the UKRI OA policy.
\end{acknowledgments}

\appendix

\section{Multinomial expansion of normalization factor}\label{app:FN}

In this section, we provide the details for the derivation of Eq.~\eqref{eqn:FN}. Using the commuting property of creation and annihilation operators $[X_i^\dagger,X_j]=0$ for $i\neq j$, we can carry out the multinomial expansion of $\hat{B}^\dagger$ as
\begin{equation*}
(\hat{B}^\dagger)^N=\sum_{n_1+\dots+n_{N_s}=N}\frac{N!N_{\mathrm{s}}^{N/2}}{n_1!\dots n_{N_{\mathrm{s}}}!}(X^\dagger_1)^{n_1}\dots(X^\dagger_{N_{\mathrm{s}}})^{n_{N_{\mathrm{s}}}}
\end{equation*}
where $n_i=0,1,2\dots$, and
\begin{equation*}
    (X_i^\dagger)^{n_i}\!=\!
    \mathbb{1}^{i-1} \otimes \left( \sum_{n=0}^{\ell-n_i-1}\!\sqrt{P(n_i)}|n+n_i+1\rangle\langle n| \right) \otimes  \mathbb{1}^{N_s-i}
\end{equation*}
with $P(n)=r_0\dots r_{n-1}$ [$P(0)\equiv1$]. Also, we note that $(X^{\dagger}_i)^n=0$ if $n>\ell$. Therefore, substituting the above into 
\begin{equation}
\mathcal{F}^{(N_{\mathrm{s}})}_N=\langle\varnothing |\hat{B}^{N}(\hat{B}^\dagger)^{N}|\varnothing\rangle,
\end{equation}
we obtain Eq.~\eqref{eqn:FN}.

\section{Microscopic derivation of many-body Pauli blockade}\label{app:Pauli}

To derive the Hamiltonian in Eq.~\eqref{eqn:model} microscopically, we may model the exciton wavefunction in a single site (emitter) as~\cite{RuizTijerina:PRB102(2020)}
\begin{equation}\label{eqn:Xeqn}
    \Big[\frac{\hat{\mathbf{Q}}^2}{2M}+\frac{\hat{\mathbf{p}}^2}{2\mu}+V_{M}(\mathbf{R})+V(\mathbf{r})\Big]\Psi(\mathbf{r}_e,\mathbf{r}_h)=\mathcal{E}\Psi(\mathbf{r}_e,\mathbf{r}_h),
\end{equation}
where $\Psi(\mathbf{r}_e,\mathbf{r}_h)$ with $\mathbf{r}_e$ and $\mathbf{r}_h$ being the positions of the electron and hole. The electron and hole masses are $m_e$ and $m_h$ with total mass $M=m_e+m_h$ and reduced mass $\mu=m_em_h/M$. The center-of-mass and relative position are $\mathbf{R}=(m_e\mathbf{r}_e+m_h\mathbf{r}_h)/M$ and $\mathbf{r}=\mathbf{r}_e-\mathbf{r}_h$ with their corresponding total momentum operator $\hat{\mathbf{Q}}$ and the relative momentum operator $\hat{\mathbf{p}}$. The exciton localized potential is $V_M(\mathbf{R})$ and the screened electron-hole interacting potential is $V(\mathbf{r}_e,\mathbf{r}_h)$. Solving Eq.~\eqref{eqn:Xeqn}, this gives the exciton creation operator of each site in Eq.~\eqref{eqn:x-composite}.

Once the excitonic field operator in Eq.~\eqref{eqn:x-composite} is specified by Eq.~\eqref{eqn:Xeqn}, we use the $n$-exciton states in Eq.~\eqref{eqn:n-states} as the emitter's excited states to construct $N_{\rm s}$-site Hilbert space with the (over) completeness relation~\cite{Combescot2008}
\begin{equation}\label{eqn:I}
    \mathbb{1}^{\otimes N_{\rm s}}=
    \bigotimes\limits_i^{N_{\rm s}}\sum_{n=0}|n\rangle\langle n|.
\end{equation}
Assuming that the light-matter interacting Hamiltonian of a single emitter as
\begin{equation}\label{eqn:Hint}
        \mathcal{H}_{int}=\sum_{i}^{N_s}\Omega_0c x^\dagger_{i0}+\rm{h.c},
\end{equation}
where the exciton in $i$-emitter is labeled by the subscript, we obtain Eq.~\eqref{eqn:model} by sandwiching \eqref{eqn:Hint} by the identity operator in Eq.~\eqref{eqn:I}. Namely, 
\begin{equation}
 X_i^\dagger=\mathbb{1}^{\otimes N_{\rm s}}x_{i0}^\dagger\mathbb{1}^{\otimes N_{\rm s}}.   
\end{equation}
This yields  
\begin{equation}
    \sqrt{r_{n}}=\langle n |x_0^\dagger|n-1\rangle=\sqrt{\mathcal{F}^{(1)}_{n}/\mathcal{F}^{(1)}_{n-1}}.
\end{equation}

To derive the transition matrix element for the Pauli blockade, we calculate
\begin{align*}
    \mathcal{F}^{(1)}_n&=\langle0|x^n_0(x^\dagger_0)^n|0\rangle\notag\\
    &=
    \langle 0|\Big[n \Delta_1x^{n-1}_0+\binom{n}{2}[x_0,\Delta_1](x_0)^{n-2}(x^\dagger_0)^{n-1}\Big]|0\rangle.
\end{align*}
Here, $\Delta_1=[x_0,x_0^\dagger]$ and we introduce the notation
\begin{equation}\label{eqn:Delta_n}
    \Delta_{n}=\begin{cases}[[x_0,\Delta_{n-1}],x^\dagger_0],& n~\text{is even}\\
    [x_0,[\Delta_{n-1},x^\dagger_0]],& n~\text{is odd}
    \end{cases}
\end{equation}
Using the fact that $[[\Delta_n,x^\dagger_0],x^\dagger_0]=0$ and $[[\Delta_n,x_0],x_0]=0$, and $\Delta_n |0\rangle=(-2)^{n-1}\sigma_n|0\rangle$, this leads to
\begin{align}\label{eqn:FN_x}
    \mathcal{F}^{(1)}_n
    &=
    n \sigma_1\mathcal{F}^{(1)}_{n-1}+\binom{n}{2}\langle 0|x_0^{n-2}[x_0,\Delta_1](x^\dagger_0)^{n-1}|0\rangle
\end{align}
We let  
\begin{equation}
    \Pi^{(n)}_m=\frac{\langle 0|x_0^{n-m-2}[x_0,\Delta_{m+1}](x^\dagger_0)^{n-m-1}|0\rangle}{2^{m+1}(n-m-1)[(n-m-2)!]^2},
\end{equation}
and rewrite Eq.\eqref{eqn:FN_x} into
\begin{align}\label{eqn:FN_x_2}
    \mathcal{F}^{(1)}_n
    &=
    \frac{n!^2}{n}\Big\{\frac{\sigma_1\mathcal{F}^{(1)}_{n-1}}{[(n-1)!]^2} +\Pi^{(n)}_0\Big\}.
\end{align}
To proceed further, we derive the iteration formula for $\Pi^{(n)}_m$, leading to
\begin{align}
   \Pi^{(n)}_0
    &=
   \frac{\binom{n-1}{1}(2\sigma_2)\mathcal{F}^{(1)}_{n-2}\!+\!\binom{n-1}{2}\langle 0|x_0^{n-2}[\Delta_2,x^\dagger_i](x^\dagger_0)^{n-3}|0\rangle}{2(n-1)[(n-2)!]^2}\notag\\
    &=
   \frac{-\sigma_2\mathcal{F}^{(1)}_{n-2}}{[(n-2)!]^2}+\frac{\langle 0|x_0^{n-2}[\Delta_2,x^\dagger_i](x^\dagger_0)^{n-3}|0\rangle}{2^2(n-2)[(n-3)!]^2}\notag\\
    &=
    \frac{-\sigma_2\mathcal{F}^{(1)}_{n-2}}{[(n-2)!]^2}+\frac{\binom{n-2}{1}2^2\sigma_3f_{n-3}}{2^2(n-2)[(n-3)!]^2}\notag\\
    &+\frac{\binom{n-2}{2}\langle 0|x_0^{n-4}[x_0,\Delta_3](x^\dagger_0)^{n-3}|0\rangle}{2^2(n-2)[(n-3)!]^2}\notag\\
    &=
    \frac{-\sigma_{2}\mathcal{F}^{(1)}_{n-2}}{[(n-2)!]^2}+\frac{\sigma_{3}\mathcal{F}^{(1)}_{n-3}}{[(n-3)!]^2}+\Pi^{(n)}_2\label{eqn:Pi_n}
\end{align}
Using this iterating formula and setting $\mathcal{F}^{(1)}_{N<0}=0$, we obtain Eq.~\eqref{eqn:fn-emitter} in the main text.

Similarly, we generalize the above calculation for the $N_{\rm s}$-site system. To calculate
\begin{equation}\label{eqn:FN_B}
    \mathcal{F}^{(N_{\mathrm{s}})}_N=\langle\varnothing|B^N(\hat{B}^\dagger)^N|\varnothing\rangle,
\end{equation}
we follow the same procedure as in calculating $\mathcal{F}^{(1)}_n$, and we define
\begin{equation}
    \tilde{\Delta}_{n}=\begin{cases}[[B,\tilde{\Delta}_{n-1}],\hat{B}^\dagger],& n~\text{is even}\\
    [B,[\tilde{\Delta}_{n-1},B^\dagger]],& n~\text{is odd}
    \end{cases},
\end{equation}
where $\tilde{\Delta}_1=\frac{1}{N_{\rm s}}\sum_{i=1}^{N_{\rm s}}\Delta^{(i)}_1$ with $\Delta^{(i)}_1=[X_i,X^\dagger_i]=[x_{i0},x^\dagger_{i0}]$ where $i=1,\dots N_{\rm s}$ is to label the exciton in each different emitter (where we ignore the identity operator $\mathbb{1}^{\otimes N_{\rm s}}$ in the commutators). It is not difficult to show that
\begin{equation}
    \tilde{\Delta}_n=\frac{1}{N_{\rm s}^n}\sum_{i=1}^{N_{\rm s}}\Delta^{(i)}_n .
\end{equation}
Using $\tilde{\Delta}_n|\varnothing\rangle=\sigma_n/N_{\rm s}^{n-1}|\varnothing\rangle$ and iterating Eq.~\eqref{eqn:FN_B} [same as Eq.~\eqref{eqn:FN_x} and Eq.~\eqref{eqn:Pi_n}], we obtain Eq.~\eqref{eqn:FN}.

\section{Calculation of $\sigma_n$}\label{app:sigma_n}

In this section, we explore the property of $\sigma_n$ for a single emitter (e.g. quantum dot \cite{Laussy:PRB73(2006)}) in Eq.~\eqref{eqn:Delta_n}. Here, we only focus on one emitter. Therefore, we drop the emitter index $i$ for simplicity. To calculate $\sigma_n$ Eq.~\eqref{eqn:Delta_n}, we use the commutation relation
\begin{equation}
    [x_\mu,x^\dagger_{\nu}]=\delta_{\mu\nu}+D_{\mu\nu},
\end{equation}
where $\nu$ is the quantum number of the exciton ($\nu=0$ being the ground state). The non-bosonicity is
\begin{align}
    D_{\mu\nu}=-\int_\mathbf{r}\Big[&\int_{\mathbf{r}_e}\int_{\mathbf{r}_e'}\Psi^\ast_{\mu}(\mathbf{r}_e,\mathbf{r})\Psi_{\nu}(\mathbf{r}_e',\mathbf{r})a^\dagger_{\mathbf{r}_e'}a_{\mathbf{r}_e}\notag\\
    &+\int_{\mathbf{r}_e}\int_{\mathbf{r}_e'}\Psi^\ast_{\mu}(\mathbf{r},\mathbf{r}_h)\Psi_{\nu}(\mathbf{r},\mathbf{r}_h')b^\dagger_{\mathbf{r}_h'}b_{\mathbf{r}_h}\Big].
\end{align}
Here, we use the notation $\int_{\mathbf{r}_1\dots\mathbf{r}_n}=\int d\mathbf{r}_1\dots d\mathbf{r}_n$. Using the completeness relation of the exciton wavefunction,
\begin{equation}    \sum_{\nu}\Psi^\ast_\nu(\mathbf{r}_e,\mathbf{r}_h)\Psi_\nu(\mathbf{r}_e',\mathbf{r}_h')=\delta_{\mathbf{r}_e,\mathbf{r}_e'}\delta_{\mathbf{r}_h,\mathbf{r}_h'},
\end{equation}
this gives
\begin{align}
    [D_{\mu\nu},x^\dagger_{\beta}]=&-\sum_{\alpha}2\Lambda^{\mu\alpha}_{\nu\beta}x^\dagger_{\alpha},\label{eqn:[D,x]}\\ [x_{\beta},D_{\mu\nu}]=&-\sum_{\alpha}2\Lambda^{\mu\alpha}_{\nu\beta}x_{\alpha},
\end{align}
where 
\begin{equation*}
    \Lambda^{\mu\alpha}_{\nu\beta}=\int_{\mathbf{r}_e\mathbf{r}_e'\mathbf{r}_h\mathbf{r}_h'}\Psi^\ast_{\mu}(\mathbf{r}_e,\mathbf{r}_h)\Psi^\ast_{\alpha}(\mathbf{r}_e',\mathbf{r}_h')\Psi_{\nu}(\mathbf{r}_e',\mathbf{r}_h)\Psi_{\beta}(\mathbf{r}_e,\mathbf{r}_h').
\end{equation*}
Therefore, using $a^\dagger_{\mathbf{r}}|0\rangle=b_{\mathbf{r}}|0\rangle=0$, we have
\begin{equation}
    \Delta_2|0\rangle
    =[-\sum_\beta2\Lambda^{00}_{0\beta}x_{\beta},x^\dagger_0]|0\rangle=-2\sigma_2|0\rangle
\end{equation}
with $\sigma_2=\Lambda^{00}_{00}$. Similarly, it is not difficult to show that
\begin{equation}
    \Delta_n|0\rangle=-2^{n-1}\sigma_n|0\rangle
\end{equation}
where $\sigma_1 = 1$, $\sigma_2 = \Lambda^{00}_{00}$, and 
\begin{align}
    \sigma_n =& \sum_{\nu_1}\dots\sum_{ \nu_{n-2}} \Lambda^{00}_{0\nu_1} \Lambda_{0,\nu_2}^{\nu_1,0} \cdots \Lambda^{ \nu_{n-2},0}_{ 0,0}, \quad n > 2. \label{eq:Sigma}
\end{align}

To do the integration in $\sigma_n$ analytically, we estimate the exciton wavefunction by using a Gaussian function. Then, the integration reduces to
\begin{align}
    \sigma_n=\int\prod_{i=1}^{n}\frac{ d \mathbf{r}_{e_i}d\mathbf{r}_{h_i}}{\pi  a_{\mathrm{X}}/L}\exp\left[-\mathbf{x}^T\mathbf{A}\mathbf{x}-\mathbf{y}^T\mathbf{A}\mathbf{y}\right],
\end{align}
where we have changed the variable $\mathbf{r}_{e_i}\to L \mathbf{r}_{e_i}$, $\mathbf{r}_{h_i}\to L\mathbf{r}_{h_i}$, and
\begin{align}
\mathbf{x}^T&=
    \begin{bmatrix}
    x_{h_1},x_{e_1},\dots,x_{h_N},x_{e_N}
    \end{bmatrix},\\
    \mathbf{y}^T&=
    \begin{bmatrix}
   y_{h_1},y_{e_1},\dots,y_{h_N},y_{e_N}
    \end{bmatrix},
\end{align}
and
\begin{equation}
    \mathbf{A}=\begin{bmatrix}
    \theta^{v}&\xi&0&0&\dots&\xi\\
    \xi&\theta^{c}&\xi&0&\dots&0\\
    0&\xi&\theta^{v}&\xi&\ddots&0\\
    \vdots&\vdots&\ddots&\ddots&\ddots&\vdots\\
    0&0&\cdots&\xi&\theta^v&\xi\\
    \xi&0&\cdots&0&\xi&\theta^c
    \end{bmatrix}.
\end{equation}
The eigenvalues of $\mathbf{A}$ can be obtained exactly by using a tight-binding approach (nearest-neighbor hopping with two sublattices). The eigenvalues are
\begin{equation*}
    \lambda^\pm_{k}=\frac{1}{2}\Big[(\theta^v+\theta^c)\pm\sqrt{(\theta^v-\theta^c)^2+16\xi^2\cos^2[k\pi/n]}\Big]
\end{equation*}
with $k=0,\dots,n-1$,
\begin{equation}
    \theta^{c,v}=\frac{m_{c,v}^2}{M^2}+\frac{L^2}{a_{\mathrm{X}}^2},\quad \xi=\frac{m_{\mathrm{c}}m_{\mathrm{v}}}{2M^2}-\frac{L^2}{2a_{\mathrm{X}}^2}.
\end{equation}
Finally, we integrate out $\mathbf{r}_{e_i}$ and $\mathbf{r}_{h_i}$. This gives
\begin{equation}
    \sigma_n=\left(\frac{L}{\pi^2 a_{\mathrm{X}} }\right)^{n}\frac{\pi^{2n}}{|\mathbf{A}|}=\prod_{k=0}^{n-1}\frac{L/a_{\mathrm{X}}}{\lambda^+_k\lambda^-_k}
\end{equation}
as a closed expression for the Pauli scatterings in the finite size system. 

\bibliography{Saturation}

\begin{thebibliography}{120}%
\makeatletter
\providecommand \@ifxundefined [1]{%
 \@ifx{#1\undefined}
}%
\providecommand \@ifnum [1]{%
 \ifnum #1\expandafter \@firstoftwo
 \else \expandafter \@secondoftwo
 \fi
}%
\providecommand \@ifx [1]{%
 \ifx #1\expandafter \@firstoftwo
 \else \expandafter \@secondoftwo
 \fi
}%
\providecommand \natexlab [1]{#1}%
\providecommand \enquote  [1]{``#1''}%
\providecommand \bibnamefont  [1]{#1}%
\providecommand \bibfnamefont [1]{#1}%
\providecommand \citenamefont [1]{#1}%
\providecommand \href@noop [0]{\@secondoftwo}%
\providecommand \href [0]{\begingroup \@sanitize@url \@href}%
\providecommand \@href[1]{\@@startlink{#1}\@@href}%
\providecommand \@@href[1]{\endgroup#1\@@endlink}%
\providecommand \@sanitize@url [0]{\catcode `\\12\catcode `\$12\catcode
  `\&12\catcode `\#12\catcode `\^12\catcode `\_12\catcode `\%12\relax}%
\providecommand \@@startlink[1]{}%
\providecommand \@@endlink[0]{}%
\providecommand \url  [0]{\begingroup\@sanitize@url \@url }%
\providecommand \@url [1]{\endgroup\@href {#1}{\urlprefix }}%
\providecommand \urlprefix  [0]{URL }%
\providecommand \Eprint [0]{\href }%
\providecommand \doibase [0]{https://doi.org/}%
\providecommand \selectlanguage [0]{\@gobble}%
\providecommand \bibinfo  [0]{\@secondoftwo}%
\providecommand \bibfield  [0]{\@secondoftwo}%
\providecommand \translation [1]{[#1]}%
\providecommand \BibitemOpen [0]{}%
\providecommand \bibitemStop [0]{}%
\providecommand \bibitemNoStop [0]{.\EOS\space}%
\providecommand \EOS [0]{\spacefactor3000\relax}%
\providecommand \BibitemShut  [1]{\csname bibitem#1\endcsname}%
\let\auto@bib@innerbib\@empty
\bibitem [{\citenamefont {Basov}\ \emph {et~al.}(2021)\citenamefont {Basov},
  \citenamefont {Asenjo-Garcia}, \citenamefont {Schuck}, \citenamefont {Zhu},\
  and\ \citenamefont {Rubio}}]{BasovAsenjo2021}%
  \BibitemOpen
  \bibfield  {author} {\bibinfo {author} {\bibfnamefont {D.~N.}\ \bibnamefont
  {Basov}}, \bibinfo {author} {\bibfnamefont {A.}~\bibnamefont
  {Asenjo-Garcia}}, \bibinfo {author} {\bibfnamefont {P.~J.}\ \bibnamefont
  {Schuck}}, \bibinfo {author} {\bibfnamefont {X.}~\bibnamefont {Zhu}},\ and\
  \bibinfo {author} {\bibfnamefont {A.}~\bibnamefont {Rubio}},\ }\bibfield
  {title} {\bibinfo {title} {Polariton panorama},\ }\href
  {https://doi.org/doi:10.1515/nanoph-2020-0449} {\bibfield  {journal}
  {\bibinfo  {journal} {Nanophotonics}\ }\textbf {\bibinfo {volume} {10}},\
  \bibinfo {pages} {549} (\bibinfo {year} {2021})}\BibitemShut {NoStop}%
\bibitem [{\citenamefont {Carusotto}\ and\ \citenamefont
  {Ciuti}(2013)}]{CarusottoCiuti2013}%
  \BibitemOpen
  \bibfield  {author} {\bibinfo {author} {\bibfnamefont {I.}~\bibnamefont
  {Carusotto}}\ and\ \bibinfo {author} {\bibfnamefont {C.}~\bibnamefont
  {Ciuti}},\ }\bibfield  {title} {\bibinfo {title} {Quantum fluids of light},\
  }\href {https://doi.org/10.1103/RevModPhys.85.299} {\bibfield  {journal}
  {\bibinfo  {journal} {Rev. Mod. Phys.}\ }\textbf {\bibinfo {volume} {85}},\
  \bibinfo {pages} {299} (\bibinfo {year} {2013})}\BibitemShut {NoStop}%
\bibitem [{\citenamefont {Deng}\ \emph {et~al.}(2010)\citenamefont {Deng},
  \citenamefont {Haug},\ and\ \citenamefont {Yamamoto}}]{Deng2010}%
  \BibitemOpen
  \bibfield  {author} {\bibinfo {author} {\bibfnamefont {H.}~\bibnamefont
  {Deng}}, \bibinfo {author} {\bibfnamefont {H.}~\bibnamefont {Haug}},\ and\
  \bibinfo {author} {\bibfnamefont {Y.}~\bibnamefont {Yamamoto}},\ }\bibfield
  {title} {\bibinfo {title} {Exciton-polariton bose-einstein condensation},\
  }\href {https://doi.org/10.1103/RevModPhys.82.1489} {\bibfield  {journal}
  {\bibinfo  {journal} {Rev. Mod. Phys.}\ }\textbf {\bibinfo {volume} {82}},\
  \bibinfo {pages} {1489} (\bibinfo {year} {2010})}\BibitemShut {NoStop}%
\bibitem [{\citenamefont {Liew}\ \emph {et~al.}(2011)\citenamefont {Liew},
  \citenamefont {Shelykh},\ and\ \citenamefont {Malpuech}}]{Liew2011}%
  \BibitemOpen
  \bibfield  {author} {\bibinfo {author} {\bibfnamefont {T.}~\bibnamefont
  {Liew}}, \bibinfo {author} {\bibfnamefont {I.}~\bibnamefont {Shelykh}},\ and\
  \bibinfo {author} {\bibfnamefont {G.}~\bibnamefont {Malpuech}},\ }\bibfield
  {title} {\bibinfo {title} {Polaritonic devices},\ }\href
  {https://doi.org/10.1016/j.physe.2011.04.003} {\bibfield  {journal} {\bibinfo
   {journal} {Physica E: Low-dimensional Systems and Nanostructures}\ }\textbf
  {\bibinfo {volume} {43}},\ \bibinfo {pages} {1543} (\bibinfo {year}
  {2011})}\BibitemShut {NoStop}%
\bibitem [{\citenamefont {Hammerer}\ \emph {et~al.}(2010)\citenamefont
  {Hammerer}, \citenamefont {S\o{}rensen},\ and\ \citenamefont
  {Polzik}}]{Hammerer2010}%
  \BibitemOpen
  \bibfield  {author} {\bibinfo {author} {\bibfnamefont {K.}~\bibnamefont
  {Hammerer}}, \bibinfo {author} {\bibfnamefont {A.~S.}\ \bibnamefont
  {S\o{}rensen}},\ and\ \bibinfo {author} {\bibfnamefont {E.~S.}\ \bibnamefont
  {Polzik}},\ }\bibfield  {title} {\bibinfo {title} {Quantum interface between
  light and atomic ensembles},\ }\href
  {https://doi.org/10.1103/RevModPhys.82.1041} {\bibfield  {journal} {\bibinfo
  {journal} {Rev. Mod. Phys.}\ }\textbf {\bibinfo {volume} {82}},\ \bibinfo
  {pages} {1041} (\bibinfo {year} {2010})}\BibitemShut {NoStop}%
\bibitem [{\citenamefont {Chang}\ \emph {et~al.}(2018)\citenamefont {Chang},
  \citenamefont {Douglas}, \citenamefont {Gonz\'alez-Tudela}, \citenamefont
  {Hung},\ and\ \citenamefont {Kimble}}]{Chang2018}%
  \BibitemOpen
  \bibfield  {author} {\bibinfo {author} {\bibfnamefont {D.~E.}\ \bibnamefont
  {Chang}}, \bibinfo {author} {\bibfnamefont {J.~S.}\ \bibnamefont {Douglas}},
  \bibinfo {author} {\bibfnamefont {A.}~\bibnamefont {Gonz\'alez-Tudela}},
  \bibinfo {author} {\bibfnamefont {C.-L.}\ \bibnamefont {Hung}},\ and\
  \bibinfo {author} {\bibfnamefont {H.~J.}\ \bibnamefont {Kimble}},\ }\bibfield
   {title} {\bibinfo {title} {Colloquium: Quantum matter built from nanoscopic
  lattices of atoms and photons},\ }\href
  {https://doi.org/10.1103/RevModPhys.90.031002} {\bibfield  {journal}
  {\bibinfo  {journal} {Rev. Mod. Phys.}\ }\textbf {\bibinfo {volume} {90}},\
  \bibinfo {pages} {031002} (\bibinfo {year} {2018})}\BibitemShut {NoStop}%
\bibitem [{\citenamefont {Fehler}\ \emph {et~al.}(2019)\citenamefont {Fehler},
  \citenamefont {Ovvyan}, \citenamefont {Gruhler}, \citenamefont {Pernice},\
  and\ \citenamefont {Kubanek}}]{Fehler2019}%
  \BibitemOpen
  \bibfield  {author} {\bibinfo {author} {\bibfnamefont {K.~G.}\ \bibnamefont
  {Fehler}}, \bibinfo {author} {\bibfnamefont {A.~P.}\ \bibnamefont {Ovvyan}},
  \bibinfo {author} {\bibfnamefont {N.}~\bibnamefont {Gruhler}}, \bibinfo
  {author} {\bibfnamefont {W.~H.~P.}\ \bibnamefont {Pernice}},\ and\ \bibinfo
  {author} {\bibfnamefont {A.}~\bibnamefont {Kubanek}},\ }\bibfield  {title}
  {\bibinfo {title} {Efficient coupling of an ensemble of nitrogen vacancy
  center to the mode of a high-q, si3n4 photonic crystal cavity},\ }\href
  {https://doi.org/10.1021/acsnano.9b01668} {\bibfield  {journal} {\bibinfo
  {journal} {ACS Nano}\ }\textbf {\bibinfo {volume} {13}},\ \bibinfo {pages}
  {6891} (\bibinfo {year} {2019})}\BibitemShut {NoStop}%
\bibitem [{\citenamefont {Radulaski}\ \emph {et~al.}(2019)\citenamefont
  {Radulaski}, \citenamefont {Zhang}, \citenamefont {Tzeng}, \citenamefont
  {Lagoudakis}, \citenamefont {Ishiwata}, \citenamefont {Dory}, \citenamefont
  {Fischer}, \citenamefont {Kelaita}, \citenamefont {Sun}, \citenamefont
  {Maurer}, \citenamefont {Alassaad}, \citenamefont {Ferro}, \citenamefont
  {Shen}, \citenamefont {Melosh}, \citenamefont {Chu},\ and\ \citenamefont
  {Vučković}}]{Radulaski2019}%
  \BibitemOpen
  \bibfield  {author} {\bibinfo {author} {\bibfnamefont {M.}~\bibnamefont
  {Radulaski}}, \bibinfo {author} {\bibfnamefont {J.~L.}\ \bibnamefont
  {Zhang}}, \bibinfo {author} {\bibfnamefont {Y.-K.}\ \bibnamefont {Tzeng}},
  \bibinfo {author} {\bibfnamefont {K.~G.}\ \bibnamefont {Lagoudakis}},
  \bibinfo {author} {\bibfnamefont {H.}~\bibnamefont {Ishiwata}}, \bibinfo
  {author} {\bibfnamefont {C.}~\bibnamefont {Dory}}, \bibinfo {author}
  {\bibfnamefont {K.~A.}\ \bibnamefont {Fischer}}, \bibinfo {author}
  {\bibfnamefont {Y.~A.}\ \bibnamefont {Kelaita}}, \bibinfo {author}
  {\bibfnamefont {S.}~\bibnamefont {Sun}}, \bibinfo {author} {\bibfnamefont
  {P.~C.}\ \bibnamefont {Maurer}}, \bibinfo {author} {\bibfnamefont
  {K.}~\bibnamefont {Alassaad}}, \bibinfo {author} {\bibfnamefont
  {G.}~\bibnamefont {Ferro}}, \bibinfo {author} {\bibfnamefont {Z.-X.}\
  \bibnamefont {Shen}}, \bibinfo {author} {\bibfnamefont {N.~A.}\ \bibnamefont
  {Melosh}}, \bibinfo {author} {\bibfnamefont {S.}~\bibnamefont {Chu}},\ and\
  \bibinfo {author} {\bibfnamefont {J.}~\bibnamefont {Vučković}},\ }\bibfield
   {title} {\bibinfo {title} {Nanodiamond integration with photonic devices},\
  }\href {https://doi.org/https://doi.org/10.1002/lpor.201800316} {\bibfield
  {journal} {\bibinfo  {journal} {Laser \& Photonics Reviews}\ }\textbf
  {\bibinfo {volume} {13}},\ \bibinfo {pages} {1800316} (\bibinfo {year}
  {2019})}\BibitemShut {NoStop}%
\bibitem [{\citenamefont {Eisenach}\ \emph {et~al.}(2021)\citenamefont
  {Eisenach}, \citenamefont {Barry}, \citenamefont {O'Keeffe}, \citenamefont
  {Schloss}, \citenamefont {Steinecker}, \citenamefont {Englund},\ and\
  \citenamefont {Braje}}]{Eisenach2021}%
  \BibitemOpen
  \bibfield  {author} {\bibinfo {author} {\bibfnamefont {E.~R.}\ \bibnamefont
  {Eisenach}}, \bibinfo {author} {\bibfnamefont {J.~F.}\ \bibnamefont {Barry}},
  \bibinfo {author} {\bibfnamefont {M.~F.}\ \bibnamefont {O'Keeffe}}, \bibinfo
  {author} {\bibfnamefont {J.~M.}\ \bibnamefont {Schloss}}, \bibinfo {author}
  {\bibfnamefont {M.~H.}\ \bibnamefont {Steinecker}}, \bibinfo {author}
  {\bibfnamefont {D.~R.}\ \bibnamefont {Englund}},\ and\ \bibinfo {author}
  {\bibfnamefont {D.~A.}\ \bibnamefont {Braje}},\ }\bibfield  {title} {\bibinfo
  {title} {Cavity-enhanced microwave readout of a solid-state spin sensor},\
  }\href {https://doi.org/10.1038/s41467-021-21256-7} {\bibfield  {journal}
  {\bibinfo  {journal} {Nature Communications}\ }\textbf {\bibinfo {volume}
  {12}},\ \bibinfo {pages} {1357} (\bibinfo {year} {2021})}\BibitemShut
  {NoStop}%
\bibitem [{\citenamefont {Diniz}\ \emph {et~al.}(2011)\citenamefont {Diniz},
  \citenamefont {Portolan}, \citenamefont {Ferreira}, \citenamefont {G\'erard},
  \citenamefont {Bertet},\ and\ \citenamefont {Auff\`eves}}]{Diniz2011}%
  \BibitemOpen
  \bibfield  {author} {\bibinfo {author} {\bibfnamefont {I.}~\bibnamefont
  {Diniz}}, \bibinfo {author} {\bibfnamefont {S.}~\bibnamefont {Portolan}},
  \bibinfo {author} {\bibfnamefont {R.}~\bibnamefont {Ferreira}}, \bibinfo
  {author} {\bibfnamefont {J.~M.}\ \bibnamefont {G\'erard}}, \bibinfo {author}
  {\bibfnamefont {P.}~\bibnamefont {Bertet}},\ and\ \bibinfo {author}
  {\bibfnamefont {A.}~\bibnamefont {Auff\`eves}},\ }\bibfield  {title}
  {\bibinfo {title} {Strongly coupling a cavity to inhomogeneous ensembles of
  emitters: Potential for long-lived solid-state quantum memories},\ }\href
  {https://doi.org/10.1103/PhysRevA.84.063810} {\bibfield  {journal} {\bibinfo
  {journal} {Phys. Rev. A}\ }\textbf {\bibinfo {volume} {84}},\ \bibinfo
  {pages} {063810} (\bibinfo {year} {2011})}\BibitemShut {NoStop}%
\bibitem [{\citenamefont {Trivedi}\ \emph {et~al.}(2019)\citenamefont
  {Trivedi}, \citenamefont {Radulaski}, \citenamefont {Fischer}, \citenamefont
  {Fan},\ and\ \citenamefont {Vu\ifmmode \check{c}\else
  \v{c}\fi{}kovi\ifmmode~\acute{c}\else \'{c}\fi{}}}]{Trivedi2019}%
  \BibitemOpen
  \bibfield  {author} {\bibinfo {author} {\bibfnamefont {R.}~\bibnamefont
  {Trivedi}}, \bibinfo {author} {\bibfnamefont {M.}~\bibnamefont {Radulaski}},
  \bibinfo {author} {\bibfnamefont {K.~A.}\ \bibnamefont {Fischer}}, \bibinfo
  {author} {\bibfnamefont {S.}~\bibnamefont {Fan}},\ and\ \bibinfo {author}
  {\bibfnamefont {J.}~\bibnamefont {Vu\ifmmode \check{c}\else
  \v{c}\fi{}kovi\ifmmode~\acute{c}\else \'{c}\fi{}}},\ }\bibfield  {title}
  {\bibinfo {title} {Photon blockade in weakly driven cavity quantum
  electrodynamics systems with many emitters},\ }\href
  {https://doi.org/10.1103/PhysRevLett.122.243602} {\bibfield  {journal}
  {\bibinfo  {journal} {Phys. Rev. Lett.}\ }\textbf {\bibinfo {volume} {122}},\
  \bibinfo {pages} {243602} (\bibinfo {year} {2019})}\BibitemShut {NoStop}%
\bibitem [{\citenamefont {Blais}\ \emph {et~al.}(2004)\citenamefont {Blais},
  \citenamefont {Huang}, \citenamefont {Wallraff}, \citenamefont {Girvin},\
  and\ \citenamefont {Schoelkopf}}]{Blais2004}%
  \BibitemOpen
  \bibfield  {author} {\bibinfo {author} {\bibfnamefont {A.}~\bibnamefont
  {Blais}}, \bibinfo {author} {\bibfnamefont {R.-S.}\ \bibnamefont {Huang}},
  \bibinfo {author} {\bibfnamefont {A.}~\bibnamefont {Wallraff}}, \bibinfo
  {author} {\bibfnamefont {S.~M.}\ \bibnamefont {Girvin}},\ and\ \bibinfo
  {author} {\bibfnamefont {R.~J.}\ \bibnamefont {Schoelkopf}},\ }\bibfield
  {title} {\bibinfo {title} {Cavity quantum electrodynamics for superconducting
  electrical circuits: An architecture for quantum computation},\ }\href
  {https://doi.org/10.1103/PhysRevA.69.062320} {\bibfield  {journal} {\bibinfo
  {journal} {Phys. Rev. A}\ }\textbf {\bibinfo {volume} {69}},\ \bibinfo
  {pages} {062320} (\bibinfo {year} {2004})}\BibitemShut {NoStop}%
\bibitem [{\citenamefont {S{\'a}nchez-Barquilla}\ \emph
  {et~al.}(2022)\citenamefont {S{\'a}nchez-Barquilla}, \citenamefont
  {Fern{\'a}ndez-Dom{\'i}nguez}, \citenamefont {Feist},\ and\ \citenamefont
  {Garc{\'i}a-Vidal}}]{Sanchez-Barquilla2022}%
  \BibitemOpen
  \bibfield  {author} {\bibinfo {author} {\bibfnamefont {M.}~\bibnamefont
  {S{\'a}nchez-Barquilla}}, \bibinfo {author} {\bibfnamefont {A.~I.}\
  \bibnamefont {Fern{\'a}ndez-Dom{\'i}nguez}}, \bibinfo {author} {\bibfnamefont
  {J.}~\bibnamefont {Feist}},\ and\ \bibinfo {author} {\bibfnamefont {F.~J.}\
  \bibnamefont {Garc{\'i}a-Vidal}},\ }\bibfield  {title} {\bibinfo {title} {A
  theoretical perspective on molecular polaritonics},\ }\href
  {https://doi.org/10.1021/acsphotonics.2c00048} {\bibfield  {journal}
  {\bibinfo  {journal} {ACS Photonics}\ }\textbf {\bibinfo {volume} {9}},\
  \bibinfo {pages} {1830} (\bibinfo {year} {2022})}\BibitemShut {NoStop}%
\bibitem [{\citenamefont {Wang}\ and\ \citenamefont {Yelin}(2021)}]{Wang2021}%
  \BibitemOpen
  \bibfield  {author} {\bibinfo {author} {\bibfnamefont {D.~S.}\ \bibnamefont
  {Wang}}\ and\ \bibinfo {author} {\bibfnamefont {S.~F.}\ \bibnamefont
  {Yelin}},\ }\bibfield  {title} {\bibinfo {title} {A roadmap toward the theory
  of vibrational polariton chemistry},\ }\href
  {https://doi.org/10.1021/acsphotonics.1c01028} {\bibfield  {journal}
  {\bibinfo  {journal} {ACS Photonics}\ }\textbf {\bibinfo {volume} {8}},\
  \bibinfo {pages} {2818} (\bibinfo {year} {2021})}\BibitemShut {NoStop}%
\bibitem [{\citenamefont {Yuen-Zhou}\ \emph {et~al.}(2022)\citenamefont
  {Yuen-Zhou}, \citenamefont {Xiong},\ and\ \citenamefont
  {Shegai}}]{Yuen-Zhou2022}%
  \BibitemOpen
  \bibfield  {author} {\bibinfo {author} {\bibfnamefont {J.}~\bibnamefont
  {Yuen-Zhou}}, \bibinfo {author} {\bibfnamefont {W.}~\bibnamefont {Xiong}},\
  and\ \bibinfo {author} {\bibfnamefont {T.}~\bibnamefont {Shegai}},\
  }\bibfield  {title} {\bibinfo {title} {Polariton chemistry: Molecules in
  cavities and plasmonic media},\ }\href {https://doi.org/10.1063/5.0080134}
  {\bibfield  {journal} {\bibinfo  {journal} {The Journal of Chemical Physics}\
  }\textbf {\bibinfo {volume} {156}},\ \bibinfo {pages} {030401} (\bibinfo
  {year} {2022})}\BibitemShut {NoStop}%
\bibitem [{\citenamefont {Brodbeck}\ \emph {et~al.}(2017)\citenamefont
  {Brodbeck}, \citenamefont {De~Liberato}, \citenamefont {Amthor},
  \citenamefont {Klaas}, \citenamefont {Kamp}, \citenamefont {Worschech},
  \citenamefont {Schneider},\ and\ \citenamefont {H\"ofling}}]{Brodbeck2017}%
  \BibitemOpen
  \bibfield  {author} {\bibinfo {author} {\bibfnamefont {S.}~\bibnamefont
  {Brodbeck}}, \bibinfo {author} {\bibfnamefont {S.}~\bibnamefont
  {De~Liberato}}, \bibinfo {author} {\bibfnamefont {M.}~\bibnamefont {Amthor}},
  \bibinfo {author} {\bibfnamefont {M.}~\bibnamefont {Klaas}}, \bibinfo
  {author} {\bibfnamefont {M.}~\bibnamefont {Kamp}}, \bibinfo {author}
  {\bibfnamefont {L.}~\bibnamefont {Worschech}}, \bibinfo {author}
  {\bibfnamefont {C.}~\bibnamefont {Schneider}},\ and\ \bibinfo {author}
  {\bibfnamefont {S.}~\bibnamefont {H\"ofling}},\ }\bibfield  {title} {\bibinfo
  {title} {Experimental verification of the very strong coupling regime in a
  gaas quantum well microcavity},\ }\href
  {https://doi.org/10.1103/PhysRevLett.119.027401} {\bibfield  {journal}
  {\bibinfo  {journal} {Phys. Rev. Lett.}\ }\textbf {\bibinfo {volume} {119}},\
  \bibinfo {pages} {027401} (\bibinfo {year} {2017})}\BibitemShut {NoStop}%
\bibitem [{\citenamefont {Ballarini}\ and\ \citenamefont
  {Liberato}(2019)}]{BallariniDeLiberato2019}%
  \BibitemOpen
  \bibfield  {author} {\bibinfo {author} {\bibfnamefont {D.}~\bibnamefont
  {Ballarini}}\ and\ \bibinfo {author} {\bibfnamefont {S.~D.}\ \bibnamefont
  {Liberato}},\ }\bibfield  {title} {\bibinfo {title} {Polaritonics: from
  microcavities to sub-wavelength confinement},\ }\href
  {https://doi.org/doi:10.1515/nanoph-2018-0188} {\bibfield  {journal}
  {\bibinfo  {journal} {Nanophotonics}\ }\textbf {\bibinfo {volume} {8}},\
  \bibinfo {pages} {641} (\bibinfo {year} {2019})}\BibitemShut {NoStop}%
\bibitem [{\citenamefont {Liu}\ \emph {et~al.}(2014)\citenamefont {Liu},
  \citenamefont {Galfsky}, \citenamefont {Sun}, \citenamefont {Xia},
  \citenamefont {chen Lin}, \citenamefont {Lee}, \citenamefont
  {K{\'{e}}na-Cohen},\ and\ \citenamefont {Menon}}]{liu2014}%
  \BibitemOpen
  \bibfield  {author} {\bibinfo {author} {\bibfnamefont {X.}~\bibnamefont
  {Liu}}, \bibinfo {author} {\bibfnamefont {T.}~\bibnamefont {Galfsky}},
  \bibinfo {author} {\bibfnamefont {Z.}~\bibnamefont {Sun}}, \bibinfo {author}
  {\bibfnamefont {F.}~\bibnamefont {Xia}}, \bibinfo {author} {\bibfnamefont
  {E.}~\bibnamefont {chen Lin}}, \bibinfo {author} {\bibfnamefont {Y.-H.}\
  \bibnamefont {Lee}}, \bibinfo {author} {\bibfnamefont {S.}~\bibnamefont
  {K{\'{e}}na-Cohen}},\ and\ \bibinfo {author} {\bibfnamefont {V.~M.}\
  \bibnamefont {Menon}},\ }\bibfield  {title} {\bibinfo {title} {Strong
  light{\textendash}matter coupling in two-dimensional atomic crystals},\
  }\href {https://doi.org/10.1038/nphoton.2014.304} {\bibfield  {journal}
  {\bibinfo  {journal} {Nature Photonics}\ }\textbf {\bibinfo {volume} {9}},\
  \bibinfo {pages} {30} (\bibinfo {year} {2014})}\BibitemShut {NoStop}%
\bibitem [{\citenamefont {Lundt}\ \emph {et~al.}(2016)\citenamefont {Lundt},
  \citenamefont {Mary{\'{n}}ski}, \citenamefont {Cherotchenko}, \citenamefont
  {Pant}, \citenamefont {Fan}, \citenamefont {Tongay}, \citenamefont {Sek},
  \citenamefont {Kavokin}, \citenamefont {H\"{o}fling},\ and\ \citenamefont
  {Schneider}}]{lundt2016}%
  \BibitemOpen
  \bibfield  {author} {\bibinfo {author} {\bibfnamefont {N.}~\bibnamefont
  {Lundt}}, \bibinfo {author} {\bibfnamefont {A.}~\bibnamefont
  {Mary{\'{n}}ski}}, \bibinfo {author} {\bibfnamefont {E.}~\bibnamefont
  {Cherotchenko}}, \bibinfo {author} {\bibfnamefont {A.}~\bibnamefont {Pant}},
  \bibinfo {author} {\bibfnamefont {X.}~\bibnamefont {Fan}}, \bibinfo {author}
  {\bibfnamefont {S.}~\bibnamefont {Tongay}}, \bibinfo {author} {\bibfnamefont
  {G.}~\bibnamefont {Sek}}, \bibinfo {author} {\bibfnamefont {A.~V.}\
  \bibnamefont {Kavokin}}, \bibinfo {author} {\bibfnamefont {S.}~\bibnamefont
  {H\"{o}fling}},\ and\ \bibinfo {author} {\bibfnamefont {C.}~\bibnamefont
  {Schneider}},\ }\bibfield  {title} {\bibinfo {title} {Monolayered {MoSe} 2 :
  a candidate for room temperature polaritonics},\ }\href
  {https://doi.org/10.1088/2053-1583/4/1/015006} {\bibfield  {journal}
  {\bibinfo  {journal} {2D Materials}\ }\textbf {\bibinfo {volume} {4}},\
  \bibinfo {pages} {015006} (\bibinfo {year} {2016})}\BibitemShut {NoStop}%
\bibitem [{\citenamefont {Sidler}\ \emph {et~al.}(2016)\citenamefont {Sidler},
  \citenamefont {Back}, \citenamefont {Cotlet}, \citenamefont {Srivastava},
  \citenamefont {Fink}, \citenamefont {Kroner}, \citenamefont {Demler},\ and\
  \citenamefont {Imamoglu}}]{Sidler:NatPhys13(2016)}%
  \BibitemOpen
  \bibfield  {author} {\bibinfo {author} {\bibfnamefont {M.}~\bibnamefont
  {Sidler}}, \bibinfo {author} {\bibfnamefont {P.}~\bibnamefont {Back}},
  \bibinfo {author} {\bibfnamefont {O.}~\bibnamefont {Cotlet}}, \bibinfo
  {author} {\bibfnamefont {A.}~\bibnamefont {Srivastava}}, \bibinfo {author}
  {\bibfnamefont {T.}~\bibnamefont {Fink}}, \bibinfo {author} {\bibfnamefont
  {M.}~\bibnamefont {Kroner}}, \bibinfo {author} {\bibfnamefont
  {E.}~\bibnamefont {Demler}},\ and\ \bibinfo {author} {\bibfnamefont
  {A.}~\bibnamefont {Imamoglu}},\ }\bibfield  {title} {\bibinfo {title} {Fermi
  polaron-polaritons in charge-tunable atomically thin semiconductors},\ }\href
  {https://doi.org/10.1038/nphys3949} {\bibfield  {journal} {\bibinfo
  {journal} {Nat. Phys.}\ }\textbf {\bibinfo {volume} {13}},\ \bibinfo {pages}
  {255} (\bibinfo {year} {2016})}\BibitemShut {NoStop}%
\bibitem [{\citenamefont {Dufferwiel}\ \emph {et~al.}(2017)\citenamefont
  {Dufferwiel}, \citenamefont {Lyons}, \citenamefont {Solnyshkov},
  \citenamefont {Trichet}, \citenamefont {Withers}, \citenamefont {Schwarz},
  \citenamefont {Malpuech}, \citenamefont {Smith}, \citenamefont {Novoselov},
  \citenamefont {Skolnick}, \citenamefont {Krizhanovskii},\ and\ \citenamefont
  {Tartakovskii}}]{Dufferwiel:NatPhoto11(2017)}%
  \BibitemOpen
  \bibfield  {author} {\bibinfo {author} {\bibfnamefont {S.}~\bibnamefont
  {Dufferwiel}}, \bibinfo {author} {\bibfnamefont {T.~P.}\ \bibnamefont
  {Lyons}}, \bibinfo {author} {\bibfnamefont {D.~D.}\ \bibnamefont
  {Solnyshkov}}, \bibinfo {author} {\bibfnamefont {A.~A.~P.}\ \bibnamefont
  {Trichet}}, \bibinfo {author} {\bibfnamefont {F.}~\bibnamefont {Withers}},
  \bibinfo {author} {\bibfnamefont {S.}~\bibnamefont {Schwarz}}, \bibinfo
  {author} {\bibfnamefont {G.}~\bibnamefont {Malpuech}}, \bibinfo {author}
  {\bibfnamefont {J.~M.}\ \bibnamefont {Smith}}, \bibinfo {author}
  {\bibfnamefont {K.~S.}\ \bibnamefont {Novoselov}}, \bibinfo {author}
  {\bibfnamefont {M.~S.}\ \bibnamefont {Skolnick}}, \bibinfo {author}
  {\bibfnamefont {D.~N.}\ \bibnamefont {Krizhanovskii}},\ and\ \bibinfo
  {author} {\bibfnamefont {A.~I.}\ \bibnamefont {Tartakovskii}},\ }\bibfield
  {title} {\bibinfo {title} {Valley-addressable polaritons in atomically thin
  semiconductors},\ }\href {https://doi.org/10.1038/nphoton.2017.125}
  {\bibfield  {journal} {\bibinfo  {journal} {Nat. Photon.}\ }\textbf {\bibinfo
  {volume} {11}},\ \bibinfo {pages} {497} (\bibinfo {year} {2017})}\BibitemShut
  {NoStop}%
\bibitem [{\citenamefont {Schneider}\ \emph {et~al.}(2018)\citenamefont
  {Schneider}, \citenamefont {Glazov}, \citenamefont {Korn}, \citenamefont
  {H{\"o}fling},\ and\ \citenamefont {Urbaszek}}]{Schneider2018}%
  \BibitemOpen
  \bibfield  {author} {\bibinfo {author} {\bibfnamefont {C.}~\bibnamefont
  {Schneider}}, \bibinfo {author} {\bibfnamefont {M.~M.}\ \bibnamefont
  {Glazov}}, \bibinfo {author} {\bibfnamefont {T.}~\bibnamefont {Korn}},
  \bibinfo {author} {\bibfnamefont {S.}~\bibnamefont {H{\"o}fling}},\ and\
  \bibinfo {author} {\bibfnamefont {B.}~\bibnamefont {Urbaszek}},\ }\bibfield
  {title} {\bibinfo {title} {Two-dimensional semiconductors in the regime of
  strong light-matter coupling},\ }\href
  {https://doi.org/10.1038/s41467-018-04866-6} {\bibfield  {journal} {\bibinfo
  {journal} {Nat. Commun.}\ }\textbf {\bibinfo {volume} {9}},\ \bibinfo {pages}
  {2695} (\bibinfo {year} {2018})}\BibitemShut {NoStop}%
\bibitem [{\citenamefont {Wang}\ \emph {et~al.}(2018)\citenamefont {Wang},
  \citenamefont {Chernikov}, \citenamefont {Glazov}, \citenamefont {Heinz},
  \citenamefont {Marie}, \citenamefont {Amand},\ and\ \citenamefont
  {Urbaszek}}]{Wang:RMP90(2018)}%
  \BibitemOpen
  \bibfield  {author} {\bibinfo {author} {\bibfnamefont {G.}~\bibnamefont
  {Wang}}, \bibinfo {author} {\bibfnamefont {A.}~\bibnamefont {Chernikov}},
  \bibinfo {author} {\bibfnamefont {M.~M.}\ \bibnamefont {Glazov}}, \bibinfo
  {author} {\bibfnamefont {T.~F.}\ \bibnamefont {Heinz}}, \bibinfo {author}
  {\bibfnamefont {X.}~\bibnamefont {Marie}}, \bibinfo {author} {\bibfnamefont
  {T.}~\bibnamefont {Amand}},\ and\ \bibinfo {author} {\bibfnamefont
  {B.}~\bibnamefont {Urbaszek}},\ }\bibfield  {title} {\bibinfo {title}
  {Colloquium : Excitons in atomically thin transition metal dichalcogenides},\
  }\href {https://doi.org/10.1103/revmodphys.90.021001} {\bibfield  {journal}
  {\bibinfo  {journal} {Reviews of Modern Physics}\ }\textbf {\bibinfo {volume}
  {90}},\ \bibinfo {pages} {021001} (\bibinfo {year} {2018})}\BibitemShut
  {NoStop}%
\bibitem [{\citenamefont {Emmanuele}\ \emph {et~al.}(2020)\citenamefont
  {Emmanuele}, \citenamefont {Sich}, \citenamefont {Kyriienko}, \citenamefont
  {Shahnazaryan}, \citenamefont {Withers}, \citenamefont {Catanzaro},
  \citenamefont {Walker}, \citenamefont {Benimetskiy}, \citenamefont
  {Skolnick}, \citenamefont {Tartakovskii}, \citenamefont {Shelykh},\ and\
  \citenamefont {Krizhanovskii}}]{Emmanuele2020}%
  \BibitemOpen
  \bibfield  {author} {\bibinfo {author} {\bibfnamefont {R.~P.~A.}\
  \bibnamefont {Emmanuele}}, \bibinfo {author} {\bibfnamefont {M.}~\bibnamefont
  {Sich}}, \bibinfo {author} {\bibfnamefont {O.}~\bibnamefont {Kyriienko}},
  \bibinfo {author} {\bibfnamefont {V.}~\bibnamefont {Shahnazaryan}}, \bibinfo
  {author} {\bibfnamefont {F.}~\bibnamefont {Withers}}, \bibinfo {author}
  {\bibfnamefont {A.}~\bibnamefont {Catanzaro}}, \bibinfo {author}
  {\bibfnamefont {P.~M.}\ \bibnamefont {Walker}}, \bibinfo {author}
  {\bibfnamefont {F.~A.}\ \bibnamefont {Benimetskiy}}, \bibinfo {author}
  {\bibfnamefont {M.~S.}\ \bibnamefont {Skolnick}}, \bibinfo {author}
  {\bibfnamefont {A.~I.}\ \bibnamefont {Tartakovskii}}, \bibinfo {author}
  {\bibfnamefont {I.~A.}\ \bibnamefont {Shelykh}},\ and\ \bibinfo {author}
  {\bibfnamefont {D.~N.}\ \bibnamefont {Krizhanovskii}},\ }\bibfield  {title}
  {\bibinfo {title} {Highly nonlinear trion-polaritons in a monolayer
  semiconductor},\ }\bibfield  {journal} {\bibinfo  {journal} {Nat. Commun.}\
  }\textbf {\bibinfo {volume} {11}},\ \href
  {https://doi.org/10.1038/s41467-020-17340-z} {10.1038/s41467-020-17340-z}
  (\bibinfo {year} {2020})\BibitemShut {NoStop}%
\bibitem [{\citenamefont {Lackner}\ \emph {et~al.}(2021)\citenamefont
  {Lackner}, \citenamefont {Dusel}, \citenamefont {Egorov}, \citenamefont
  {Han}, \citenamefont {Knopf}, \citenamefont {Eilenberger}, \citenamefont
  {Schr{\"o}der}, \citenamefont {Watanabe}, \citenamefont {Taniguchi},
  \citenamefont {Tongay}, \citenamefont {Anton-Solanas}, \citenamefont
  {H{\"o}fling},\ and\ \citenamefont {Schneider}}]{Lackner2021}%
  \BibitemOpen
  \bibfield  {author} {\bibinfo {author} {\bibfnamefont {L.}~\bibnamefont
  {Lackner}}, \bibinfo {author} {\bibfnamefont {M.}~\bibnamefont {Dusel}},
  \bibinfo {author} {\bibfnamefont {O.~A.}\ \bibnamefont {Egorov}}, \bibinfo
  {author} {\bibfnamefont {B.}~\bibnamefont {Han}}, \bibinfo {author}
  {\bibfnamefont {H.}~\bibnamefont {Knopf}}, \bibinfo {author} {\bibfnamefont
  {F.}~\bibnamefont {Eilenberger}}, \bibinfo {author} {\bibfnamefont
  {S.}~\bibnamefont {Schr{\"o}der}}, \bibinfo {author} {\bibfnamefont
  {K.}~\bibnamefont {Watanabe}}, \bibinfo {author} {\bibfnamefont
  {T.}~\bibnamefont {Taniguchi}}, \bibinfo {author} {\bibfnamefont
  {S.}~\bibnamefont {Tongay}}, \bibinfo {author} {\bibfnamefont
  {C.}~\bibnamefont {Anton-Solanas}}, \bibinfo {author} {\bibfnamefont
  {S.}~\bibnamefont {H{\"o}fling}},\ and\ \bibinfo {author} {\bibfnamefont
  {C.}~\bibnamefont {Schneider}},\ }\bibfield  {title} {\bibinfo {title}
  {Tunable exciton-polaritons emerging from ws2 monolayer excitons in a
  photonic lattice at room temperature},\ }\href
  {https://doi.org/10.1038/s41467-021-24925-9} {\bibfield  {journal} {\bibinfo
  {journal} {Nat. Comm.}\ }\textbf {\bibinfo {volume} {12}},\ \bibinfo {pages}
  {4933} (\bibinfo {year} {2021})}\BibitemShut {NoStop}%
\bibitem [{\citenamefont {Anton-Solanas}\ \emph {et~al.}(2021)\citenamefont
  {Anton-Solanas}, \citenamefont {Waldherr}, \citenamefont {Klaas},
  \citenamefont {Suchomel}, \citenamefont {Harder}, \citenamefont {Cai},
  \citenamefont {Sedov}, \citenamefont {Klembt}, \citenamefont {Kavokin},
  \citenamefont {Tongay}, \citenamefont {Watanabe}, \citenamefont {Taniguchi},
  \citenamefont {H{\"o}fling},\ and\ \citenamefont
  {Schneider}}]{Anton-Solanas2021}%
  \BibitemOpen
  \bibfield  {author} {\bibinfo {author} {\bibfnamefont {C.}~\bibnamefont
  {Anton-Solanas}}, \bibinfo {author} {\bibfnamefont {M.}~\bibnamefont
  {Waldherr}}, \bibinfo {author} {\bibfnamefont {M.}~\bibnamefont {Klaas}},
  \bibinfo {author} {\bibfnamefont {H.}~\bibnamefont {Suchomel}}, \bibinfo
  {author} {\bibfnamefont {T.~H.}\ \bibnamefont {Harder}}, \bibinfo {author}
  {\bibfnamefont {H.}~\bibnamefont {Cai}}, \bibinfo {author} {\bibfnamefont
  {E.}~\bibnamefont {Sedov}}, \bibinfo {author} {\bibfnamefont
  {S.}~\bibnamefont {Klembt}}, \bibinfo {author} {\bibfnamefont {A.~V.}\
  \bibnamefont {Kavokin}}, \bibinfo {author} {\bibfnamefont {S.}~\bibnamefont
  {Tongay}}, \bibinfo {author} {\bibfnamefont {K.}~\bibnamefont {Watanabe}},
  \bibinfo {author} {\bibfnamefont {T.}~\bibnamefont {Taniguchi}}, \bibinfo
  {author} {\bibfnamefont {S.}~\bibnamefont {H{\"o}fling}},\ and\ \bibinfo
  {author} {\bibfnamefont {C.}~\bibnamefont {Schneider}},\ }\bibfield  {title}
  {\bibinfo {title} {Bosonic condensation of exciton--polaritons in an
  atomically thin crystal},\ }\href
  {https://doi.org/10.1038/s41563-021-01000-8} {\bibfield  {journal} {\bibinfo
  {journal} {Nat. Mater.}\ }\textbf {\bibinfo {volume} {20}},\ \bibinfo {pages}
  {1233} (\bibinfo {year} {2021})}\BibitemShut {NoStop}%
\bibitem [{\citenamefont {Zhumagulov}\ \emph {et~al.}(2022)\citenamefont
  {Zhumagulov}, \citenamefont {Chiavazzo}, \citenamefont {Gulevich},
  \citenamefont {Perebeinos}, \citenamefont {Shelykh},\ and\ \citenamefont
  {Kyriienko}}]{Zhumagulov2022}%
  \BibitemOpen
  \bibfield  {author} {\bibinfo {author} {\bibfnamefont {Y.~V.}\ \bibnamefont
  {Zhumagulov}}, \bibinfo {author} {\bibfnamefont {S.}~\bibnamefont
  {Chiavazzo}}, \bibinfo {author} {\bibfnamefont {D.~R.}\ \bibnamefont
  {Gulevich}}, \bibinfo {author} {\bibfnamefont {V.}~\bibnamefont
  {Perebeinos}}, \bibinfo {author} {\bibfnamefont {I.~A.}\ \bibnamefont
  {Shelykh}},\ and\ \bibinfo {author} {\bibfnamefont {O.}~\bibnamefont
  {Kyriienko}},\ }\bibfield  {title} {\bibinfo {title} {Microscopic theory of
  exciton and trion polaritons in doped monolayers of transition metal
  dichalcogenides},\ }\href@noop {} {\bibfield  {journal} {\bibinfo  {journal}
  {npj Computational Materials}\ }\textbf {\bibinfo {volume} {8}},\ \bibinfo
  {pages} {1} (\bibinfo {year} {2022})}\BibitemShut {NoStop}%
\bibitem [{\citenamefont {Chang}\ \emph {et~al.}(2014)\citenamefont {Chang},
  \citenamefont {Vuleti{\'{c}}},\ and\ \citenamefont {Lukin}}]{Chang2014}%
  \BibitemOpen
  \bibfield  {author} {\bibinfo {author} {\bibfnamefont {D.~E.}\ \bibnamefont
  {Chang}}, \bibinfo {author} {\bibfnamefont {V.}~\bibnamefont
  {Vuleti{\'{c}}}},\ and\ \bibinfo {author} {\bibfnamefont {M.~D.}\
  \bibnamefont {Lukin}},\ }\bibfield  {title} {\bibinfo {title} {Quantum
  nonlinear optics~{\textemdash}~photon by photon},\ }\href
  {https://doi.org/10.1038/nphoton.2014.192} {\bibfield  {journal} {\bibinfo
  {journal} {Nat. Photon.}\ }\textbf {\bibinfo {volume} {8}},\ \bibinfo {pages}
  {685} (\bibinfo {year} {2014})}\BibitemShut {NoStop}%
\bibitem [{\citenamefont {Mu{\~{n}}oz-Matutano}\ \emph
  {et~al.}(2019)\citenamefont {Mu{\~{n}}oz-Matutano}, \citenamefont {Wood},
  \citenamefont {Johnsson}, \citenamefont {Vidal}, \citenamefont {Baragiola},
  \citenamefont {Reinhard}, \citenamefont {Lema{\^i}tre}, \citenamefont
  {Bloch}, \citenamefont {Amo}, \citenamefont {Nogues}, \citenamefont {Besga},
  \citenamefont {Richard},\ and\ \citenamefont {Volz}}]{Munoz-Matutano2019}%
  \BibitemOpen
  \bibfield  {author} {\bibinfo {author} {\bibfnamefont {G.}~\bibnamefont
  {Mu{\~{n}}oz-Matutano}}, \bibinfo {author} {\bibfnamefont {A.}~\bibnamefont
  {Wood}}, \bibinfo {author} {\bibfnamefont {M.}~\bibnamefont {Johnsson}},
  \bibinfo {author} {\bibfnamefont {X.}~\bibnamefont {Vidal}}, \bibinfo
  {author} {\bibfnamefont {B.~Q.}\ \bibnamefont {Baragiola}}, \bibinfo {author}
  {\bibfnamefont {A.}~\bibnamefont {Reinhard}}, \bibinfo {author}
  {\bibfnamefont {A.}~\bibnamefont {Lema{\^i}tre}}, \bibinfo {author}
  {\bibfnamefont {J.}~\bibnamefont {Bloch}}, \bibinfo {author} {\bibfnamefont
  {A.}~\bibnamefont {Amo}}, \bibinfo {author} {\bibfnamefont {G.}~\bibnamefont
  {Nogues}}, \bibinfo {author} {\bibfnamefont {B.}~\bibnamefont {Besga}},
  \bibinfo {author} {\bibfnamefont {M.}~\bibnamefont {Richard}},\ and\ \bibinfo
  {author} {\bibfnamefont {T.}~\bibnamefont {Volz}},\ }\bibfield  {title}
  {\bibinfo {title} {Emergence of quantum correlations from interacting
  fibre-cavity polaritons},\ }\href {https://doi.org/10.1038/s41563-019-0281-z}
  {\bibfield  {journal} {\bibinfo  {journal} {Nature Materials}\ }\textbf
  {\bibinfo {volume} {18}},\ \bibinfo {pages} {213} (\bibinfo {year}
  {2019})}\BibitemShut {NoStop}%
\bibitem [{\citenamefont {Delteil}\ \emph {et~al.}(2019)\citenamefont
  {Delteil}, \citenamefont {Fink}, \citenamefont {Schade}, \citenamefont
  {H{\"o}fling}, \citenamefont {Schneider},\ and\ \citenamefont
  {{\.{I}}mamo{\u{g}}lu}}]{Delteil2019}%
  \BibitemOpen
  \bibfield  {author} {\bibinfo {author} {\bibfnamefont {A.}~\bibnamefont
  {Delteil}}, \bibinfo {author} {\bibfnamefont {T.}~\bibnamefont {Fink}},
  \bibinfo {author} {\bibfnamefont {A.}~\bibnamefont {Schade}}, \bibinfo
  {author} {\bibfnamefont {S.}~\bibnamefont {H{\"o}fling}}, \bibinfo {author}
  {\bibfnamefont {C.}~\bibnamefont {Schneider}},\ and\ \bibinfo {author}
  {\bibfnamefont {A.}~\bibnamefont {{\.{I}}mamo{\u{g}}lu}},\ }\bibfield
  {title} {\bibinfo {title} {Towards polariton blockade of confined
  exciton--polaritons},\ }\href {https://doi.org/10.1038/s41563-019-0282-y}
  {\bibfield  {journal} {\bibinfo  {journal} {Nature Materials}\ }\textbf
  {\bibinfo {volume} {18}},\ \bibinfo {pages} {219} (\bibinfo {year}
  {2019})}\BibitemShut {NoStop}%
\bibitem [{\citenamefont {Zasedatelev}\ \emph {et~al.}(2021)\citenamefont
  {Zasedatelev}, \citenamefont {Baranikov}, \citenamefont {Sannikov},
  \citenamefont {Urbonas}, \citenamefont {Scafirimuto}, \citenamefont
  {Shishkov}, \citenamefont {Andrianov}, \citenamefont {Lozovik}, \citenamefont
  {Scherf}, \citenamefont {St{\"o}ferle}, \citenamefont {Mahrt},\ and\
  \citenamefont {Lagoudakis}}]{Zasedatelev2021}%
  \BibitemOpen
  \bibfield  {author} {\bibinfo {author} {\bibfnamefont {A.~V.}\ \bibnamefont
  {Zasedatelev}}, \bibinfo {author} {\bibfnamefont {A.~V.}\ \bibnamefont
  {Baranikov}}, \bibinfo {author} {\bibfnamefont {D.}~\bibnamefont {Sannikov}},
  \bibinfo {author} {\bibfnamefont {D.}~\bibnamefont {Urbonas}}, \bibinfo
  {author} {\bibfnamefont {F.}~\bibnamefont {Scafirimuto}}, \bibinfo {author}
  {\bibfnamefont {V.~Y.}\ \bibnamefont {Shishkov}}, \bibinfo {author}
  {\bibfnamefont {E.~S.}\ \bibnamefont {Andrianov}}, \bibinfo {author}
  {\bibfnamefont {Y.~E.}\ \bibnamefont {Lozovik}}, \bibinfo {author}
  {\bibfnamefont {U.}~\bibnamefont {Scherf}}, \bibinfo {author} {\bibfnamefont
  {T.}~\bibnamefont {St{\"o}ferle}}, \bibinfo {author} {\bibfnamefont {R.~F.}\
  \bibnamefont {Mahrt}},\ and\ \bibinfo {author} {\bibfnamefont {P.~G.}\
  \bibnamefont {Lagoudakis}},\ }\bibfield  {title} {\bibinfo {title}
  {Single-photon nonlinearity at room temperature},\ }\href
  {https://doi.org/10.1038/s41586-021-03866-9} {\bibfield  {journal} {\bibinfo
  {journal} {Nature}\ }\textbf {\bibinfo {volume} {597}},\ \bibinfo {pages}
  {493} (\bibinfo {year} {2021})}\BibitemShut {NoStop}%
\bibitem [{\citenamefont {Kuriakose}\ \emph {et~al.}(2022)\citenamefont
  {Kuriakose}, \citenamefont {Walker}, \citenamefont {Dowling}, \citenamefont
  {Kyriienko}, \citenamefont {Shelykh}, \citenamefont {St-Jean}, \citenamefont
  {Zambon}, \citenamefont {Lema{\^i}tre}, \citenamefont {Sagnes}, \citenamefont
  {Legratiet}, \citenamefont {Harouri}, \citenamefont {Ravets}, \citenamefont
  {Skolnick}, \citenamefont {Amo}, \citenamefont {Bloch},\ and\ \citenamefont
  {Krizhanovskii}}]{Kuriakose2022}%
  \BibitemOpen
  \bibfield  {author} {\bibinfo {author} {\bibfnamefont {T.}~\bibnamefont
  {Kuriakose}}, \bibinfo {author} {\bibfnamefont {P.~M.}\ \bibnamefont
  {Walker}}, \bibinfo {author} {\bibfnamefont {T.}~\bibnamefont {Dowling}},
  \bibinfo {author} {\bibfnamefont {O.}~\bibnamefont {Kyriienko}}, \bibinfo
  {author} {\bibfnamefont {I.~A.}\ \bibnamefont {Shelykh}}, \bibinfo {author}
  {\bibfnamefont {P.}~\bibnamefont {St-Jean}}, \bibinfo {author} {\bibfnamefont
  {N.~C.}\ \bibnamefont {Zambon}}, \bibinfo {author} {\bibfnamefont
  {A.}~\bibnamefont {Lema{\^i}tre}}, \bibinfo {author} {\bibfnamefont
  {I.}~\bibnamefont {Sagnes}}, \bibinfo {author} {\bibfnamefont
  {L.}~\bibnamefont {Legratiet}}, \bibinfo {author} {\bibfnamefont
  {A.}~\bibnamefont {Harouri}}, \bibinfo {author} {\bibfnamefont
  {S.}~\bibnamefont {Ravets}}, \bibinfo {author} {\bibfnamefont {M.~S.}\
  \bibnamefont {Skolnick}}, \bibinfo {author} {\bibfnamefont {A.}~\bibnamefont
  {Amo}}, \bibinfo {author} {\bibfnamefont {J.}~\bibnamefont {Bloch}},\ and\
  \bibinfo {author} {\bibfnamefont {D.~N.}\ \bibnamefont {Krizhanovskii}},\
  }\bibfield  {title} {\bibinfo {title} {Few-photon all-optical phase rotation
  in a quantum-well micropillar cavity},\ }\bibfield  {journal} {\bibinfo
  {journal} {Nature Photonics}\ }\href
  {https://doi.org/10.1038/s41566-022-01019-6} {10.1038/s41566-022-01019-6}
  (\bibinfo {year} {2022})\BibitemShut {NoStop}%
\bibitem [{\citenamefont {Sich}\ \emph {et~al.}(2012)\citenamefont {Sich},
  \citenamefont {Krizhanovskii}, \citenamefont {Skolnick}, \citenamefont
  {Gorbach}, \citenamefont {Hartley}, \citenamefont {Skryabin}, \citenamefont
  {Cerda-M{\'e}ndez}, \citenamefont {Biermann}, \citenamefont {Hey},\ and\
  \citenamefont {Santos}}]{Sich2012}%
  \BibitemOpen
  \bibfield  {author} {\bibinfo {author} {\bibfnamefont {M.}~\bibnamefont
  {Sich}}, \bibinfo {author} {\bibfnamefont {D.~N.}\ \bibnamefont
  {Krizhanovskii}}, \bibinfo {author} {\bibfnamefont {M.~S.}\ \bibnamefont
  {Skolnick}}, \bibinfo {author} {\bibfnamefont {A.~V.}\ \bibnamefont
  {Gorbach}}, \bibinfo {author} {\bibfnamefont {R.}~\bibnamefont {Hartley}},
  \bibinfo {author} {\bibfnamefont {D.~V.}\ \bibnamefont {Skryabin}}, \bibinfo
  {author} {\bibfnamefont {E.~A.}\ \bibnamefont {Cerda-M{\'e}ndez}}, \bibinfo
  {author} {\bibfnamefont {K.}~\bibnamefont {Biermann}}, \bibinfo {author}
  {\bibfnamefont {R.}~\bibnamefont {Hey}},\ and\ \bibinfo {author}
  {\bibfnamefont {P.~V.}\ \bibnamefont {Santos}},\ }\bibfield  {title}
  {\bibinfo {title} {Observation of bright polariton solitons in a
  semiconductor microcavity},\ }\href
  {https://doi.org/10.1038/nphoton.2011.267} {\bibfield  {journal} {\bibinfo
  {journal} {Nature Photonics}\ }\textbf {\bibinfo {volume} {6}},\ \bibinfo
  {pages} {50} (\bibinfo {year} {2012})}\BibitemShut {NoStop}%
\bibitem [{\citenamefont {Walker}\ \emph {et~al.}(2017)\citenamefont {Walker},
  \citenamefont {Tinkler}, \citenamefont {Royall}, \citenamefont {Skryabin},
  \citenamefont {Farrer}, \citenamefont {Ritchie}, \citenamefont {Skolnick},\
  and\ \citenamefont {Krizhanovskii}}]{Walker2017}%
  \BibitemOpen
  \bibfield  {author} {\bibinfo {author} {\bibfnamefont {P.~M.}\ \bibnamefont
  {Walker}}, \bibinfo {author} {\bibfnamefont {L.}~\bibnamefont {Tinkler}},
  \bibinfo {author} {\bibfnamefont {B.}~\bibnamefont {Royall}}, \bibinfo
  {author} {\bibfnamefont {D.~V.}\ \bibnamefont {Skryabin}}, \bibinfo {author}
  {\bibfnamefont {I.}~\bibnamefont {Farrer}}, \bibinfo {author} {\bibfnamefont
  {D.~A.}\ \bibnamefont {Ritchie}}, \bibinfo {author} {\bibfnamefont {M.~S.}\
  \bibnamefont {Skolnick}},\ and\ \bibinfo {author} {\bibfnamefont {D.~N.}\
  \bibnamefont {Krizhanovskii}},\ }\bibfield  {title} {\bibinfo {title} {Dark
  solitons in high velocity waveguide polariton fluids},\ }\href
  {https://doi.org/10.1103/PhysRevLett.119.097403} {\bibfield  {journal}
  {\bibinfo  {journal} {Phys. Rev. Lett.}\ }\textbf {\bibinfo {volume} {119}},\
  \bibinfo {pages} {097403} (\bibinfo {year} {2017})}\BibitemShut {NoStop}%
\bibitem [{\citenamefont {Ma\^{\i}tre}\ \emph {et~al.}(2020)\citenamefont
  {Ma\^{\i}tre}, \citenamefont {Lerario}, \citenamefont {Medeiros},
  \citenamefont {Claude}, \citenamefont {Glorieux}, \citenamefont {Giacobino},
  \citenamefont {Pigeon},\ and\ \citenamefont {Bramati}}]{Maitre2020}%
  \BibitemOpen
  \bibfield  {author} {\bibinfo {author} {\bibfnamefont {A.}~\bibnamefont
  {Ma\^{\i}tre}}, \bibinfo {author} {\bibfnamefont {G.}~\bibnamefont
  {Lerario}}, \bibinfo {author} {\bibfnamefont {A.}~\bibnamefont {Medeiros}},
  \bibinfo {author} {\bibfnamefont {F.}~\bibnamefont {Claude}}, \bibinfo
  {author} {\bibfnamefont {Q.}~\bibnamefont {Glorieux}}, \bibinfo {author}
  {\bibfnamefont {E.}~\bibnamefont {Giacobino}}, \bibinfo {author}
  {\bibfnamefont {S.}~\bibnamefont {Pigeon}},\ and\ \bibinfo {author}
  {\bibfnamefont {A.}~\bibnamefont {Bramati}},\ }\bibfield  {title} {\bibinfo
  {title} {Dark-soliton molecules in an exciton-polariton superfluid},\ }\href
  {https://doi.org/10.1103/PhysRevX.10.041028} {\bibfield  {journal} {\bibinfo
  {journal} {Phys. Rev. X}\ }\textbf {\bibinfo {volume} {10}},\ \bibinfo
  {pages} {041028} (\bibinfo {year} {2020})}\BibitemShut {NoStop}%
\bibitem [{\citenamefont {Sanvitto}\ and\ \citenamefont
  {K{\'e}na-Cohen}(2016)}]{Sanvitto2016}%
  \BibitemOpen
  \bibfield  {author} {\bibinfo {author} {\bibfnamefont {D.}~\bibnamefont
  {Sanvitto}}\ and\ \bibinfo {author} {\bibfnamefont {S.}~\bibnamefont
  {K{\'e}na-Cohen}},\ }\bibfield  {title} {\bibinfo {title} {The road towards
  polaritonic devices},\ }\href {https://doi.org/10.1038/nmat4668} {\bibfield
  {journal} {\bibinfo  {journal} {Nature Materials}\ }\textbf {\bibinfo
  {volume} {15}},\ \bibinfo {pages} {1061} (\bibinfo {year}
  {2016})}\BibitemShut {NoStop}%
\bibitem [{\citenamefont {Kyriienko}\ and\ \citenamefont
  {Liew}(2016)}]{KyriienkoLiew2016}%
  \BibitemOpen
  \bibfield  {author} {\bibinfo {author} {\bibfnamefont {O.}~\bibnamefont
  {Kyriienko}}\ and\ \bibinfo {author} {\bibfnamefont {T.~C.~H.}\ \bibnamefont
  {Liew}},\ }\bibfield  {title} {\bibinfo {title} {Exciton-polariton quantum
  gates based on continuous variables},\ }\href
  {https://doi.org/10.1103/PhysRevB.93.035301} {\bibfield  {journal} {\bibinfo
  {journal} {Phys. Rev. B}\ }\textbf {\bibinfo {volume} {93}},\ \bibinfo
  {pages} {035301} (\bibinfo {year} {2016})}\BibitemShut {NoStop}%
\bibitem [{\citenamefont {Ghosh}\ and\ \citenamefont {Liew}(2020)}]{Ghosh2020}%
  \BibitemOpen
  \bibfield  {author} {\bibinfo {author} {\bibfnamefont {S.}~\bibnamefont
  {Ghosh}}\ and\ \bibinfo {author} {\bibfnamefont {T.~C.~H.}\ \bibnamefont
  {Liew}},\ }\bibfield  {title} {\bibinfo {title} {Quantum computing with
  exciton-polariton condensates},\ }\href
  {https://doi.org/10.1038/s41534-020-0244-x} {\bibfield  {journal} {\bibinfo
  {journal} {npj Quantum Information}\ }\textbf {\bibinfo {volume} {6}},\
  \bibinfo {pages} {16} (\bibinfo {year} {2020})}\BibitemShut {NoStop}%
\bibitem [{\citenamefont {Kavokin}\ \emph {et~al.}(2022)\citenamefont
  {Kavokin}, \citenamefont {Liew}, \citenamefont {Schneider}, \citenamefont
  {Lagoudakis}, \citenamefont {Klembt},\ and\ \citenamefont
  {Hoefling}}]{Kavokin2022}%
  \BibitemOpen
  \bibfield  {author} {\bibinfo {author} {\bibfnamefont {A.}~\bibnamefont
  {Kavokin}}, \bibinfo {author} {\bibfnamefont {T.~C.~H.}\ \bibnamefont
  {Liew}}, \bibinfo {author} {\bibfnamefont {C.}~\bibnamefont {Schneider}},
  \bibinfo {author} {\bibfnamefont {P.~G.}\ \bibnamefont {Lagoudakis}},
  \bibinfo {author} {\bibfnamefont {S.}~\bibnamefont {Klembt}},\ and\ \bibinfo
  {author} {\bibfnamefont {S.}~\bibnamefont {Hoefling}},\ }\bibfield  {title}
  {\bibinfo {title} {Polariton condensates for classical and quantum
  computing},\ }\href {https://doi.org/10.1038/s42254-022-00447-1} {\bibfield
  {journal} {\bibinfo  {journal} {Nature Reviews Physics}\ }\textbf {\bibinfo
  {volume} {4}},\ \bibinfo {pages} {435} (\bibinfo {year} {2022})}\BibitemShut
  {NoStop}%
\bibitem [{\citenamefont {Amo}\ \emph {et~al.}(2009)\citenamefont {Amo},
  \citenamefont {Sanvitto}, \citenamefont {Laussy}, \citenamefont {Ballarini},
  \citenamefont {Valle}, \citenamefont {Martin}, \citenamefont {Lema{\^i}tre},
  \citenamefont {Bloch}, \citenamefont {Krizhanovskii}, \citenamefont
  {Skolnick}, \citenamefont {Tejedor},\ and\ \citenamefont
  {Vi{\~{n}}a}}]{Amo2009}%
  \BibitemOpen
  \bibfield  {author} {\bibinfo {author} {\bibfnamefont {A.}~\bibnamefont
  {Amo}}, \bibinfo {author} {\bibfnamefont {D.}~\bibnamefont {Sanvitto}},
  \bibinfo {author} {\bibfnamefont {F.~P.}\ \bibnamefont {Laussy}}, \bibinfo
  {author} {\bibfnamefont {D.}~\bibnamefont {Ballarini}}, \bibinfo {author}
  {\bibfnamefont {E.~d.}\ \bibnamefont {Valle}}, \bibinfo {author}
  {\bibfnamefont {M.~D.}\ \bibnamefont {Martin}}, \bibinfo {author}
  {\bibfnamefont {A.}~\bibnamefont {Lema{\^i}tre}}, \bibinfo {author}
  {\bibfnamefont {J.}~\bibnamefont {Bloch}}, \bibinfo {author} {\bibfnamefont
  {D.~N.}\ \bibnamefont {Krizhanovskii}}, \bibinfo {author} {\bibfnamefont
  {M.~S.}\ \bibnamefont {Skolnick}}, \bibinfo {author} {\bibfnamefont
  {C.}~\bibnamefont {Tejedor}},\ and\ \bibinfo {author} {\bibfnamefont
  {L.}~\bibnamefont {Vi{\~{n}}a}},\ }\bibfield  {title} {\bibinfo {title}
  {Collective fluid dynamics of a polariton condensate in a semiconductor
  microcavity},\ }\href {https://doi.org/10.1038/nature07640} {\bibfield
  {journal} {\bibinfo  {journal} {Nature}\ }\textbf {\bibinfo {volume} {457}},\
  \bibinfo {pages} {291} (\bibinfo {year} {2009})}\BibitemShut {NoStop}%
\bibitem [{\citenamefont {Amo}\ \emph {et~al.}(2011)\citenamefont {Amo},
  \citenamefont {Pigeon}, \citenamefont {Sanvitto}, \citenamefont {Sala},
  \citenamefont {Hivet}, \citenamefont {Carusotto}, \citenamefont {Pisanello},
  \citenamefont {Leménager}, \citenamefont {Houdré}, \citenamefont
  {Giacobino}, \citenamefont {Ciuti},\ and\ \citenamefont {Bramati}}]{Amo2011}%
  \BibitemOpen
  \bibfield  {author} {\bibinfo {author} {\bibfnamefont {A.}~\bibnamefont
  {Amo}}, \bibinfo {author} {\bibfnamefont {S.}~\bibnamefont {Pigeon}},
  \bibinfo {author} {\bibfnamefont {D.}~\bibnamefont {Sanvitto}}, \bibinfo
  {author} {\bibfnamefont {V.~G.}\ \bibnamefont {Sala}}, \bibinfo {author}
  {\bibfnamefont {R.}~\bibnamefont {Hivet}}, \bibinfo {author} {\bibfnamefont
  {I.}~\bibnamefont {Carusotto}}, \bibinfo {author} {\bibfnamefont
  {F.}~\bibnamefont {Pisanello}}, \bibinfo {author} {\bibfnamefont
  {G.}~\bibnamefont {Leménager}}, \bibinfo {author} {\bibfnamefont
  {R.}~\bibnamefont {Houdré}}, \bibinfo {author} {\bibfnamefont
  {E.}~\bibnamefont {Giacobino}}, \bibinfo {author} {\bibfnamefont
  {C.}~\bibnamefont {Ciuti}},\ and\ \bibinfo {author} {\bibfnamefont
  {A.}~\bibnamefont {Bramati}},\ }\bibfield  {title} {\bibinfo {title}
  {Polariton superfluids reveal quantum hydrodynamic solitons},\ }\href
  {https://doi.org/10.1126/science.1202307} {\bibfield  {journal} {\bibinfo
  {journal} {Science}\ }\textbf {\bibinfo {volume} {332}},\ \bibinfo {pages}
  {1167} (\bibinfo {year} {2011})}\BibitemShut {NoStop}%
\bibitem [{\citenamefont {Liew}\ \emph {et~al.}(2008)\citenamefont {Liew},
  \citenamefont {Kavokin},\ and\ \citenamefont {Shelykh}}]{Liew2008}%
  \BibitemOpen
  \bibfield  {author} {\bibinfo {author} {\bibfnamefont {T.~C.~H.}\
  \bibnamefont {Liew}}, \bibinfo {author} {\bibfnamefont {A.~V.}\ \bibnamefont
  {Kavokin}},\ and\ \bibinfo {author} {\bibfnamefont {I.~A.}\ \bibnamefont
  {Shelykh}},\ }\bibfield  {title} {\bibinfo {title} {Optical circuits based on
  polariton neurons in semiconductor microcavities},\ }\href
  {https://doi.org/10.1103/PhysRevLett.101.016402} {\bibfield  {journal}
  {\bibinfo  {journal} {Phys. Rev. Lett.}\ }\textbf {\bibinfo {volume} {101}},\
  \bibinfo {pages} {016402} (\bibinfo {year} {2008})}\BibitemShut {NoStop}%
\bibitem [{\citenamefont {Yang}\ and\ \citenamefont {Kim}(2022)}]{YangKim2022}%
  \BibitemOpen
  \bibfield  {author} {\bibinfo {author} {\bibfnamefont {H.}~\bibnamefont
  {Yang}}\ and\ \bibinfo {author} {\bibfnamefont {N.~Y.}\ \bibnamefont {Kim}},\
  }\bibfield  {title} {\bibinfo {title} {Microcavity exciton-polariton quantum
  spin fluids},\ }\href
  {https://doi.org/https://doi.org/10.1002/qute.202100137} {\bibfield
  {journal} {\bibinfo  {journal} {Advanced Quantum Technologies}\ }\textbf
  {\bibinfo {volume} {5}},\ \bibinfo {pages} {2100137} (\bibinfo {year}
  {2022})}\BibitemShut {NoStop}%
\bibitem [{\citenamefont {Kim}\ \emph {et~al.}(2013)\citenamefont {Kim},
  \citenamefont {Kusudo}, \citenamefont {Löffler}, \citenamefont {Höfling},
  \citenamefont {Forchel},\ and\ \citenamefont {Yamamoto}}]{Kim2013}%
  \BibitemOpen
  \bibfield  {author} {\bibinfo {author} {\bibfnamefont {N.~Y.}\ \bibnamefont
  {Kim}}, \bibinfo {author} {\bibfnamefont {K.}~\bibnamefont {Kusudo}},
  \bibinfo {author} {\bibfnamefont {A.}~\bibnamefont {Löffler}}, \bibinfo
  {author} {\bibfnamefont {S.}~\bibnamefont {Höfling}}, \bibinfo {author}
  {\bibfnamefont {A.}~\bibnamefont {Forchel}},\ and\ \bibinfo {author}
  {\bibfnamefont {Y.}~\bibnamefont {Yamamoto}},\ }\bibfield  {title} {\bibinfo
  {title} {Exciton–polariton condensates near the dirac point in a triangular
  lattice},\ }\href {https://doi.org/10.1088/1367-2630/15/3/035032} {\bibfield
  {journal} {\bibinfo  {journal} {New Journal of Physics}\ }\textbf {\bibinfo
  {volume} {15}},\ \bibinfo {pages} {035032} (\bibinfo {year}
  {2013})}\BibitemShut {NoStop}%
\bibitem [{\citenamefont {Jacqmin}\ \emph {et~al.}(2014)\citenamefont
  {Jacqmin}, \citenamefont {Carusotto}, \citenamefont {Sagnes}, \citenamefont
  {Abbarchi}, \citenamefont {Solnyshkov}, \citenamefont {Malpuech},
  \citenamefont {Galopin}, \citenamefont {Lema\^{\i}tre}, \citenamefont
  {Bloch},\ and\ \citenamefont {Amo}}]{Jacqmin2014}%
  \BibitemOpen
  \bibfield  {author} {\bibinfo {author} {\bibfnamefont {T.}~\bibnamefont
  {Jacqmin}}, \bibinfo {author} {\bibfnamefont {I.}~\bibnamefont {Carusotto}},
  \bibinfo {author} {\bibfnamefont {I.}~\bibnamefont {Sagnes}}, \bibinfo
  {author} {\bibfnamefont {M.}~\bibnamefont {Abbarchi}}, \bibinfo {author}
  {\bibfnamefont {D.~D.}\ \bibnamefont {Solnyshkov}}, \bibinfo {author}
  {\bibfnamefont {G.}~\bibnamefont {Malpuech}}, \bibinfo {author}
  {\bibfnamefont {E.}~\bibnamefont {Galopin}}, \bibinfo {author} {\bibfnamefont
  {A.}~\bibnamefont {Lema\^{\i}tre}}, \bibinfo {author} {\bibfnamefont
  {J.}~\bibnamefont {Bloch}},\ and\ \bibinfo {author} {\bibfnamefont
  {A.}~\bibnamefont {Amo}},\ }\bibfield  {title} {\bibinfo {title} {Direct
  observation of dirac cones and a flatband in a honeycomb lattice for
  polaritons},\ }\href {https://doi.org/10.1103/PhysRevLett.112.116402}
  {\bibfield  {journal} {\bibinfo  {journal} {Phys. Rev. Lett.}\ }\textbf
  {\bibinfo {volume} {112}},\ \bibinfo {pages} {116402} (\bibinfo {year}
  {2014})}\BibitemShut {NoStop}%
\bibitem [{\citenamefont {Ohadi}\ \emph {et~al.}(2017)\citenamefont {Ohadi},
  \citenamefont {Ramsay}, \citenamefont {Sigurdsson}, \citenamefont {del
  Valle-Inclan~Redondo}, \citenamefont {Tsintzos}, \citenamefont {Hatzopoulos},
  \citenamefont {Liew}, \citenamefont {Shelykh}, \citenamefont {Rubo},
  \citenamefont {Savvidis},\ and\ \citenamefont {Baumberg}}]{Ohadi2017}%
  \BibitemOpen
  \bibfield  {author} {\bibinfo {author} {\bibfnamefont {H.}~\bibnamefont
  {Ohadi}}, \bibinfo {author} {\bibfnamefont {A.~J.}\ \bibnamefont {Ramsay}},
  \bibinfo {author} {\bibfnamefont {H.}~\bibnamefont {Sigurdsson}}, \bibinfo
  {author} {\bibfnamefont {Y.}~\bibnamefont {del Valle-Inclan~Redondo}},
  \bibinfo {author} {\bibfnamefont {S.~I.}\ \bibnamefont {Tsintzos}}, \bibinfo
  {author} {\bibfnamefont {Z.}~\bibnamefont {Hatzopoulos}}, \bibinfo {author}
  {\bibfnamefont {T.~C.~H.}\ \bibnamefont {Liew}}, \bibinfo {author}
  {\bibfnamefont {I.~A.}\ \bibnamefont {Shelykh}}, \bibinfo {author}
  {\bibfnamefont {Y.~G.}\ \bibnamefont {Rubo}}, \bibinfo {author}
  {\bibfnamefont {P.~G.}\ \bibnamefont {Savvidis}},\ and\ \bibinfo {author}
  {\bibfnamefont {J.~J.}\ \bibnamefont {Baumberg}},\ }\bibfield  {title}
  {\bibinfo {title} {Spin order and phase transitions in chains of polariton
  condensates},\ }\href {https://doi.org/10.1103/PhysRevLett.119.067401}
  {\bibfield  {journal} {\bibinfo  {journal} {Phys. Rev. Lett.}\ }\textbf
  {\bibinfo {volume} {119}},\ \bibinfo {pages} {067401} (\bibinfo {year}
  {2017})}\BibitemShut {NoStop}%
\bibitem [{\citenamefont {Whittaker}\ \emph {et~al.}(2018)\citenamefont
  {Whittaker}, \citenamefont {Cancellieri}, \citenamefont {Walker},
  \citenamefont {Gulevich}, \citenamefont {Schomerus}, \citenamefont
  {Vaitiekus}, \citenamefont {Royall}, \citenamefont {Whittaker}, \citenamefont
  {Clarke}, \citenamefont {Iorsh}, \citenamefont {Shelykh}, \citenamefont
  {Skolnick},\ and\ \citenamefont {Krizhanovskii}}]{Whittaker2018}%
  \BibitemOpen
  \bibfield  {author} {\bibinfo {author} {\bibfnamefont {C.~E.}\ \bibnamefont
  {Whittaker}}, \bibinfo {author} {\bibfnamefont {E.}~\bibnamefont
  {Cancellieri}}, \bibinfo {author} {\bibfnamefont {P.~M.}\ \bibnamefont
  {Walker}}, \bibinfo {author} {\bibfnamefont {D.~R.}\ \bibnamefont
  {Gulevich}}, \bibinfo {author} {\bibfnamefont {H.}~\bibnamefont {Schomerus}},
  \bibinfo {author} {\bibfnamefont {D.}~\bibnamefont {Vaitiekus}}, \bibinfo
  {author} {\bibfnamefont {B.}~\bibnamefont {Royall}}, \bibinfo {author}
  {\bibfnamefont {D.~M.}\ \bibnamefont {Whittaker}}, \bibinfo {author}
  {\bibfnamefont {E.}~\bibnamefont {Clarke}}, \bibinfo {author} {\bibfnamefont
  {I.~V.}\ \bibnamefont {Iorsh}}, \bibinfo {author} {\bibfnamefont {I.~A.}\
  \bibnamefont {Shelykh}}, \bibinfo {author} {\bibfnamefont {M.~S.}\
  \bibnamefont {Skolnick}},\ and\ \bibinfo {author} {\bibfnamefont {D.~N.}\
  \bibnamefont {Krizhanovskii}},\ }\bibfield  {title} {\bibinfo {title}
  {Exciton polaritons in a two-dimensional lieb lattice with spin-orbit
  coupling},\ }\href {https://doi.org/10.1103/PhysRevLett.120.097401}
  {\bibfield  {journal} {\bibinfo  {journal} {Phys. Rev. Lett.}\ }\textbf
  {\bibinfo {volume} {120}},\ \bibinfo {pages} {097401} (\bibinfo {year}
  {2018})}\BibitemShut {NoStop}%
\bibitem [{\citenamefont {Cerda-Méndez}\ \emph {et~al.}(2017)\citenamefont
  {Cerda-Méndez}, \citenamefont {Krizhanovskii}, \citenamefont {Skolnick},\
  and\ \citenamefont {Santos}}]{Cerda-Mendez2018}%
  \BibitemOpen
  \bibfield  {author} {\bibinfo {author} {\bibfnamefont {E.~A.}\ \bibnamefont
  {Cerda-Méndez}}, \bibinfo {author} {\bibfnamefont {D.~N.}\ \bibnamefont
  {Krizhanovskii}}, \bibinfo {author} {\bibfnamefont {M.~S.}\ \bibnamefont
  {Skolnick}},\ and\ \bibinfo {author} {\bibfnamefont {P.~V.}\ \bibnamefont
  {Santos}},\ }\bibfield  {title} {\bibinfo {title} {Quantum fluids of light in
  acoustic lattices},\ }\href {https://doi.org/10.1088/1361-6463/aa9ec7}
  {\bibfield  {journal} {\bibinfo  {journal} {Journal of Physics D: Applied
  Physics}\ }\textbf {\bibinfo {volume} {51}},\ \bibinfo {pages} {033001}
  (\bibinfo {year} {2017})}\BibitemShut {NoStop}%
\bibitem [{\citenamefont {Scafirimuto}\ \emph {et~al.}(2021)\citenamefont
  {Scafirimuto}, \citenamefont {Urbonas}, \citenamefont {Becker}, \citenamefont
  {Scherf}, \citenamefont {Mahrt},\ and\ \citenamefont
  {St{\"o}ferle}}]{Scafirimuto2021}%
  \BibitemOpen
  \bibfield  {author} {\bibinfo {author} {\bibfnamefont {F.}~\bibnamefont
  {Scafirimuto}}, \bibinfo {author} {\bibfnamefont {D.}~\bibnamefont
  {Urbonas}}, \bibinfo {author} {\bibfnamefont {M.~A.}\ \bibnamefont {Becker}},
  \bibinfo {author} {\bibfnamefont {U.}~\bibnamefont {Scherf}}, \bibinfo
  {author} {\bibfnamefont {R.~F.}\ \bibnamefont {Mahrt}},\ and\ \bibinfo
  {author} {\bibfnamefont {T.}~\bibnamefont {St{\"o}ferle}},\ }\bibfield
  {title} {\bibinfo {title} {Tunable exciton--polariton condensation in a
  two-dimensional lieb lattice at room temperature},\ }\href
  {https://doi.org/10.1038/s42005-021-00548-w} {\bibfield  {journal} {\bibinfo
  {journal} {Communications Physics}\ }\textbf {\bibinfo {volume} {4}},\
  \bibinfo {pages} {39} (\bibinfo {year} {2021})}\BibitemShut {NoStop}%
\bibitem [{\citenamefont {Alyatkin}\ \emph {et~al.}(2021)\citenamefont
  {Alyatkin}, \citenamefont {Sigurdsson}, \citenamefont {Askitopoulos},
  \citenamefont {T{\"o}pfer},\ and\ \citenamefont {Lagoudakis}}]{Alyatkin2021}%
  \BibitemOpen
  \bibfield  {author} {\bibinfo {author} {\bibfnamefont {S.}~\bibnamefont
  {Alyatkin}}, \bibinfo {author} {\bibfnamefont {H.}~\bibnamefont
  {Sigurdsson}}, \bibinfo {author} {\bibfnamefont {A.}~\bibnamefont
  {Askitopoulos}}, \bibinfo {author} {\bibfnamefont {J.~D.}\ \bibnamefont
  {T{\"o}pfer}},\ and\ \bibinfo {author} {\bibfnamefont {P.~G.}\ \bibnamefont
  {Lagoudakis}},\ }\bibfield  {title} {\bibinfo {title} {Quantum fluids of
  light in all-optical scatterer lattices},\ }\href
  {https://doi.org/10.1038/s41467-021-25845-4} {\bibfield  {journal} {\bibinfo
  {journal} {Nature Communications}\ }\textbf {\bibinfo {volume} {12}},\
  \bibinfo {pages} {5571} (\bibinfo {year} {2021})}\BibitemShut {NoStop}%
\bibitem [{\citenamefont {Cookson}\ \emph {et~al.}(2021)\citenamefont
  {Cookson}, \citenamefont {Kalinin}, \citenamefont {Sigurdsson}, \citenamefont
  {T{\"o}pfer}, \citenamefont {Alyatkin}, \citenamefont {Silva}, \citenamefont
  {Langbein}, \citenamefont {Berloff},\ and\ \citenamefont
  {Lagoudakis}}]{Cookson2021}%
  \BibitemOpen
  \bibfield  {author} {\bibinfo {author} {\bibfnamefont {T.}~\bibnamefont
  {Cookson}}, \bibinfo {author} {\bibfnamefont {K.}~\bibnamefont {Kalinin}},
  \bibinfo {author} {\bibfnamefont {H.}~\bibnamefont {Sigurdsson}}, \bibinfo
  {author} {\bibfnamefont {J.~D.}\ \bibnamefont {T{\"o}pfer}}, \bibinfo
  {author} {\bibfnamefont {S.}~\bibnamefont {Alyatkin}}, \bibinfo {author}
  {\bibfnamefont {M.}~\bibnamefont {Silva}}, \bibinfo {author} {\bibfnamefont
  {W.}~\bibnamefont {Langbein}}, \bibinfo {author} {\bibfnamefont {N.~G.}\
  \bibnamefont {Berloff}},\ and\ \bibinfo {author} {\bibfnamefont {P.~G.}\
  \bibnamefont {Lagoudakis}},\ }\bibfield  {title} {\bibinfo {title} {Geometric
  frustration in polygons of polariton condensates creating vortices of varying
  topological charge},\ }\href {https://doi.org/10.1038/s41467-021-22121-3}
  {\bibfield  {journal} {\bibinfo  {journal} {Nature Communications}\ }\textbf
  {\bibinfo {volume} {12}},\ \bibinfo {pages} {2120} (\bibinfo {year}
  {2021})}\BibitemShut {NoStop}%
\bibitem [{\citenamefont {T\"{o}pfer}\ \emph {et~al.}(2021)\citenamefont
  {T\"{o}pfer}, \citenamefont {Chatzopoulos}, \citenamefont {Sigurdsson},
  \citenamefont {Cookson}, \citenamefont {Rubo},\ and\ \citenamefont
  {Lagoudakis}}]{Topfer2021}%
  \BibitemOpen
  \bibfield  {author} {\bibinfo {author} {\bibfnamefont {J.~D.}\ \bibnamefont
  {T\"{o}pfer}}, \bibinfo {author} {\bibfnamefont {I.}~\bibnamefont
  {Chatzopoulos}}, \bibinfo {author} {\bibfnamefont {H.}~\bibnamefont
  {Sigurdsson}}, \bibinfo {author} {\bibfnamefont {T.}~\bibnamefont {Cookson}},
  \bibinfo {author} {\bibfnamefont {Y.~G.}\ \bibnamefont {Rubo}},\ and\
  \bibinfo {author} {\bibfnamefont {P.~G.}\ \bibnamefont {Lagoudakis}},\
  }\bibfield  {title} {\bibinfo {title} {Engineering spatial coherence in
  lattices of polariton condensates},\ }\href
  {https://doi.org/10.1364/OPTICA.409976} {\bibfield  {journal} {\bibinfo
  {journal} {Optica}\ }\textbf {\bibinfo {volume} {8}},\ \bibinfo {pages} {106}
  (\bibinfo {year} {2021})}\BibitemShut {NoStop}%
\bibitem [{\citenamefont {Harder}\ \emph {et~al.}(2021)\citenamefont {Harder},
  \citenamefont {Egorov}, \citenamefont {Krause}, \citenamefont {Beierlein},
  \citenamefont {Gagel}, \citenamefont {Emmerling}, \citenamefont {Schneider},
  \citenamefont {Peschel}, \citenamefont {H{\"o}fling},\ and\ \citenamefont
  {Klembt}}]{Harder2021}%
  \BibitemOpen
  \bibfield  {author} {\bibinfo {author} {\bibfnamefont {T.~H.}\ \bibnamefont
  {Harder}}, \bibinfo {author} {\bibfnamefont {O.~A.}\ \bibnamefont {Egorov}},
  \bibinfo {author} {\bibfnamefont {C.}~\bibnamefont {Krause}}, \bibinfo
  {author} {\bibfnamefont {J.}~\bibnamefont {Beierlein}}, \bibinfo {author}
  {\bibfnamefont {P.}~\bibnamefont {Gagel}}, \bibinfo {author} {\bibfnamefont
  {M.}~\bibnamefont {Emmerling}}, \bibinfo {author} {\bibfnamefont
  {C.}~\bibnamefont {Schneider}}, \bibinfo {author} {\bibfnamefont
  {U.}~\bibnamefont {Peschel}}, \bibinfo {author} {\bibfnamefont
  {S.}~\bibnamefont {H{\"o}fling}},\ and\ \bibinfo {author} {\bibfnamefont
  {S.}~\bibnamefont {Klembt}},\ }\bibfield  {title} {\bibinfo {title} {Kagome
  flatbands for coherent exciton-polariton lasing},\ }\href
  {https://doi.org/10.1021/acsphotonics.1c00950} {\bibfield  {journal}
  {\bibinfo  {journal} {ACS Photonics}\ }\textbf {\bibinfo {volume} {8}},\
  \bibinfo {pages} {3193} (\bibinfo {year} {2021})}\BibitemShut {NoStop}%
\bibitem [{\citenamefont {Zvyagintseva}\ \emph {et~al.}(2022)\citenamefont
  {Zvyagintseva}, \citenamefont {Sigurdsson}, \citenamefont {Kozin},
  \citenamefont {Iorsh}, \citenamefont {Shelykh}, \citenamefont {Ulyantsev},\
  and\ \citenamefont {Kyriienko}}]{Zvyagintseva2022}%
  \BibitemOpen
  \bibfield  {author} {\bibinfo {author} {\bibfnamefont {D.}~\bibnamefont
  {Zvyagintseva}}, \bibinfo {author} {\bibfnamefont {H.}~\bibnamefont
  {Sigurdsson}}, \bibinfo {author} {\bibfnamefont {V.~K.}\ \bibnamefont
  {Kozin}}, \bibinfo {author} {\bibfnamefont {I.}~\bibnamefont {Iorsh}},
  \bibinfo {author} {\bibfnamefont {I.~A.}\ \bibnamefont {Shelykh}}, \bibinfo
  {author} {\bibfnamefont {V.}~\bibnamefont {Ulyantsev}},\ and\ \bibinfo
  {author} {\bibfnamefont {O.}~\bibnamefont {Kyriienko}},\ }\bibfield  {title}
  {\bibinfo {title} {Machine learning of phase transitions in nonlinear
  polariton lattices},\ }\href {https://doi.org/10.1038/s42005-021-00755-5}
  {\bibfield  {journal} {\bibinfo  {journal} {Communications Physics}\ }\textbf
  {\bibinfo {volume} {5}},\ \bibinfo {pages} {8} (\bibinfo {year}
  {2022})}\BibitemShut {NoStop}%
\bibitem [{\citenamefont {Ciuti}\ \emph {et~al.}(1998)\citenamefont {Ciuti},
  \citenamefont {Savona}, \citenamefont {Piermarocchi}, \citenamefont
  {Quattropani},\ and\ \citenamefont {Schwendimann}}]{Ciuti1998}%
  \BibitemOpen
  \bibfield  {author} {\bibinfo {author} {\bibfnamefont {C.}~\bibnamefont
  {Ciuti}}, \bibinfo {author} {\bibfnamefont {V.}~\bibnamefont {Savona}},
  \bibinfo {author} {\bibfnamefont {C.}~\bibnamefont {Piermarocchi}}, \bibinfo
  {author} {\bibfnamefont {A.}~\bibnamefont {Quattropani}},\ and\ \bibinfo
  {author} {\bibfnamefont {P.}~\bibnamefont {Schwendimann}},\ }\bibfield
  {title} {\bibinfo {title} {Role of the exchange of carriers in elastic
  exciton-exciton scattering in quantum wells},\ }\href
  {https://doi.org/10.1103/PhysRevB.58.7926} {\bibfield  {journal} {\bibinfo
  {journal} {Phys. Rev. B}\ }\textbf {\bibinfo {volume} {58}},\ \bibinfo
  {pages} {7926} (\bibinfo {year} {1998})}\BibitemShut {NoStop}%
\bibitem [{\citenamefont {Tassone}\ and\ \citenamefont
  {Yamamoto}(1999)}]{Tassone1999}%
  \BibitemOpen
  \bibfield  {author} {\bibinfo {author} {\bibfnamefont {F.}~\bibnamefont
  {Tassone}}\ and\ \bibinfo {author} {\bibfnamefont {Y.}~\bibnamefont
  {Yamamoto}},\ }\bibfield  {title} {\bibinfo {title} {Exciton-exciton
  scattering dynamics in a semiconductor microcavity and stimulated scattering
  into polaritons},\ }\href {https://doi.org/10.1103/PhysRevB.59.10830}
  {\bibfield  {journal} {\bibinfo  {journal} {Phys. Rev. B}\ }\textbf {\bibinfo
  {volume} {59}},\ \bibinfo {pages} {10830} (\bibinfo {year}
  {1999})}\BibitemShut {NoStop}%
\bibitem [{\citenamefont {Combescot}\ \emph {et~al.}(2007)\citenamefont
  {Combescot}, \citenamefont {Dupertuis},\ and\ \citenamefont
  {Betbeder-Matibet}}]{Combescot2007}%
  \BibitemOpen
  \bibfield  {author} {\bibinfo {author} {\bibfnamefont {M.}~\bibnamefont
  {Combescot}}, \bibinfo {author} {\bibfnamefont {M.~A.}\ \bibnamefont
  {Dupertuis}},\ and\ \bibinfo {author} {\bibfnamefont {O.}~\bibnamefont
  {Betbeder-Matibet}},\ }\bibfield  {title} {\bibinfo {title}
  {Polariton-polariton scattering: Exact results through a novel approach},\
  }\href {https://doi.org/10.1209/0295-5075/79/17001} {\bibfield  {journal}
  {\bibinfo  {journal} {Europhysics Letters}\ }\textbf {\bibinfo {volume}
  {79}},\ \bibinfo {pages} {17001} (\bibinfo {year} {2007})}\BibitemShut
  {NoStop}%
\bibitem [{\citenamefont {Glazov}\ \emph {et~al.}(2009)\citenamefont {Glazov},
  \citenamefont {Ouerdane}, \citenamefont {Pilozzi}, \citenamefont {Malpuech},
  \citenamefont {Kavokin},\ and\ \citenamefont {D'Andrea}}]{Glazov2009}%
  \BibitemOpen
  \bibfield  {author} {\bibinfo {author} {\bibfnamefont {M.~M.}\ \bibnamefont
  {Glazov}}, \bibinfo {author} {\bibfnamefont {H.}~\bibnamefont {Ouerdane}},
  \bibinfo {author} {\bibfnamefont {L.}~\bibnamefont {Pilozzi}}, \bibinfo
  {author} {\bibfnamefont {G.}~\bibnamefont {Malpuech}}, \bibinfo {author}
  {\bibfnamefont {A.~V.}\ \bibnamefont {Kavokin}},\ and\ \bibinfo {author}
  {\bibfnamefont {A.}~\bibnamefont {D'Andrea}},\ }\bibfield  {title} {\bibinfo
  {title} {Polariton-polariton scattering in microcavities: A microscopic
  theory},\ }\href {https://doi.org/10.1103/PhysRevB.80.155306} {\bibfield
  {journal} {\bibinfo  {journal} {Phys. Rev. B}\ }\textbf {\bibinfo {volume}
  {80}},\ \bibinfo {pages} {155306} (\bibinfo {year} {2009})}\BibitemShut
  {NoStop}%
\bibitem [{\citenamefont {Brichkin}\ \emph {et~al.}(2011)\citenamefont
  {Brichkin}, \citenamefont {Novikov}, \citenamefont {Larionov}, \citenamefont
  {Kulakovskii}, \citenamefont {Glazov}, \citenamefont {Schneider},
  \citenamefont {H\"ofling}, \citenamefont {Kamp},\ and\ \citenamefont
  {Forchel}}]{Brichkin2011}%
  \BibitemOpen
  \bibfield  {author} {\bibinfo {author} {\bibfnamefont {A.~S.}\ \bibnamefont
  {Brichkin}}, \bibinfo {author} {\bibfnamefont {S.~I.}\ \bibnamefont
  {Novikov}}, \bibinfo {author} {\bibfnamefont {A.~V.}\ \bibnamefont
  {Larionov}}, \bibinfo {author} {\bibfnamefont {V.~D.}\ \bibnamefont
  {Kulakovskii}}, \bibinfo {author} {\bibfnamefont {M.~M.}\ \bibnamefont
  {Glazov}}, \bibinfo {author} {\bibfnamefont {C.}~\bibnamefont {Schneider}},
  \bibinfo {author} {\bibfnamefont {S.}~\bibnamefont {H\"ofling}}, \bibinfo
  {author} {\bibfnamefont {M.}~\bibnamefont {Kamp}},\ and\ \bibinfo {author}
  {\bibfnamefont {A.}~\bibnamefont {Forchel}},\ }\bibfield  {title} {\bibinfo
  {title} {Effect of coulomb interaction on exciton-polariton condensates in
  gaas pillar microcavities},\ }\href
  {https://doi.org/10.1103/PhysRevB.84.195301} {\bibfield  {journal} {\bibinfo
  {journal} {Phys. Rev. B}\ }\textbf {\bibinfo {volume} {84}},\ \bibinfo
  {pages} {195301} (\bibinfo {year} {2011})}\BibitemShut {NoStop}%
\bibitem [{\citenamefont {Shahnazaryan}\ \emph {et~al.}(2016)\citenamefont
  {Shahnazaryan}, \citenamefont {Shelykh},\ and\ \citenamefont
  {Kyriienko}}]{Shahnazaryan2016}%
  \BibitemOpen
  \bibfield  {author} {\bibinfo {author} {\bibfnamefont {V.}~\bibnamefont
  {Shahnazaryan}}, \bibinfo {author} {\bibfnamefont {I.~A.}\ \bibnamefont
  {Shelykh}},\ and\ \bibinfo {author} {\bibfnamefont {O.}~\bibnamefont
  {Kyriienko}},\ }\bibfield  {title} {\bibinfo {title} {Attractive coulomb
  interaction of two-dimensional rydberg excitons},\ }\href
  {https://doi.org/10.1103/PhysRevB.93.245302} {\bibfield  {journal} {\bibinfo
  {journal} {Phys. Rev. B}\ }\textbf {\bibinfo {volume} {93}},\ \bibinfo
  {pages} {245302} (\bibinfo {year} {2016})}\BibitemShut {NoStop}%
\bibitem [{\citenamefont {Shahnazaryan}\ \emph {et~al.}(2017)\citenamefont
  {Shahnazaryan}, \citenamefont {Iorsh}, \citenamefont {Shelykh},\ and\
  \citenamefont {Kyriienko}}]{Shahnazaryan2017}%
  \BibitemOpen
  \bibfield  {author} {\bibinfo {author} {\bibfnamefont {V.}~\bibnamefont
  {Shahnazaryan}}, \bibinfo {author} {\bibfnamefont {I.}~\bibnamefont {Iorsh}},
  \bibinfo {author} {\bibfnamefont {I.~A.}\ \bibnamefont {Shelykh}},\ and\
  \bibinfo {author} {\bibfnamefont {O.}~\bibnamefont {Kyriienko}},\ }\bibfield
  {title} {\bibinfo {title} {Exciton-exciton interaction in transition-metal
  dichalcogenide monolayers},\ }\href
  {https://doi.org/10.1103/physrevb.96.115409} {\bibfield  {journal} {\bibinfo
  {journal} {Phys. Rev. B}\ }\textbf {\bibinfo {volume} {96}},\ \bibinfo
  {pages} {115409} (\bibinfo {year} {2017})}\BibitemShut {NoStop}%
\bibitem [{\citenamefont {Barachati}\ \emph {et~al.}(2018)\citenamefont
  {Barachati}, \citenamefont {Fieramosca}, \citenamefont {Hafezian},
  \citenamefont {Gu}, \citenamefont {Chakraborty}, \citenamefont {Ballarini},
  \citenamefont {Martinu}, \citenamefont {Menon}, \citenamefont {Sanvitto},\
  and\ \citenamefont {K{\'{e}}na-Cohen}}]{Barachati2018}%
  \BibitemOpen
  \bibfield  {author} {\bibinfo {author} {\bibfnamefont {F.}~\bibnamefont
  {Barachati}}, \bibinfo {author} {\bibfnamefont {A.}~\bibnamefont
  {Fieramosca}}, \bibinfo {author} {\bibfnamefont {S.}~\bibnamefont
  {Hafezian}}, \bibinfo {author} {\bibfnamefont {J.}~\bibnamefont {Gu}},
  \bibinfo {author} {\bibfnamefont {B.}~\bibnamefont {Chakraborty}}, \bibinfo
  {author} {\bibfnamefont {D.}~\bibnamefont {Ballarini}}, \bibinfo {author}
  {\bibfnamefont {L.}~\bibnamefont {Martinu}}, \bibinfo {author} {\bibfnamefont
  {V.}~\bibnamefont {Menon}}, \bibinfo {author} {\bibfnamefont
  {D.}~\bibnamefont {Sanvitto}},\ and\ \bibinfo {author} {\bibfnamefont
  {S.}~\bibnamefont {K{\'{e}}na-Cohen}},\ }\bibfield  {title} {\bibinfo {title}
  {Interacting polariton fluids in a monolayer of tungsten disulfide},\ }\href
  {https://doi.org/10.1038/s41565-018-0219-7} {\bibfield  {journal} {\bibinfo
  {journal} {Nat. Nanotechnol.}\ }\textbf {\bibinfo {volume} {13}},\ \bibinfo
  {pages} {906} (\bibinfo {year} {2018})}\BibitemShut {NoStop}%
\bibitem [{\citenamefont {Bleu}\ \emph {et~al.}(2020)\citenamefont {Bleu},
  \citenamefont {Li}, \citenamefont {Levinsen},\ and\ \citenamefont
  {Parish}}]{Bleu2020}%
  \BibitemOpen
  \bibfield  {author} {\bibinfo {author} {\bibfnamefont {O.}~\bibnamefont
  {Bleu}}, \bibinfo {author} {\bibfnamefont {G.}~\bibnamefont {Li}}, \bibinfo
  {author} {\bibfnamefont {J.}~\bibnamefont {Levinsen}},\ and\ \bibinfo
  {author} {\bibfnamefont {M.~M.}\ \bibnamefont {Parish}},\ }\bibfield  {title}
  {\bibinfo {title} {Polariton interactions in microcavities with atomically
  thin semiconductor layers},\ }\href
  {https://doi.org/10.1103/PhysRevResearch.2.043185} {\bibfield  {journal}
  {\bibinfo  {journal} {Phys. Rev. Research}\ }\textbf {\bibinfo {volume}
  {2}},\ \bibinfo {pages} {043185} (\bibinfo {year} {2020})}\BibitemShut
  {NoStop}%
\bibitem [{\citenamefont {Estrecho}\ \emph {et~al.}(2019)\citenamefont
  {Estrecho}, \citenamefont {Gao}, \citenamefont {Bobrovska}, \citenamefont
  {Comber-Todd}, \citenamefont {Fraser}, \citenamefont {Steger}, \citenamefont
  {West}, \citenamefont {Pfeiffer}, \citenamefont {Levinsen}, \citenamefont
  {Parish}, \citenamefont {Liew}, \citenamefont {Matuszewski}, \citenamefont
  {Snoke}, \citenamefont {Truscott},\ and\ \citenamefont
  {Ostrovskaya}}]{Estrecho2019}%
  \BibitemOpen
  \bibfield  {author} {\bibinfo {author} {\bibfnamefont {E.}~\bibnamefont
  {Estrecho}}, \bibinfo {author} {\bibfnamefont {T.}~\bibnamefont {Gao}},
  \bibinfo {author} {\bibfnamefont {N.}~\bibnamefont {Bobrovska}}, \bibinfo
  {author} {\bibfnamefont {D.}~\bibnamefont {Comber-Todd}}, \bibinfo {author}
  {\bibfnamefont {M.~D.}\ \bibnamefont {Fraser}}, \bibinfo {author}
  {\bibfnamefont {M.}~\bibnamefont {Steger}}, \bibinfo {author} {\bibfnamefont
  {K.}~\bibnamefont {West}}, \bibinfo {author} {\bibfnamefont {L.~N.}\
  \bibnamefont {Pfeiffer}}, \bibinfo {author} {\bibfnamefont {J.}~\bibnamefont
  {Levinsen}}, \bibinfo {author} {\bibfnamefont {M.~M.}\ \bibnamefont
  {Parish}}, \bibinfo {author} {\bibfnamefont {T.~C.~H.}\ \bibnamefont {Liew}},
  \bibinfo {author} {\bibfnamefont {M.}~\bibnamefont {Matuszewski}}, \bibinfo
  {author} {\bibfnamefont {D.~W.}\ \bibnamefont {Snoke}}, \bibinfo {author}
  {\bibfnamefont {A.~G.}\ \bibnamefont {Truscott}},\ and\ \bibinfo {author}
  {\bibfnamefont {E.~A.}\ \bibnamefont {Ostrovskaya}},\ }\bibfield  {title}
  {\bibinfo {title} {Direct measurement of polariton-polariton interaction
  strength in the thomas-fermi regime of exciton-polariton condensation},\
  }\href {https://doi.org/10.1103/PhysRevB.100.035306} {\bibfield  {journal}
  {\bibinfo  {journal} {Phys. Rev. B}\ }\textbf {\bibinfo {volume} {100}},\
  \bibinfo {pages} {035306} (\bibinfo {year} {2019})}\BibitemShut {NoStop}%
\bibitem [{\citenamefont {Butov}\ \emph {et~al.}(2001)\citenamefont {Butov},
  \citenamefont {Ivanov}, \citenamefont {Imamoglu}, \citenamefont {Littlewood},
  \citenamefont {Shashkin}, \citenamefont {Dolgopolov}, \citenamefont
  {Campman},\ and\ \citenamefont {Gossard}}]{Butov2001}%
  \BibitemOpen
  \bibfield  {author} {\bibinfo {author} {\bibfnamefont {L.~V.}\ \bibnamefont
  {Butov}}, \bibinfo {author} {\bibfnamefont {A.~L.}\ \bibnamefont {Ivanov}},
  \bibinfo {author} {\bibfnamefont {A.}~\bibnamefont {Imamoglu}}, \bibinfo
  {author} {\bibfnamefont {P.~B.}\ \bibnamefont {Littlewood}}, \bibinfo
  {author} {\bibfnamefont {A.~A.}\ \bibnamefont {Shashkin}}, \bibinfo {author}
  {\bibfnamefont {V.~T.}\ \bibnamefont {Dolgopolov}}, \bibinfo {author}
  {\bibfnamefont {K.~L.}\ \bibnamefont {Campman}},\ and\ \bibinfo {author}
  {\bibfnamefont {A.~C.}\ \bibnamefont {Gossard}},\ }\bibfield  {title}
  {\bibinfo {title} {Stimulated scattering of indirect excitons in coupled
  quantum wells: Signature of a degenerate bose-gas of excitons},\ }\href
  {https://doi.org/10.1103/PhysRevLett.86.5608} {\bibfield  {journal} {\bibinfo
   {journal} {Phys. Rev. Lett.}\ }\textbf {\bibinfo {volume} {86}},\ \bibinfo
  {pages} {5608} (\bibinfo {year} {2001})}\BibitemShut {NoStop}%
\bibitem [{\citenamefont {Kyriienko}\ \emph {et~al.}(2012)\citenamefont
  {Kyriienko}, \citenamefont {Magnusson},\ and\ \citenamefont
  {Shelykh}}]{Kyriienko2012}%
  \BibitemOpen
  \bibfield  {author} {\bibinfo {author} {\bibfnamefont {O.}~\bibnamefont
  {Kyriienko}}, \bibinfo {author} {\bibfnamefont {E.~B.}\ \bibnamefont
  {Magnusson}},\ and\ \bibinfo {author} {\bibfnamefont {I.~A.}\ \bibnamefont
  {Shelykh}},\ }\bibfield  {title} {\bibinfo {title} {Spin dynamics of cold
  exciton condensates},\ }\href {https://doi.org/10.1103/PhysRevB.86.115324}
  {\bibfield  {journal} {\bibinfo  {journal} {Phys. Rev. B}\ }\textbf {\bibinfo
  {volume} {86}},\ \bibinfo {pages} {115324} (\bibinfo {year}
  {2012})}\BibitemShut {NoStop}%
\bibitem [{\citenamefont {Kyriienko}\ \emph {et~al.}(2014)\citenamefont
  {Kyriienko}, \citenamefont {Shelykh},\ and\ \citenamefont
  {Liew}}]{Kyriienko2014a}%
  \BibitemOpen
  \bibfield  {author} {\bibinfo {author} {\bibfnamefont {O.}~\bibnamefont
  {Kyriienko}}, \bibinfo {author} {\bibfnamefont {I.~A.}\ \bibnamefont
  {Shelykh}},\ and\ \bibinfo {author} {\bibfnamefont {T.~C.~H.}\ \bibnamefont
  {Liew}},\ }\bibfield  {title} {\bibinfo {title} {Tunable single-photon
  emission from dipolaritons},\ }\href
  {https://doi.org/10.1103/PhysRevA.90.033807} {\bibfield  {journal} {\bibinfo
  {journal} {Phys. Rev. A}\ }\textbf {\bibinfo {volume} {90}},\ \bibinfo
  {pages} {033807} (\bibinfo {year} {2014})}\BibitemShut {NoStop}%
\bibitem [{\citenamefont {Togan}\ \emph {et~al.}(2018)\citenamefont {Togan},
  \citenamefont {Lim}, \citenamefont {Faelt}, \citenamefont {Wegscheider},\
  and\ \citenamefont {Imamoglu}}]{Togan:PRL2018}%
  \BibitemOpen
  \bibfield  {author} {\bibinfo {author} {\bibfnamefont {E.}~\bibnamefont
  {Togan}}, \bibinfo {author} {\bibfnamefont {H.-T.}\ \bibnamefont {Lim}},
  \bibinfo {author} {\bibfnamefont {S.}~\bibnamefont {Faelt}}, \bibinfo
  {author} {\bibfnamefont {W.}~\bibnamefont {Wegscheider}},\ and\ \bibinfo
  {author} {\bibfnamefont {A.}~\bibnamefont {Imamoglu}},\ }\bibfield  {title}
  {\bibinfo {title} {Enhanced interactions between dipolar polaritons},\ }\href
  {https://doi.org/10.1103/PhysRevLett.121.227402} {\bibfield  {journal}
  {\bibinfo  {journal} {Phys. Rev. Lett.}\ }\textbf {\bibinfo {volume} {121}},\
  \bibinfo {pages} {227402} (\bibinfo {year} {2018})}\BibitemShut {NoStop}%
\bibitem [{\citenamefont {Hubert}\ \emph {et~al.}(2019)\citenamefont {Hubert},
  \citenamefont {Baruchi}, \citenamefont {Mazuz-Harpaz}, \citenamefont {Cohen},
  \citenamefont {Biermann}, \citenamefont {Lemeshko}, \citenamefont {West},
  \citenamefont {Pfeiffer}, \citenamefont {Rapaport},\ and\ \citenamefont
  {Santos}}]{Hubert2019}%
  \BibitemOpen
  \bibfield  {author} {\bibinfo {author} {\bibfnamefont {C.}~\bibnamefont
  {Hubert}}, \bibinfo {author} {\bibfnamefont {Y.}~\bibnamefont {Baruchi}},
  \bibinfo {author} {\bibfnamefont {Y.}~\bibnamefont {Mazuz-Harpaz}}, \bibinfo
  {author} {\bibfnamefont {K.}~\bibnamefont {Cohen}}, \bibinfo {author}
  {\bibfnamefont {K.}~\bibnamefont {Biermann}}, \bibinfo {author}
  {\bibfnamefont {M.}~\bibnamefont {Lemeshko}}, \bibinfo {author}
  {\bibfnamefont {K.}~\bibnamefont {West}}, \bibinfo {author} {\bibfnamefont
  {L.}~\bibnamefont {Pfeiffer}}, \bibinfo {author} {\bibfnamefont
  {R.}~\bibnamefont {Rapaport}},\ and\ \bibinfo {author} {\bibfnamefont
  {P.}~\bibnamefont {Santos}},\ }\bibfield  {title} {\bibinfo {title}
  {Attractive dipolar coupling between stacked exciton fluids},\ }\href
  {https://doi.org/10.1103/PhysRevX.9.021026} {\bibfield  {journal} {\bibinfo
  {journal} {Phys. Rev. X}\ }\textbf {\bibinfo {volume} {9}},\ \bibinfo {pages}
  {021026} (\bibinfo {year} {2019})}\BibitemShut {NoStop}%
\bibitem [{\citenamefont {Browaeys}\ \emph {et~al.}(2016)\citenamefont
  {Browaeys}, \citenamefont {Barredo},\ and\ \citenamefont
  {Lahaye}}]{Browaeys2016}%
  \BibitemOpen
  \bibfield  {author} {\bibinfo {author} {\bibfnamefont {A.}~\bibnamefont
  {Browaeys}}, \bibinfo {author} {\bibfnamefont {D.}~\bibnamefont {Barredo}},\
  and\ \bibinfo {author} {\bibfnamefont {T.}~\bibnamefont {Lahaye}},\
  }\bibfield  {title} {\bibinfo {title} {Experimental investigations of
  dipole–dipole interactions between a few rydberg atoms},\ }\href
  {https://doi.org/10.1088/0953-4075/49/15/152001} {\bibfield  {journal}
  {\bibinfo  {journal} {Journal of Physics B: Atomic, Molecular and Optical
  Physics}\ }\textbf {\bibinfo {volume} {49}},\ \bibinfo {pages} {152001}
  (\bibinfo {year} {2016})}\BibitemShut {NoStop}%
\bibitem [{\citenamefont {Lukin}\ \emph {et~al.}(2001)\citenamefont {Lukin},
  \citenamefont {Fleischhauer}, \citenamefont {Cote}, \citenamefont {Duan},
  \citenamefont {Jaksch}, \citenamefont {Cirac},\ and\ \citenamefont
  {Zoller}}]{Lukin2001}%
  \BibitemOpen
  \bibfield  {author} {\bibinfo {author} {\bibfnamefont {M.~D.}\ \bibnamefont
  {Lukin}}, \bibinfo {author} {\bibfnamefont {M.}~\bibnamefont {Fleischhauer}},
  \bibinfo {author} {\bibfnamefont {R.}~\bibnamefont {Cote}}, \bibinfo {author}
  {\bibfnamefont {L.~M.}\ \bibnamefont {Duan}}, \bibinfo {author}
  {\bibfnamefont {D.}~\bibnamefont {Jaksch}}, \bibinfo {author} {\bibfnamefont
  {J.~I.}\ \bibnamefont {Cirac}},\ and\ \bibinfo {author} {\bibfnamefont
  {P.}~\bibnamefont {Zoller}},\ }\bibfield  {title} {\bibinfo {title} {Dipole
  blockade and quantum information processing in mesoscopic atomic ensembles},\
  }\href {https://doi.org/10.1103/PhysRevLett.87.037901} {\bibfield  {journal}
  {\bibinfo  {journal} {Phys. Rev. Lett.}\ }\textbf {\bibinfo {volume} {87}},\
  \bibinfo {pages} {037901} (\bibinfo {year} {2001})}\BibitemShut {NoStop}%
\bibitem [{\citenamefont {Sevin\ifmmode~\mbox{\c{c}}\else \c{c}\fi{}li}\ \emph
  {et~al.}(2011)\citenamefont {Sevin\ifmmode~\mbox{\c{c}}\else \c{c}\fi{}li},
  \citenamefont {Henkel}, \citenamefont {Ates},\ and\ \citenamefont
  {Pohl}}]{Sevincli2011}%
  \BibitemOpen
  \bibfield  {author} {\bibinfo {author} {\bibfnamefont {S.}~\bibnamefont
  {Sevin\ifmmode~\mbox{\c{c}}\else \c{c}\fi{}li}}, \bibinfo {author}
  {\bibfnamefont {N.}~\bibnamefont {Henkel}}, \bibinfo {author} {\bibfnamefont
  {C.}~\bibnamefont {Ates}},\ and\ \bibinfo {author} {\bibfnamefont
  {T.}~\bibnamefont {Pohl}},\ }\bibfield  {title} {\bibinfo {title} {Nonlocal
  nonlinear optics in cold rydberg gases},\ }\href
  {https://doi.org/10.1103/PhysRevLett.107.153001} {\bibfield  {journal}
  {\bibinfo  {journal} {Phys. Rev. Lett.}\ }\textbf {\bibinfo {volume} {107}},\
  \bibinfo {pages} {153001} (\bibinfo {year} {2011})}\BibitemShut {NoStop}%
\bibitem [{\citenamefont {Gorshkov}\ \emph {et~al.}(2011)\citenamefont
  {Gorshkov}, \citenamefont {Otterbach}, \citenamefont {Fleischhauer},
  \citenamefont {Pohl},\ and\ \citenamefont {Lukin}}]{Gorshkov2011}%
  \BibitemOpen
  \bibfield  {author} {\bibinfo {author} {\bibfnamefont {A.~V.}\ \bibnamefont
  {Gorshkov}}, \bibinfo {author} {\bibfnamefont {J.}~\bibnamefont {Otterbach}},
  \bibinfo {author} {\bibfnamefont {M.}~\bibnamefont {Fleischhauer}}, \bibinfo
  {author} {\bibfnamefont {T.}~\bibnamefont {Pohl}},\ and\ \bibinfo {author}
  {\bibfnamefont {M.~D.}\ \bibnamefont {Lukin}},\ }\bibfield  {title} {\bibinfo
  {title} {Photon-photon interactions via rydberg blockade},\ }\href
  {https://doi.org/10.1103/PhysRevLett.107.133602} {\bibfield  {journal}
  {\bibinfo  {journal} {Phys. Rev. Lett.}\ }\textbf {\bibinfo {volume} {107}},\
  \bibinfo {pages} {133602} (\bibinfo {year} {2011})}\BibitemShut {NoStop}%
\bibitem [{\citenamefont {Labuhn}\ \emph {et~al.}(2016)\citenamefont {Labuhn},
  \citenamefont {Barredo}, \citenamefont {Ravets}, \citenamefont
  {de~L{\'e}s{\'e}leuc}, \citenamefont {Macr{\`i}}, \citenamefont {Lahaye},\
  and\ \citenamefont {Browaeys}}]{Labuhn2016}%
  \BibitemOpen
  \bibfield  {author} {\bibinfo {author} {\bibfnamefont {H.}~\bibnamefont
  {Labuhn}}, \bibinfo {author} {\bibfnamefont {D.}~\bibnamefont {Barredo}},
  \bibinfo {author} {\bibfnamefont {S.}~\bibnamefont {Ravets}}, \bibinfo
  {author} {\bibfnamefont {S.}~\bibnamefont {de~L{\'e}s{\'e}leuc}}, \bibinfo
  {author} {\bibfnamefont {T.}~\bibnamefont {Macr{\`i}}}, \bibinfo {author}
  {\bibfnamefont {T.}~\bibnamefont {Lahaye}},\ and\ \bibinfo {author}
  {\bibfnamefont {A.}~\bibnamefont {Browaeys}},\ }\bibfield  {title} {\bibinfo
  {title} {Tunable two-dimensional arrays of single rydberg atoms for realizing
  quantum ising models},\ }\href {https://doi.org/10.1038/nature18274}
  {\bibfield  {journal} {\bibinfo  {journal} {Nature}\ }\textbf {\bibinfo
  {volume} {534}},\ \bibinfo {pages} {667} (\bibinfo {year}
  {2016})}\BibitemShut {NoStop}%
\bibitem [{\citenamefont {Bernien}\ \emph {et~al.}(2017)\citenamefont
  {Bernien}, \citenamefont {Schwartz}, \citenamefont {Keesling}, \citenamefont
  {Levine}, \citenamefont {Omran}, \citenamefont {Pichler}, \citenamefont
  {Choi}, \citenamefont {Zibrov}, \citenamefont {Endres}, \citenamefont
  {Greiner}, \citenamefont {Vuleti{\'{c}}},\ and\ \citenamefont
  {Lukin}}]{Bernien2017}%
  \BibitemOpen
  \bibfield  {author} {\bibinfo {author} {\bibfnamefont {H.}~\bibnamefont
  {Bernien}}, \bibinfo {author} {\bibfnamefont {S.}~\bibnamefont {Schwartz}},
  \bibinfo {author} {\bibfnamefont {A.}~\bibnamefont {Keesling}}, \bibinfo
  {author} {\bibfnamefont {H.}~\bibnamefont {Levine}}, \bibinfo {author}
  {\bibfnamefont {A.}~\bibnamefont {Omran}}, \bibinfo {author} {\bibfnamefont
  {H.}~\bibnamefont {Pichler}}, \bibinfo {author} {\bibfnamefont
  {S.}~\bibnamefont {Choi}}, \bibinfo {author} {\bibfnamefont {A.~S.}\
  \bibnamefont {Zibrov}}, \bibinfo {author} {\bibfnamefont {M.}~\bibnamefont
  {Endres}}, \bibinfo {author} {\bibfnamefont {M.}~\bibnamefont {Greiner}},
  \bibinfo {author} {\bibfnamefont {V.}~\bibnamefont {Vuleti{\'{c}}}},\ and\
  \bibinfo {author} {\bibfnamefont {M.~D.}\ \bibnamefont {Lukin}},\ }\bibfield
  {title} {\bibinfo {title} {Probing many-body dynamics on a 51-atom quantum
  simulator},\ }\href {https://doi.org/10.1038/nature24622} {\bibfield
  {journal} {\bibinfo  {journal} {Nature}\ }\textbf {\bibinfo {volume} {551}},\
  \bibinfo {pages} {579} (\bibinfo {year} {2017})}\BibitemShut {NoStop}%
\bibitem [{\citenamefont {Henriet}\ \emph {et~al.}(2020)\citenamefont
  {Henriet}, \citenamefont {Beguin}, \citenamefont {Signoles}, \citenamefont
  {Lahaye}, \citenamefont {Browaeys}, \citenamefont {Reymond},\ and\
  \citenamefont {Jurczak}}]{Henriet2020}%
  \BibitemOpen
  \bibfield  {author} {\bibinfo {author} {\bibfnamefont {L.}~\bibnamefont
  {Henriet}}, \bibinfo {author} {\bibfnamefont {L.}~\bibnamefont {Beguin}},
  \bibinfo {author} {\bibfnamefont {A.}~\bibnamefont {Signoles}}, \bibinfo
  {author} {\bibfnamefont {T.}~\bibnamefont {Lahaye}}, \bibinfo {author}
  {\bibfnamefont {A.}~\bibnamefont {Browaeys}}, \bibinfo {author}
  {\bibfnamefont {G.-O.}\ \bibnamefont {Reymond}},\ and\ \bibinfo {author}
  {\bibfnamefont {C.}~\bibnamefont {Jurczak}},\ }\bibfield  {title} {\bibinfo
  {title} {Quantum computing with neutral atoms},\ }\href
  {https://doi.org/10.22331/q-2020-09-21-327} {\bibfield  {journal} {\bibinfo
  {journal} {{Quantum}}\ }\textbf {\bibinfo {volume} {4}},\ \bibinfo {pages}
  {327} (\bibinfo {year} {2020})}\BibitemShut {NoStop}%
\bibitem [{\citenamefont {Kazimierczuk}\ \emph {et~al.}(2014)\citenamefont
  {Kazimierczuk}, \citenamefont {Fr{\"o}hlich}, \citenamefont {Scheel},
  \citenamefont {Stolz},\ and\ \citenamefont {Bayer}}]{Kazimierczuk2014}%
  \BibitemOpen
  \bibfield  {author} {\bibinfo {author} {\bibfnamefont {T.}~\bibnamefont
  {Kazimierczuk}}, \bibinfo {author} {\bibfnamefont {D.}~\bibnamefont
  {Fr{\"o}hlich}}, \bibinfo {author} {\bibfnamefont {S.}~\bibnamefont
  {Scheel}}, \bibinfo {author} {\bibfnamefont {H.}~\bibnamefont {Stolz}},\ and\
  \bibinfo {author} {\bibfnamefont {M.}~\bibnamefont {Bayer}},\ }\bibfield
  {title} {\bibinfo {title} {Giant rydberg excitons in the copper oxide cu2o},\
  }\href {https://doi.org/10.1038/nature13832} {\bibfield  {journal} {\bibinfo
  {journal} {Nature}\ }\textbf {\bibinfo {volume} {514}},\ \bibinfo {pages}
  {343} (\bibinfo {year} {2014})}\BibitemShut {NoStop}%
\bibitem [{\citenamefont {Versteegh}\ \emph {et~al.}(2021)\citenamefont
  {Versteegh}, \citenamefont {Steinhauer}, \citenamefont {Bajo}, \citenamefont
  {Lettner}, \citenamefont {Soro}, \citenamefont {Romanova}, \citenamefont
  {Gyger}, \citenamefont {Schweickert}, \citenamefont {Mysyrowicz},\ and\
  \citenamefont {Zwiller}}]{Versteegh2021}%
  \BibitemOpen
  \bibfield  {author} {\bibinfo {author} {\bibfnamefont {M.~A.~M.}\
  \bibnamefont {Versteegh}}, \bibinfo {author} {\bibfnamefont {S.}~\bibnamefont
  {Steinhauer}}, \bibinfo {author} {\bibfnamefont {J.}~\bibnamefont {Bajo}},
  \bibinfo {author} {\bibfnamefont {T.}~\bibnamefont {Lettner}}, \bibinfo
  {author} {\bibfnamefont {A.}~\bibnamefont {Soro}}, \bibinfo {author}
  {\bibfnamefont {A.}~\bibnamefont {Romanova}}, \bibinfo {author}
  {\bibfnamefont {S.}~\bibnamefont {Gyger}}, \bibinfo {author} {\bibfnamefont
  {L.}~\bibnamefont {Schweickert}}, \bibinfo {author} {\bibfnamefont
  {A.}~\bibnamefont {Mysyrowicz}},\ and\ \bibinfo {author} {\bibfnamefont
  {V.}~\bibnamefont {Zwiller}},\ }\bibfield  {title} {\bibinfo {title} {Giant
  rydberg excitons in ${\mathrm{cu}}_{2}\mathrm{O}$ probed by photoluminescence
  excitation spectroscopy},\ }\href
  {https://doi.org/10.1103/PhysRevB.104.245206} {\bibfield  {journal} {\bibinfo
   {journal} {Phys. Rev. B}\ }\textbf {\bibinfo {volume} {104}},\ \bibinfo
  {pages} {245206} (\bibinfo {year} {2021})}\BibitemShut {NoStop}%
\bibitem [{\citenamefont {Gallagher}\ \emph {et~al.}(2022)\citenamefont
  {Gallagher}, \citenamefont {Rogers}, \citenamefont {Pritchett}, \citenamefont
  {Mistry}, \citenamefont {Pizzey}, \citenamefont {Adams}, \citenamefont
  {Jones}, \citenamefont {Gr\"unwald}, \citenamefont {Walther}, \citenamefont
  {Hodges}, \citenamefont {Langbein},\ and\ \citenamefont
  {Lynch}}]{Gallagher2022}%
  \BibitemOpen
  \bibfield  {author} {\bibinfo {author} {\bibfnamefont {L.~A.~P.}\
  \bibnamefont {Gallagher}}, \bibinfo {author} {\bibfnamefont {J.~P.}\
  \bibnamefont {Rogers}}, \bibinfo {author} {\bibfnamefont {J.~D.}\
  \bibnamefont {Pritchett}}, \bibinfo {author} {\bibfnamefont {R.~A.}\
  \bibnamefont {Mistry}}, \bibinfo {author} {\bibfnamefont {D.}~\bibnamefont
  {Pizzey}}, \bibinfo {author} {\bibfnamefont {C.~S.}\ \bibnamefont {Adams}},
  \bibinfo {author} {\bibfnamefont {M.~P.~A.}\ \bibnamefont {Jones}}, \bibinfo
  {author} {\bibfnamefont {P.}~\bibnamefont {Gr\"unwald}}, \bibinfo {author}
  {\bibfnamefont {V.}~\bibnamefont {Walther}}, \bibinfo {author} {\bibfnamefont
  {C.}~\bibnamefont {Hodges}}, \bibinfo {author} {\bibfnamefont
  {W.}~\bibnamefont {Langbein}},\ and\ \bibinfo {author} {\bibfnamefont
  {S.~A.}\ \bibnamefont {Lynch}},\ }\bibfield  {title} {\bibinfo {title}
  {Microwave-optical coupling via rydberg excitons in cuprous oxide},\ }\href
  {https://doi.org/10.1103/PhysRevResearch.4.013031} {\bibfield  {journal}
  {\bibinfo  {journal} {Phys. Rev. Research}\ }\textbf {\bibinfo {volume}
  {4}},\ \bibinfo {pages} {013031} (\bibinfo {year} {2022})}\BibitemShut
  {NoStop}%
\bibitem [{\citenamefont {Orfanakis}\ \emph {et~al.}(2022)\citenamefont
  {Orfanakis}, \citenamefont {Rajendran}, \citenamefont {Walther},
  \citenamefont {Volz}, \citenamefont {Pohl},\ and\ \citenamefont
  {Ohadi}}]{Orfanakis2022}%
  \BibitemOpen
  \bibfield  {author} {\bibinfo {author} {\bibfnamefont {K.}~\bibnamefont
  {Orfanakis}}, \bibinfo {author} {\bibfnamefont {S.~K.}\ \bibnamefont
  {Rajendran}}, \bibinfo {author} {\bibfnamefont {V.}~\bibnamefont {Walther}},
  \bibinfo {author} {\bibfnamefont {T.}~\bibnamefont {Volz}}, \bibinfo {author}
  {\bibfnamefont {T.}~\bibnamefont {Pohl}},\ and\ \bibinfo {author}
  {\bibfnamefont {H.}~\bibnamefont {Ohadi}},\ }\bibfield  {title} {\bibinfo
  {title} {Rydberg exciton--polaritons in a cu2o microcavity},\ }\href
  {https://doi.org/10.1038/s41563-022-01230-4} {\bibfield  {journal} {\bibinfo
  {journal} {Nature Materials}\ }\textbf {\bibinfo {volume} {21}},\ \bibinfo
  {pages} {767} (\bibinfo {year} {2022})}\BibitemShut {NoStop}%
\bibitem [{\citenamefont {Alhassid}(2000)}]{Alhassid2000}%
  \BibitemOpen
  \bibfield  {author} {\bibinfo {author} {\bibfnamefont {Y.}~\bibnamefont
  {Alhassid}},\ }\bibfield  {title} {\bibinfo {title} {The statistical theory
  of quantum dots},\ }\href {https://doi.org/10.1103/RevModPhys.72.895}
  {\bibfield  {journal} {\bibinfo  {journal} {Rev. Mod. Phys.}\ }\textbf
  {\bibinfo {volume} {72}},\ \bibinfo {pages} {895} (\bibinfo {year}
  {2000})}\BibitemShut {NoStop}%
\bibitem [{\citenamefont {Imamo\ifmmode~\bar{g}\else \={g}\fi{}lu}\ \emph
  {et~al.}(1997)\citenamefont {Imamo\ifmmode~\bar{g}\else \={g}\fi{}lu},
  \citenamefont {Schmidt}, \citenamefont {Woods},\ and\ \citenamefont
  {Deutsch}}]{Imamoglu1997}%
  \BibitemOpen
  \bibfield  {author} {\bibinfo {author} {\bibfnamefont {A.}~\bibnamefont
  {Imamo\ifmmode~\bar{g}\else \={g}\fi{}lu}}, \bibinfo {author} {\bibfnamefont
  {H.}~\bibnamefont {Schmidt}}, \bibinfo {author} {\bibfnamefont
  {G.}~\bibnamefont {Woods}},\ and\ \bibinfo {author} {\bibfnamefont
  {M.}~\bibnamefont {Deutsch}},\ }\bibfield  {title} {\bibinfo {title}
  {Strongly interacting photons in a nonlinear cavity},\ }\href
  {https://doi.org/10.1103/PhysRevLett.79.1467} {\bibfield  {journal} {\bibinfo
   {journal} {Phys. Rev. Lett.}\ }\textbf {\bibinfo {volume} {79}},\ \bibinfo
  {pages} {1467} (\bibinfo {year} {1997})}\BibitemShut {NoStop}%
\bibitem [{\citenamefont {Verger}\ \emph {et~al.}(2006)\citenamefont {Verger},
  \citenamefont {Ciuti},\ and\ \citenamefont {Carusotto}}]{Verger2006}%
  \BibitemOpen
  \bibfield  {author} {\bibinfo {author} {\bibfnamefont {A.}~\bibnamefont
  {Verger}}, \bibinfo {author} {\bibfnamefont {C.}~\bibnamefont {Ciuti}},\ and\
  \bibinfo {author} {\bibfnamefont {I.}~\bibnamefont {Carusotto}},\ }\bibfield
  {title} {\bibinfo {title} {Polariton quantum blockade in a photonic dot},\
  }\href {https://doi.org/10.1103/PhysRevB.73.193306} {\bibfield  {journal}
  {\bibinfo  {journal} {Phys. Rev. B}\ }\textbf {\bibinfo {volume} {73}},\
  \bibinfo {pages} {193306} (\bibinfo {year} {2006})}\BibitemShut {NoStop}%
\bibitem [{\citenamefont {Schmitt-Rink}\ \emph {et~al.}(1985)\citenamefont
  {Schmitt-Rink}, \citenamefont {Chemla},\ and\ \citenamefont
  {Miller}}]{Schmitt-Rink1985}%
  \BibitemOpen
  \bibfield  {author} {\bibinfo {author} {\bibfnamefont {S.}~\bibnamefont
  {Schmitt-Rink}}, \bibinfo {author} {\bibfnamefont {D.~S.}\ \bibnamefont
  {Chemla}},\ and\ \bibinfo {author} {\bibfnamefont {D.~A.~B.}\ \bibnamefont
  {Miller}},\ }\bibfield  {title} {\bibinfo {title} {Theory of transient
  excitonic optical nonlinearities in semiconductor quantum-well structures},\
  }\href {https://doi.org/10.1103/PhysRevB.32.6601} {\bibfield  {journal}
  {\bibinfo  {journal} {Phys. Rev. B}\ }\textbf {\bibinfo {volume} {32}},\
  \bibinfo {pages} {6601} (\bibinfo {year} {1985})}\BibitemShut {NoStop}%
\bibitem [{\citenamefont {Combescot}\ \emph {et~al.}(2008)\citenamefont
  {Combescot}, \citenamefont {Betbeder-Matibet},\ and\ \citenamefont
  {Dubin}}]{Combescot2008}%
  \BibitemOpen
  \bibfield  {author} {\bibinfo {author} {\bibfnamefont {M.}~\bibnamefont
  {Combescot}}, \bibinfo {author} {\bibfnamefont {O.}~\bibnamefont
  {Betbeder-Matibet}},\ and\ \bibinfo {author} {\bibfnamefont {F.}~\bibnamefont
  {Dubin}},\ }\bibfield  {title} {\bibinfo {title} {The many-body physics of
  composite bosons},\ }\href@noop {} {\bibfield  {journal} {\bibinfo  {journal}
  {Physics Reports}\ }\textbf {\bibinfo {volume} {463}},\ \bibinfo {pages}
  {215} (\bibinfo {year} {2008})}\BibitemShut {NoStop}%
\bibitem [{\citenamefont {Combescot}\ and\ \citenamefont
  {Pogosov}(2009)}]{Combescot:EPJB68(2009)}%
  \BibitemOpen
  \bibfield  {author} {\bibinfo {author} {\bibfnamefont {M.}~\bibnamefont
  {Combescot}}\ and\ \bibinfo {author} {\bibfnamefont {W.}~\bibnamefont
  {Pogosov}},\ }\bibfield  {title} {\bibinfo {title} {Composite boson many-body
  theory for frenkel excitons},\ }\href
  {https://doi.org/10.1140/epjb/e2009-00086-6} {\bibfield  {journal} {\bibinfo
  {journal} {The European Physical Journal B}\ }\textbf {\bibinfo {volume}
  {68}},\ \bibinfo {pages} {161} (\bibinfo {year} {2009})}\BibitemShut
  {NoStop}%
\bibitem [{\citenamefont {Betzold}\ \emph {et~al.}(2020)\citenamefont
  {Betzold}, \citenamefont {Dusel}, \citenamefont {Kyriienko}, \citenamefont
  {Dietrich}, \citenamefont {Klembt}, \citenamefont {Ohmer}, \citenamefont
  {Fischer}, \citenamefont {Shelykh}, \citenamefont {Schneider},\ and\
  \citenamefont {H{\"o}fling}}]{Betzold2020}%
  \BibitemOpen
  \bibfield  {author} {\bibinfo {author} {\bibfnamefont {S.}~\bibnamefont
  {Betzold}}, \bibinfo {author} {\bibfnamefont {M.}~\bibnamefont {Dusel}},
  \bibinfo {author} {\bibfnamefont {O.}~\bibnamefont {Kyriienko}}, \bibinfo
  {author} {\bibfnamefont {C.~P.}\ \bibnamefont {Dietrich}}, \bibinfo {author}
  {\bibfnamefont {S.}~\bibnamefont {Klembt}}, \bibinfo {author} {\bibfnamefont
  {J.}~\bibnamefont {Ohmer}}, \bibinfo {author} {\bibfnamefont
  {U.}~\bibnamefont {Fischer}}, \bibinfo {author} {\bibfnamefont {I.~A.}\
  \bibnamefont {Shelykh}}, \bibinfo {author} {\bibfnamefont {C.}~\bibnamefont
  {Schneider}},\ and\ \bibinfo {author} {\bibfnamefont {S.}~\bibnamefont
  {H{\"o}fling}},\ }\bibfield  {title} {\bibinfo {title} {Coherence and
  interaction in confined room-temperature polariton condensates with frenkel
  excitons},\ }\href {https://doi.org/10.1021/acsphotonics.9b01300} {\bibfield
  {journal} {\bibinfo  {journal} {ACS Photonics}\ }\textbf {\bibinfo {volume}
  {7}},\ \bibinfo {pages} {384} (\bibinfo {year} {2020})}\BibitemShut {NoStop}%
\bibitem [{\citenamefont {Arnardottir}\ \emph {et~al.}(2020)\citenamefont
  {Arnardottir}, \citenamefont {Moilanen}, \citenamefont {Strashko},
  \citenamefont {T\"orm\"a},\ and\ \citenamefont {Keeling}}]{Arnardottir2020}%
  \BibitemOpen
  \bibfield  {author} {\bibinfo {author} {\bibfnamefont {K.~B.}\ \bibnamefont
  {Arnardottir}}, \bibinfo {author} {\bibfnamefont {A.~J.}\ \bibnamefont
  {Moilanen}}, \bibinfo {author} {\bibfnamefont {A.}~\bibnamefont {Strashko}},
  \bibinfo {author} {\bibfnamefont {P.}~\bibnamefont {T\"orm\"a}},\ and\
  \bibinfo {author} {\bibfnamefont {J.}~\bibnamefont {Keeling}},\ }\bibfield
  {title} {\bibinfo {title} {Multimode organic polariton lasing},\ }\href
  {https://doi.org/10.1103/PhysRevLett.125.233603} {\bibfield  {journal}
  {\bibinfo  {journal} {Phys. Rev. Lett.}\ }\textbf {\bibinfo {volume} {125}},\
  \bibinfo {pages} {233603} (\bibinfo {year} {2020})}\BibitemShut {NoStop}%
\bibitem [{\citenamefont {Daskalakis}\ \emph {et~al.}(2014)\citenamefont
  {Daskalakis}, \citenamefont {Maier}, \citenamefont {Murray},\ and\
  \citenamefont {K{\'e}na-Cohen}}]{Daskalakis2014}%
  \BibitemOpen
  \bibfield  {author} {\bibinfo {author} {\bibfnamefont {K.~S.}\ \bibnamefont
  {Daskalakis}}, \bibinfo {author} {\bibfnamefont {S.~A.}\ \bibnamefont
  {Maier}}, \bibinfo {author} {\bibfnamefont {R.}~\bibnamefont {Murray}},\ and\
  \bibinfo {author} {\bibfnamefont {S.}~\bibnamefont {K{\'e}na-Cohen}},\
  }\bibfield  {title} {\bibinfo {title} {Nonlinear interactions in an organic
  polariton condensate},\ }\href {https://doi.org/10.1038/nmat3874} {\bibfield
  {journal} {\bibinfo  {journal} {Nature Materials}\ }\textbf {\bibinfo
  {volume} {13}},\ \bibinfo {pages} {271} (\bibinfo {year} {2014})}\BibitemShut
  {NoStop}%
\bibitem [{\citenamefont {Keeling}\ and\ \citenamefont
  {K\'{e}na-Cohen}(2020)}]{Keeling2020}%
  \BibitemOpen
  \bibfield  {author} {\bibinfo {author} {\bibfnamefont {J.}~\bibnamefont
  {Keeling}}\ and\ \bibinfo {author} {\bibfnamefont {S.}~\bibnamefont
  {K\'{e}na-Cohen}},\ }\bibfield  {title} {\bibinfo {title} {Bose–einstein
  condensation of exciton-polaritons in organic microcavities},\ }\href
  {https://doi.org/10.1146/annurev-physchem-010920-102509} {\bibfield
  {journal} {\bibinfo  {journal} {Annual Review of Physical Chemistry}\
  }\textbf {\bibinfo {volume} {71}},\ \bibinfo {pages} {435} (\bibinfo {year}
  {2020})},\ \bibinfo {note} {pMID: 32126177}\BibitemShut {NoStop}%
\bibitem [{\citenamefont {Yagafarov}\ \emph {et~al.}(2020)\citenamefont
  {Yagafarov}, \citenamefont {Sannikov}, \citenamefont {Zasedatelev},
  \citenamefont {Georgiou}, \citenamefont {Baranikov}, \citenamefont
  {Kyriienko}, \citenamefont {Shelykh}, \citenamefont {Gai}, \citenamefont
  {Shen}, \citenamefont {Lidzey},\ and\ \citenamefont
  {Lagoudakis}}]{Yagafarov2020}%
  \BibitemOpen
  \bibfield  {author} {\bibinfo {author} {\bibfnamefont {T.}~\bibnamefont
  {Yagafarov}}, \bibinfo {author} {\bibfnamefont {D.}~\bibnamefont {Sannikov}},
  \bibinfo {author} {\bibfnamefont {A.}~\bibnamefont {Zasedatelev}}, \bibinfo
  {author} {\bibfnamefont {K.}~\bibnamefont {Georgiou}}, \bibinfo {author}
  {\bibfnamefont {A.}~\bibnamefont {Baranikov}}, \bibinfo {author}
  {\bibfnamefont {O.}~\bibnamefont {Kyriienko}}, \bibinfo {author}
  {\bibfnamefont {I.}~\bibnamefont {Shelykh}}, \bibinfo {author} {\bibfnamefont
  {L.}~\bibnamefont {Gai}}, \bibinfo {author} {\bibfnamefont {Z.}~\bibnamefont
  {Shen}}, \bibinfo {author} {\bibfnamefont {D.}~\bibnamefont {Lidzey}},\ and\
  \bibinfo {author} {\bibfnamefont {P.}~\bibnamefont {Lagoudakis}},\ }\bibfield
   {title} {\bibinfo {title} {Mechanisms of blueshifts in organic polariton
  condensates},\ }\href {https://doi.org/10.1038/s42005-019-0278-6} {\bibfield
  {journal} {\bibinfo  {journal} {Commun. Phys.}\ }\textbf {\bibinfo {volume}
  {3}},\ \bibinfo {pages} {18} (\bibinfo {year} {2020})}\BibitemShut {NoStop}%
\bibitem [{\citenamefont {Tan}\ \emph {et~al.}(2020)\citenamefont {Tan},
  \citenamefont {Cotlet}, \citenamefont {Bergschneider}, \citenamefont
  {Schmidt}, \citenamefont {Back}, \citenamefont {Shimazaki}, \citenamefont
  {Kroner},\ and\ \citenamefont {\ifmmode \dot{I}\else
  \.{I}\fi{}mamo\ifmmode~\breve{g}\else \u{g}\fi{}lu}}]{Tan2020}%
  \BibitemOpen
  \bibfield  {author} {\bibinfo {author} {\bibfnamefont {L.~B.}\ \bibnamefont
  {Tan}}, \bibinfo {author} {\bibfnamefont {O.}~\bibnamefont {Cotlet}},
  \bibinfo {author} {\bibfnamefont {A.}~\bibnamefont {Bergschneider}}, \bibinfo
  {author} {\bibfnamefont {R.}~\bibnamefont {Schmidt}}, \bibinfo {author}
  {\bibfnamefont {P.}~\bibnamefont {Back}}, \bibinfo {author} {\bibfnamefont
  {Y.}~\bibnamefont {Shimazaki}}, \bibinfo {author} {\bibfnamefont
  {M.}~\bibnamefont {Kroner}},\ and\ \bibinfo {author} {\bibfnamefont
  {A.~m.~c.}\ \bibnamefont {\ifmmode \dot{I}\else
  \.{I}\fi{}mamo\ifmmode~\breve{g}\else \u{g}\fi{}lu}},\ }\bibfield  {title}
  {\bibinfo {title} {Interacting polaron-polaritons},\ }\href
  {https://doi.org/10.1103/PhysRevX.10.021011} {\bibfield  {journal} {\bibinfo
  {journal} {Phys. Rev. X}\ }\textbf {\bibinfo {volume} {10}},\ \bibinfo
  {pages} {021011} (\bibinfo {year} {2020})}\BibitemShut {NoStop}%
\bibitem [{\citenamefont {Kravtsov}\ \emph {et~al.}(2020)\citenamefont
  {Kravtsov}, \citenamefont {Khestanova}, \citenamefont {Benimetskiy},
  \citenamefont {Ivanova}, \citenamefont {Samusev}, \citenamefont {Sinev},
  \citenamefont {Pidgayko}, \citenamefont {Mozharov}, \citenamefont {Mukhin},
  \citenamefont {Lozhkin}, \citenamefont {Kapitonov}, \citenamefont {Brichkin},
  \citenamefont {Kulakovskii}, \citenamefont {Shelykh}, \citenamefont
  {Tartakovskii}, \citenamefont {Walker}, \citenamefont {Skolnick},
  \citenamefont {Krizhanovskii},\ and\ \citenamefont {Iorsh}}]{Kravtsov2020}%
  \BibitemOpen
  \bibfield  {author} {\bibinfo {author} {\bibfnamefont {V.}~\bibnamefont
  {Kravtsov}}, \bibinfo {author} {\bibfnamefont {E.}~\bibnamefont
  {Khestanova}}, \bibinfo {author} {\bibfnamefont {F.~A.}\ \bibnamefont
  {Benimetskiy}}, \bibinfo {author} {\bibfnamefont {T.}~\bibnamefont
  {Ivanova}}, \bibinfo {author} {\bibfnamefont {A.~K.}\ \bibnamefont
  {Samusev}}, \bibinfo {author} {\bibfnamefont {I.~S.}\ \bibnamefont {Sinev}},
  \bibinfo {author} {\bibfnamefont {D.}~\bibnamefont {Pidgayko}}, \bibinfo
  {author} {\bibfnamefont {A.~M.}\ \bibnamefont {Mozharov}}, \bibinfo {author}
  {\bibfnamefont {I.~S.}\ \bibnamefont {Mukhin}}, \bibinfo {author}
  {\bibfnamefont {M.~S.}\ \bibnamefont {Lozhkin}}, \bibinfo {author}
  {\bibfnamefont {Y.~V.}\ \bibnamefont {Kapitonov}}, \bibinfo {author}
  {\bibfnamefont {A.~S.}\ \bibnamefont {Brichkin}}, \bibinfo {author}
  {\bibfnamefont {V.~D.}\ \bibnamefont {Kulakovskii}}, \bibinfo {author}
  {\bibfnamefont {I.~A.}\ \bibnamefont {Shelykh}}, \bibinfo {author}
  {\bibfnamefont {A.~I.}\ \bibnamefont {Tartakovskii}}, \bibinfo {author}
  {\bibfnamefont {P.~M.}\ \bibnamefont {Walker}}, \bibinfo {author}
  {\bibfnamefont {M.~S.}\ \bibnamefont {Skolnick}}, \bibinfo {author}
  {\bibfnamefont {D.~N.}\ \bibnamefont {Krizhanovskii}},\ and\ \bibinfo
  {author} {\bibfnamefont {I.~V.}\ \bibnamefont {Iorsh}},\ }\bibfield  {title}
  {\bibinfo {title} {Nonlinear polaritons in a monolayer semiconductor coupled
  to optical bound states in the continuum},\ }\href
  {https://doi.org/10.1038/s41377-020-0286-z} {\bibfield  {journal} {\bibinfo
  {journal} {Light Sci Appl}\ }\textbf {\bibinfo {volume} {9}} (\bibinfo {year}
  {2020})}\BibitemShut {NoStop}%
\bibitem [{\citenamefont {Kyriienko}\ \emph {et~al.}(2020)\citenamefont
  {Kyriienko}, \citenamefont {Krizhanovskii},\ and\ \citenamefont
  {Shelykh}}]{Kyriienko:PRL125(2020)}%
  \BibitemOpen
  \bibfield  {author} {\bibinfo {author} {\bibfnamefont {O.}~\bibnamefont
  {Kyriienko}}, \bibinfo {author} {\bibfnamefont {D.~N.}\ \bibnamefont
  {Krizhanovskii}},\ and\ \bibinfo {author} {\bibfnamefont {I.~A.}\
  \bibnamefont {Shelykh}},\ }\bibfield  {title} {\bibinfo {title} {Nonlinear
  quantum optics with trion polaritons in 2d monolayers: Conventional and
  unconventional photon blockade},\ }\href
  {https://doi.org/10.1103/PhysRevLett.125.197402} {\bibfield  {journal}
  {\bibinfo  {journal} {Phys. Rev. Lett.}\ }\textbf {\bibinfo {volume} {125}},\
  \bibinfo {pages} {197402} (\bibinfo {year} {2020})}\BibitemShut {NoStop}%
\bibitem [{\citenamefont {Zhang}\ \emph {et~al.}(2021)\citenamefont {Zhang},
  \citenamefont {Wu}, \citenamefont {Hou}, \citenamefont {Zhang}, \citenamefont
  {Chou}, \citenamefont {Watanabe}, \citenamefont {Taniguchi}, \citenamefont
  {Forrest},\ and\ \citenamefont {Deng}}]{Zhang:Nature591(2021)}%
  \BibitemOpen
  \bibfield  {author} {\bibinfo {author} {\bibfnamefont {L.}~\bibnamefont
  {Zhang}}, \bibinfo {author} {\bibfnamefont {F.}~\bibnamefont {Wu}}, \bibinfo
  {author} {\bibfnamefont {S.}~\bibnamefont {Hou}}, \bibinfo {author}
  {\bibfnamefont {Z.}~\bibnamefont {Zhang}}, \bibinfo {author} {\bibfnamefont
  {Y.-H.}\ \bibnamefont {Chou}}, \bibinfo {author} {\bibfnamefont
  {K.}~\bibnamefont {Watanabe}}, \bibinfo {author} {\bibfnamefont
  {T.}~\bibnamefont {Taniguchi}}, \bibinfo {author} {\bibfnamefont {S.~R.}\
  \bibnamefont {Forrest}},\ and\ \bibinfo {author} {\bibfnamefont
  {H.}~\bibnamefont {Deng}},\ }\bibfield  {title} {\bibinfo {title} {Van der
  waals heterostructure polaritons with moir{\'{e}}-induced nonlinearity},\
  }\href {https://doi.org/10.1038/s41586-021-03228-5} {\bibfield  {journal}
  {\bibinfo  {journal} {Nature}\ }\textbf {\bibinfo {volume} {591}},\ \bibinfo
  {pages} {61} (\bibinfo {year} {2021})}\BibitemShut {NoStop}%
\bibitem [{\citenamefont {Stepanov}\ \emph {et~al.}(2021)\citenamefont
  {Stepanov}, \citenamefont {Vashisht}, \citenamefont {Klaas}, \citenamefont
  {Lundt}, \citenamefont {Tongay}, \citenamefont {Blei}, \citenamefont
  {Höfling}, \citenamefont {Volz}, \citenamefont {Minguzzi}, \citenamefont
  {Renard}, \citenamefont {Schneider},\ and\ \citenamefont
  {Richard}}]{Stepanov:PRL126(2021)}%
  \BibitemOpen
  \bibfield  {author} {\bibinfo {author} {\bibfnamefont {P.}~\bibnamefont
  {Stepanov}}, \bibinfo {author} {\bibfnamefont {A.}~\bibnamefont {Vashisht}},
  \bibinfo {author} {\bibfnamefont {M.}~\bibnamefont {Klaas}}, \bibinfo
  {author} {\bibfnamefont {N.}~\bibnamefont {Lundt}}, \bibinfo {author}
  {\bibfnamefont {S.}~\bibnamefont {Tongay}}, \bibinfo {author} {\bibfnamefont
  {M.}~\bibnamefont {Blei}}, \bibinfo {author} {\bibfnamefont {S.}~\bibnamefont
  {Höfling}}, \bibinfo {author} {\bibfnamefont {T.}~\bibnamefont {Volz}},
  \bibinfo {author} {\bibfnamefont {A.}~\bibnamefont {Minguzzi}}, \bibinfo
  {author} {\bibfnamefont {J.}~\bibnamefont {Renard}}, \bibinfo {author}
  {\bibfnamefont {C.}~\bibnamefont {Schneider}},\ and\ \bibinfo {author}
  {\bibfnamefont {M.}~\bibnamefont {Richard}},\ }\bibfield  {title} {\bibinfo
  {title} {Exciton-exciton interaction beyond the hydrogenic picture in a
  {MoSe}2 monolayer in the strong light-matter coupling regime},\ }\href
  {https://doi.org/10.1103/physrevlett.126.167401} {\bibfield  {journal}
  {\bibinfo  {journal} {Phys. Rev. Lett.}\ }\textbf {\bibinfo {volume} {126}},\
  \bibinfo {pages} {167401} (\bibinfo {year} {2021})}\BibitemShut {NoStop}%
\bibitem [{\citenamefont {Bastarrachea-Magnani}\ \emph
  {et~al.}(2021)\citenamefont {Bastarrachea-Magnani}, \citenamefont
  {Camacho-Guardian},\ and\ \citenamefont
  {Bruun}}]{BastarracheaMagnani:PRL126(2021)}%
  \BibitemOpen
  \bibfield  {author} {\bibinfo {author} {\bibfnamefont {M.~A.}\ \bibnamefont
  {Bastarrachea-Magnani}}, \bibinfo {author} {\bibfnamefont {A.}~\bibnamefont
  {Camacho-Guardian}},\ and\ \bibinfo {author} {\bibfnamefont {G.~M.}\
  \bibnamefont {Bruun}},\ }\bibfield  {title} {\bibinfo {title} {Attractive and
  repulsive exciton-polariton interactions mediated by an electron gas},\
  }\href {https://doi.org/10.1103/physrevlett.126.127405} {\bibfield  {journal}
  {\bibinfo  {journal} {Phys. Rev. Lett.}\ }\textbf {\bibinfo {volume} {126}},\
  \bibinfo {pages} {127405} (\bibinfo {year} {2021})}\BibitemShut {NoStop}%
\bibitem [{\citenamefont {Denning}\ \emph
  {et~al.}(2022{\natexlab{a}})\citenamefont {Denning}, \citenamefont {Wubs},
  \citenamefont {Stenger}, \citenamefont {M{\o}rk},\ and\ \citenamefont
  {Kristensen}}]{Denning:PRB105(2022)}%
  \BibitemOpen
  \bibfield  {author} {\bibinfo {author} {\bibfnamefont {E.~V.}\ \bibnamefont
  {Denning}}, \bibinfo {author} {\bibfnamefont {M.}~\bibnamefont {Wubs}},
  \bibinfo {author} {\bibfnamefont {N.}~\bibnamefont {Stenger}}, \bibinfo
  {author} {\bibfnamefont {J.}~\bibnamefont {M{\o}rk}},\ and\ \bibinfo {author}
  {\bibfnamefont {P.~T.}\ \bibnamefont {Kristensen}},\ }\bibfield  {title}
  {\bibinfo {title} {Quantum theory of two-dimensional materials coupled to
  electromagnetic resonators},\ }\href
  {https://doi.org/10.1103/physrevb.105.085306} {\bibfield  {journal} {\bibinfo
   {journal} {Phys. Rev. B}\ }\textbf {\bibinfo {volume} {105}},\ \bibinfo
  {pages} {085306} (\bibinfo {year} {2022}{\natexlab{a}})}\BibitemShut
  {NoStop}%
\bibitem [{\citenamefont {Denning}\ \emph
  {et~al.}(2022{\natexlab{b}})\citenamefont {Denning}, \citenamefont {Wubs},
  \citenamefont {Stenger}, \citenamefont {M\o{}rk},\ and\ \citenamefont
  {Kristensen}}]{Denning:PhysRevResearch(2022)}%
  \BibitemOpen
  \bibfield  {author} {\bibinfo {author} {\bibfnamefont {E.~V.}\ \bibnamefont
  {Denning}}, \bibinfo {author} {\bibfnamefont {M.}~\bibnamefont {Wubs}},
  \bibinfo {author} {\bibfnamefont {N.}~\bibnamefont {Stenger}}, \bibinfo
  {author} {\bibfnamefont {J.}~\bibnamefont {M\o{}rk}},\ and\ \bibinfo {author}
  {\bibfnamefont {P.~T.}\ \bibnamefont {Kristensen}},\ }\bibfield  {title}
  {\bibinfo {title} {Cavity-induced exciton localization and polariton blockade
  in two-dimensional semiconductors coupled to an electromagnetic resonator},\
  }\href {https://doi.org/10.1103/PhysRevResearch.4.L012020} {\bibfield
  {journal} {\bibinfo  {journal} {Phys. Rev. Research}\ }\textbf {\bibinfo
  {volume} {4}},\ \bibinfo {pages} {L012020} (\bibinfo {year}
  {2022}{\natexlab{b}})}\BibitemShut {NoStop}%
\bibitem [{\citenamefont {Datta}\ \emph {et~al.}(2022)\citenamefont {Datta},
  \citenamefont {Khatoniar}, \citenamefont {Deshmukh}, \citenamefont {Thouin},
  \citenamefont {Bushati}, \citenamefont {De~Liberato}, \citenamefont {Cohen},\
  and\ \citenamefont {Menon}}]{Datta2022}%
  \BibitemOpen
  \bibfield  {author} {\bibinfo {author} {\bibfnamefont {B.}~\bibnamefont
  {Datta}}, \bibinfo {author} {\bibfnamefont {M.}~\bibnamefont {Khatoniar}},
  \bibinfo {author} {\bibfnamefont {P.}~\bibnamefont {Deshmukh}}, \bibinfo
  {author} {\bibfnamefont {F.}~\bibnamefont {Thouin}}, \bibinfo {author}
  {\bibfnamefont {R.}~\bibnamefont {Bushati}}, \bibinfo {author} {\bibfnamefont
  {S.}~\bibnamefont {De~Liberato}}, \bibinfo {author} {\bibfnamefont {S.~K.}\
  \bibnamefont {Cohen}},\ and\ \bibinfo {author} {\bibfnamefont {V.~M.}\
  \bibnamefont {Menon}},\ }\bibfield  {title} {\bibinfo {title} {Highly
  nonlinear dipolar exciton-polaritons in bilayer mos2},\ }\href
  {https://doi.org/10.1038/s41467-022-33940-3} {\bibfield  {journal} {\bibinfo
  {journal} {Nature Communications}\ }\textbf {\bibinfo {volume} {13}},\
  \bibinfo {pages} {6341} (\bibinfo {year} {2022})}\BibitemShut {NoStop}%
\bibitem [{\citenamefont {Louca}\ \emph {et~al.}(2022)\citenamefont {Louca},
  \citenamefont {Genco}, \citenamefont {Chiavazzo}, \citenamefont {Lyons},
  \citenamefont {Randerson}, \citenamefont {Trovatello}, \citenamefont
  {Claronino}, \citenamefont {Jayaprakash}, \citenamefont {Watanabe},
  \citenamefont {Taniguchi}, \citenamefont {Conte}, \citenamefont {Lidzey},
  \citenamefont {Cerullo}, \citenamefont {Kyriienko},\ and\ \citenamefont
  {Tartakovskii}}]{Louca2022}%
  \BibitemOpen
  \bibfield  {author} {\bibinfo {author} {\bibfnamefont {C.}~\bibnamefont
  {Louca}}, \bibinfo {author} {\bibfnamefont {A.}~\bibnamefont {Genco}},
  \bibinfo {author} {\bibfnamefont {S.}~\bibnamefont {Chiavazzo}}, \bibinfo
  {author} {\bibfnamefont {T.~P.}\ \bibnamefont {Lyons}}, \bibinfo {author}
  {\bibfnamefont {S.}~\bibnamefont {Randerson}}, \bibinfo {author}
  {\bibfnamefont {C.}~\bibnamefont {Trovatello}}, \bibinfo {author}
  {\bibfnamefont {P.}~\bibnamefont {Claronino}}, \bibinfo {author}
  {\bibfnamefont {R.}~\bibnamefont {Jayaprakash}}, \bibinfo {author}
  {\bibfnamefont {K.}~\bibnamefont {Watanabe}}, \bibinfo {author}
  {\bibfnamefont {T.}~\bibnamefont {Taniguchi}}, \bibinfo {author}
  {\bibfnamefont {S.~D.}\ \bibnamefont {Conte}}, \bibinfo {author}
  {\bibfnamefont {D.~G.}\ \bibnamefont {Lidzey}}, \bibinfo {author}
  {\bibfnamefont {G.}~\bibnamefont {Cerullo}}, \bibinfo {author} {\bibfnamefont
  {O.}~\bibnamefont {Kyriienko}},\ and\ \bibinfo {author} {\bibfnamefont
  {A.~I.}\ \bibnamefont {Tartakovskii}},\ }\href
  {https://doi.org/10.48550/arXiv.2204.00485} {\bibinfo {title} {Nonlinear
  interactions of dipolar excitons and polaritons in mos2 bilayers}} (\bibinfo
  {year} {2022}),\ \Eprint {https://arxiv.org/abs/arXiv:2204.00485}
  {arXiv:2204.00485} \BibitemShut {NoStop}%
\bibitem [{\citenamefont {Combescot}\ \emph {et~al.}(2003)\citenamefont
  {Combescot}, \citenamefont {Leyronas},\ and\ \citenamefont
  {Tanguy}}]{Combescot:EurPhysJB(2003)}%
  \BibitemOpen
  \bibfield  {author} {\bibinfo {author} {\bibfnamefont {M.}~\bibnamefont
  {Combescot}}, \bibinfo {author} {\bibfnamefont {X.}~\bibnamefont
  {Leyronas}},\ and\ \bibinfo {author} {\bibfnamefont {C.}~\bibnamefont
  {Tanguy}},\ }\bibfield  {title} {\bibinfo {title} {On the n-exciton
  normalization factor},\ }\href@noop {} {\bibfield  {journal} {\bibinfo
  {journal} {The European Physical Journal B - Condensed Matter and Complex
  Systems}\ }\textbf {\bibinfo {volume} {31}},\ \bibinfo {pages} {17} (\bibinfo
  {year} {2003})}\BibitemShut {NoStop}%
\bibitem [{\citenamefont {Laussy}\ \emph {et~al.}(2006)\citenamefont {Laussy},
  \citenamefont {Glazov}, \citenamefont {Kavokin}, \citenamefont {Whittaker},\
  and\ \citenamefont {Malpuech}}]{Laussy:PRB73(2006)}%
  \BibitemOpen
  \bibfield  {author} {\bibinfo {author} {\bibfnamefont {F.~P.}\ \bibnamefont
  {Laussy}}, \bibinfo {author} {\bibfnamefont {M.~M.}\ \bibnamefont {Glazov}},
  \bibinfo {author} {\bibfnamefont {A.}~\bibnamefont {Kavokin}}, \bibinfo
  {author} {\bibfnamefont {D.~M.}\ \bibnamefont {Whittaker}},\ and\ \bibinfo
  {author} {\bibfnamefont {G.}~\bibnamefont {Malpuech}},\ }\bibfield  {title}
  {\bibinfo {title} {Statistics of excitons in quantum dots and their effect on
  the optical emission spectra of microcavities},\ }\href
  {https://doi.org/10.1103/physrevb.73.115343} {\bibfield  {journal} {\bibinfo
  {journal} {Physical Review B}\ }\textbf {\bibinfo {volume} {73}},\ \bibinfo
  {pages} {115343} (\bibinfo {year} {2006})}\BibitemShut {NoStop}%
\bibitem [{\citenamefont {Makhonin}\ \emph {et~al.}(2024)\citenamefont
  {Makhonin}, \citenamefont {Delphan}, \citenamefont {Song}, \citenamefont
  {Walker}, \citenamefont {Isoniemi}, \citenamefont {Claronino}, \citenamefont
  {Orfanakis}, \citenamefont {Rajendran}, \citenamefont {Ohadi}, \citenamefont
  {Heck{\"o}tter}, \citenamefont {Assmann}, \citenamefont {Bayer},
  \citenamefont {Tartakovskii}, \citenamefont {Skolnick}, \citenamefont
  {Kyriienko},\ and\ \citenamefont
  {Krizhanovskii}}]{Makhonin:LightSciAppl13(2024)}%
  \BibitemOpen
  \bibfield  {author} {\bibinfo {author} {\bibfnamefont {M.}~\bibnamefont
  {Makhonin}}, \bibinfo {author} {\bibfnamefont {A.}~\bibnamefont {Delphan}},
  \bibinfo {author} {\bibfnamefont {K.~W.}\ \bibnamefont {Song}}, \bibinfo
  {author} {\bibfnamefont {P.}~\bibnamefont {Walker}}, \bibinfo {author}
  {\bibfnamefont {T.}~\bibnamefont {Isoniemi}}, \bibinfo {author}
  {\bibfnamefont {P.}~\bibnamefont {Claronino}}, \bibinfo {author}
  {\bibfnamefont {K.}~\bibnamefont {Orfanakis}}, \bibinfo {author}
  {\bibfnamefont {S.~K.}\ \bibnamefont {Rajendran}}, \bibinfo {author}
  {\bibfnamefont {H.}~\bibnamefont {Ohadi}}, \bibinfo {author} {\bibfnamefont
  {J.}~\bibnamefont {Heck{\"o}tter}}, \bibinfo {author} {\bibfnamefont
  {M.}~\bibnamefont {Assmann}}, \bibinfo {author} {\bibfnamefont
  {M.}~\bibnamefont {Bayer}}, \bibinfo {author} {\bibfnamefont
  {A.}~\bibnamefont {Tartakovskii}}, \bibinfo {author} {\bibfnamefont
  {M.}~\bibnamefont {Skolnick}}, \bibinfo {author} {\bibfnamefont
  {O.}~\bibnamefont {Kyriienko}},\ and\ \bibinfo {author} {\bibfnamefont
  {D.}~\bibnamefont {Krizhanovskii}},\ }\bibfield  {title} {\bibinfo {title}
  {Nonlinear rydberg exciton-polaritons in cu2o microcavities},\ }\href
  {https://doi.org/10.1038/s41377-024-01382-9} {\bibfield  {journal} {\bibinfo
  {journal} {Light: Science \& Applications}\ }\textbf {\bibinfo {volume}
  {13}},\ \bibinfo {pages} {47} (\bibinfo {year} {2024})}\BibitemShut {NoStop}%
\bibitem [{\citenamefont {Shahnazaryan}\ \emph {et~al.}(2020)\citenamefont
  {Shahnazaryan}, \citenamefont {Kozin}, \citenamefont {Shelykh}, \citenamefont
  {Iorsh},\ and\ \citenamefont {Kyriienko}}]{Shahnazaryan:PRB102(2020)}%
  \BibitemOpen
  \bibfield  {author} {\bibinfo {author} {\bibfnamefont {V.}~\bibnamefont
  {Shahnazaryan}}, \bibinfo {author} {\bibfnamefont {V.~K.}\ \bibnamefont
  {Kozin}}, \bibinfo {author} {\bibfnamefont {I.~A.}\ \bibnamefont {Shelykh}},
  \bibinfo {author} {\bibfnamefont {I.~V.}\ \bibnamefont {Iorsh}},\ and\
  \bibinfo {author} {\bibfnamefont {O.}~\bibnamefont {Kyriienko}},\ }\bibfield
  {title} {\bibinfo {title} {Tunable optical nonlinearity for transition metal
  dichalcogenide polaritons dressed by a fermi sea},\ }\href
  {https://doi.org/10.1103/physrevb.102.115310} {\bibfield  {journal} {\bibinfo
   {journal} {Phys. Rev. B}\ }\textbf {\bibinfo {volume} {102}},\ \bibinfo
  {pages} {115310} (\bibinfo {year} {2020})}\BibitemShut {NoStop}%
\bibitem [{\citenamefont {Alexeev}\ \emph {et~al.}(2019)\citenamefont
  {Alexeev}, \citenamefont {Ruiz-Tijerina}, \citenamefont {Danovich},
  \citenamefont {Hamer}, \citenamefont {Terry}, \citenamefont {Nayak},
  \citenamefont {Ahn}, \citenamefont {Pak}, \citenamefont {Lee}, \citenamefont
  {Sohn}, \citenamefont {Molas}, \citenamefont {Koperski}, \citenamefont
  {Watanabe}, \citenamefont {Taniguchi}, \citenamefont {Novoselov},
  \citenamefont {Gorbachev}, \citenamefont {Shin}, \citenamefont {Fal'ko},\
  and\ \citenamefont {Tartakovskii}}]{Alexeev:Nature567(2019)}%
  \BibitemOpen
  \bibfield  {author} {\bibinfo {author} {\bibfnamefont {E.~M.}\ \bibnamefont
  {Alexeev}}, \bibinfo {author} {\bibfnamefont {D.~A.}\ \bibnamefont
  {Ruiz-Tijerina}}, \bibinfo {author} {\bibfnamefont {M.}~\bibnamefont
  {Danovich}}, \bibinfo {author} {\bibfnamefont {M.~J.}\ \bibnamefont {Hamer}},
  \bibinfo {author} {\bibfnamefont {D.~J.}\ \bibnamefont {Terry}}, \bibinfo
  {author} {\bibfnamefont {P.~K.}\ \bibnamefont {Nayak}}, \bibinfo {author}
  {\bibfnamefont {S.}~\bibnamefont {Ahn}}, \bibinfo {author} {\bibfnamefont
  {S.}~\bibnamefont {Pak}}, \bibinfo {author} {\bibfnamefont {J.}~\bibnamefont
  {Lee}}, \bibinfo {author} {\bibfnamefont {J.~I.}\ \bibnamefont {Sohn}},
  \bibinfo {author} {\bibfnamefont {M.~R.}\ \bibnamefont {Molas}}, \bibinfo
  {author} {\bibfnamefont {M.}~\bibnamefont {Koperski}}, \bibinfo {author}
  {\bibfnamefont {K.}~\bibnamefont {Watanabe}}, \bibinfo {author}
  {\bibfnamefont {T.}~\bibnamefont {Taniguchi}}, \bibinfo {author}
  {\bibfnamefont {K.~S.}\ \bibnamefont {Novoselov}}, \bibinfo {author}
  {\bibfnamefont {R.~V.}\ \bibnamefont {Gorbachev}}, \bibinfo {author}
  {\bibfnamefont {H.~S.}\ \bibnamefont {Shin}}, \bibinfo {author}
  {\bibfnamefont {V.~I.}\ \bibnamefont {Fal'ko}},\ and\ \bibinfo {author}
  {\bibfnamefont {A.~I.}\ \bibnamefont {Tartakovskii}},\ }\bibfield  {title}
  {\bibinfo {title} {Resonantly hybridized excitons in moir{\'{e}}
  superlattices in van der waals heterostructures},\ }\href
  {https://doi.org/10.1038/s41586-019-0986-9} {\bibfield  {journal} {\bibinfo
  {journal} {Nature}\ }\textbf {\bibinfo {volume} {567}},\ \bibinfo {pages}
  {81} (\bibinfo {year} {2019})}\BibitemShut {NoStop}%
\bibitem [{\citenamefont {Tran}\ \emph {et~al.}(2019)\citenamefont {Tran},
  \citenamefont {Moody}, \citenamefont {Wu}, \citenamefont {Lu}, \citenamefont
  {Choi}, \citenamefont {Kim}, \citenamefont {Rai}, \citenamefont {Sanchez},
  \citenamefont {Quan}, \citenamefont {Singh}, \citenamefont {Embley},
  \citenamefont {Zepeda}, \citenamefont {Campbell}, \citenamefont {Autry},
  \citenamefont {Taniguchi}, \citenamefont {Watanabe}, \citenamefont {Lu},
  \citenamefont {Banerjee}, \citenamefont {Silverman}, \citenamefont {Kim},
  \citenamefont {Tutuc}, \citenamefont {Yang}, \citenamefont {MacDonald},\ and\
  \citenamefont {Li}}]{Tran:Nature567(2019)}%
  \BibitemOpen
  \bibfield  {author} {\bibinfo {author} {\bibfnamefont {K.}~\bibnamefont
  {Tran}}, \bibinfo {author} {\bibfnamefont {G.}~\bibnamefont {Moody}},
  \bibinfo {author} {\bibfnamefont {F.}~\bibnamefont {Wu}}, \bibinfo {author}
  {\bibfnamefont {X.}~\bibnamefont {Lu}}, \bibinfo {author} {\bibfnamefont
  {J.}~\bibnamefont {Choi}}, \bibinfo {author} {\bibfnamefont {K.}~\bibnamefont
  {Kim}}, \bibinfo {author} {\bibfnamefont {A.}~\bibnamefont {Rai}}, \bibinfo
  {author} {\bibfnamefont {D.~A.}\ \bibnamefont {Sanchez}}, \bibinfo {author}
  {\bibfnamefont {J.}~\bibnamefont {Quan}}, \bibinfo {author} {\bibfnamefont
  {A.}~\bibnamefont {Singh}}, \bibinfo {author} {\bibfnamefont
  {J.}~\bibnamefont {Embley}}, \bibinfo {author} {\bibfnamefont
  {A.}~\bibnamefont {Zepeda}}, \bibinfo {author} {\bibfnamefont
  {M.}~\bibnamefont {Campbell}}, \bibinfo {author} {\bibfnamefont
  {T.}~\bibnamefont {Autry}}, \bibinfo {author} {\bibfnamefont
  {T.}~\bibnamefont {Taniguchi}}, \bibinfo {author} {\bibfnamefont
  {K.}~\bibnamefont {Watanabe}}, \bibinfo {author} {\bibfnamefont
  {N.}~\bibnamefont {Lu}}, \bibinfo {author} {\bibfnamefont {S.~K.}\
  \bibnamefont {Banerjee}}, \bibinfo {author} {\bibfnamefont {K.~L.}\
  \bibnamefont {Silverman}}, \bibinfo {author} {\bibfnamefont {S.}~\bibnamefont
  {Kim}}, \bibinfo {author} {\bibfnamefont {E.}~\bibnamefont {Tutuc}}, \bibinfo
  {author} {\bibfnamefont {L.}~\bibnamefont {Yang}}, \bibinfo {author}
  {\bibfnamefont {A.~H.}\ \bibnamefont {MacDonald}},\ and\ \bibinfo {author}
  {\bibfnamefont {X.}~\bibnamefont {Li}},\ }\bibfield  {title} {\bibinfo
  {title} {Evidence for moir{\'{e}} excitons in van der waals
  heterostructures},\ }\href {https://doi.org/10.1038/s41586-019-0975-z}
  {\bibfield  {journal} {\bibinfo  {journal} {Nature}\ }\textbf {\bibinfo
  {volume} {567}},\ \bibinfo {pages} {71} (\bibinfo {year} {2019})}\BibitemShut
  {NoStop}%
\bibitem [{\citenamefont {Tran}\ \emph {et~al.}(2020)\citenamefont {Tran},
  \citenamefont {Choi},\ and\ \citenamefont {Singh}}]{Tran:2DMat8(2020)}%
  \BibitemOpen
  \bibfield  {author} {\bibinfo {author} {\bibfnamefont {K.}~\bibnamefont
  {Tran}}, \bibinfo {author} {\bibfnamefont {J.}~\bibnamefont {Choi}},\ and\
  \bibinfo {author} {\bibfnamefont {A.}~\bibnamefont {Singh}},\ }\bibfield
  {title} {\bibinfo {title} {Moir{\'{e}} and beyond in transition metal
  dichalcogenide twisted bilayers},\ }\href
  {https://doi.org/10.1088/2053-1583/abd3e7} {\bibfield  {journal} {\bibinfo
  {journal} {2D Materials}\ }\textbf {\bibinfo {volume} {8}},\ \bibinfo {pages}
  {022002} (\bibinfo {year} {2020})}\BibitemShut {NoStop}%
\bibitem [{\citenamefont {Wu}\ \emph {et~al.}(2018)\citenamefont {Wu},
  \citenamefont {Lovorn},\ and\ \citenamefont {MacDonald}}]{Wu:PRB97(2018)}%
  \BibitemOpen
  \bibfield  {author} {\bibinfo {author} {\bibfnamefont {F.}~\bibnamefont
  {Wu}}, \bibinfo {author} {\bibfnamefont {T.}~\bibnamefont {Lovorn}},\ and\
  \bibinfo {author} {\bibfnamefont {A.~H.}\ \bibnamefont {MacDonald}},\
  }\bibfield  {title} {\bibinfo {title} {Theory of optical absorption by
  interlayer excitons in transition metal dichalcogenide heterobilayers},\
  }\href {https://doi.org/10.1103/physrevb.97.035306} {\bibfield  {journal}
  {\bibinfo  {journal} {Physical Review B}\ }\textbf {\bibinfo {volume} {97}},\
  \bibinfo {pages} {035306} (\bibinfo {year} {2018})}\BibitemShut {NoStop}%
\bibitem [{\citenamefont {Ruiz-Tijerina}\ \emph {et~al.}(2020)\citenamefont
  {Ruiz-Tijerina}, \citenamefont {Soltero},\ and\ \citenamefont
  {Mireles}}]{RuizTijerina:PRB102(2020)}%
  \BibitemOpen
  \bibfield  {author} {\bibinfo {author} {\bibfnamefont {D.~A.}\ \bibnamefont
  {Ruiz-Tijerina}}, \bibinfo {author} {\bibfnamefont {I.}~\bibnamefont
  {Soltero}},\ and\ \bibinfo {author} {\bibfnamefont {F.}~\bibnamefont
  {Mireles}},\ }\bibfield  {title} {\bibinfo {title} {Theory of moir{\'{e}}
  localized excitons in transition metal dichalcogenide heterobilayers},\
  }\href {https://doi.org/10.1103/physrevb.102.195403} {\bibfield  {journal}
  {\bibinfo  {journal} {Physical Review B}\ }\textbf {\bibinfo {volume}
  {102}},\ \bibinfo {pages} {195403} (\bibinfo {year} {2020})}\BibitemShut
  {NoStop}%
\bibitem [{\citenamefont {Baek}\ \emph {et~al.}(2020)\citenamefont {Baek},
  \citenamefont {Brotons-Gisbert}, \citenamefont {Koong}, \citenamefont
  {Campbell}, \citenamefont {Rambach}, \citenamefont {Watanabe}, \citenamefont
  {Taniguchi},\ and\ \citenamefont {Gerardot}}]{Baek2020}%
  \BibitemOpen
  \bibfield  {author} {\bibinfo {author} {\bibfnamefont {H.}~\bibnamefont
  {Baek}}, \bibinfo {author} {\bibfnamefont {M.}~\bibnamefont
  {Brotons-Gisbert}}, \bibinfo {author} {\bibfnamefont {Z.~X.}\ \bibnamefont
  {Koong}}, \bibinfo {author} {\bibfnamefont {A.}~\bibnamefont {Campbell}},
  \bibinfo {author} {\bibfnamefont {M.}~\bibnamefont {Rambach}}, \bibinfo
  {author} {\bibfnamefont {K.}~\bibnamefont {Watanabe}}, \bibinfo {author}
  {\bibfnamefont {T.}~\bibnamefont {Taniguchi}},\ and\ \bibinfo {author}
  {\bibfnamefont {B.~D.}\ \bibnamefont {Gerardot}},\ }\bibfield  {title}
  {\bibinfo {title} {Highly energy-tunable quantum light from moir\'{e}-trapped
  excitons},\ }\href {https://doi.org/10.1126/sciadv.aba8526} {\bibfield
  {journal} {\bibinfo  {journal} {Science Advances}\ }\textbf {\bibinfo
  {volume} {6}},\ \bibinfo {pages} {eaba8526} (\bibinfo {year}
  {2020})}\BibitemShut {NoStop}%
\bibitem [{\citenamefont {Kremser}\ \emph {et~al.}(2020)\citenamefont
  {Kremser}, \citenamefont {Brotons-Gisbert}, \citenamefont {Kn{\"o}rzer},
  \citenamefont {G{\"u}ckelhorn}, \citenamefont {Meyer}, \citenamefont
  {Barbone}, \citenamefont {Stier}, \citenamefont {Gerardot}, \citenamefont
  {M{\"u}ller},\ and\ \citenamefont {Finley}}]{Kremser2020}%
  \BibitemOpen
  \bibfield  {author} {\bibinfo {author} {\bibfnamefont {M.}~\bibnamefont
  {Kremser}}, \bibinfo {author} {\bibfnamefont {M.}~\bibnamefont
  {Brotons-Gisbert}}, \bibinfo {author} {\bibfnamefont {J.}~\bibnamefont
  {Kn{\"o}rzer}}, \bibinfo {author} {\bibfnamefont {J.}~\bibnamefont
  {G{\"u}ckelhorn}}, \bibinfo {author} {\bibfnamefont {M.}~\bibnamefont
  {Meyer}}, \bibinfo {author} {\bibfnamefont {M.}~\bibnamefont {Barbone}},
  \bibinfo {author} {\bibfnamefont {A.~V.}\ \bibnamefont {Stier}}, \bibinfo
  {author} {\bibfnamefont {B.~D.}\ \bibnamefont {Gerardot}}, \bibinfo {author}
  {\bibfnamefont {K.}~\bibnamefont {M{\"u}ller}},\ and\ \bibinfo {author}
  {\bibfnamefont {J.~J.}\ \bibnamefont {Finley}},\ }\bibfield  {title}
  {\bibinfo {title} {Discrete interactions between a few interlayer excitons
  trapped at a mose2--wse2 heterointerface},\ }\href
  {https://doi.org/10.1038/s41699-020-0141-3} {\bibfield  {journal} {\bibinfo
  {journal} {npj 2D Materials and Applications}\ }\textbf {\bibinfo {volume}
  {4}},\ \bibinfo {pages} {8} (\bibinfo {year} {2020})}\BibitemShut {NoStop}%
\bibitem [{\citenamefont {Andersen}\ \emph {et~al.}(2021)\citenamefont
  {Andersen}, \citenamefont {Scuri}, \citenamefont {Sushko}, \citenamefont
  {De~Greve}, \citenamefont {Sung}, \citenamefont {Zhou}, \citenamefont {Wild},
  \citenamefont {Gelly}, \citenamefont {Heo}, \citenamefont {B{\'e}rub{\'e}},
  \citenamefont {Joe}, \citenamefont {Jauregui}, \citenamefont {Watanabe},
  \citenamefont {Taniguchi}, \citenamefont {Kim}, \citenamefont {Park},\ and\
  \citenamefont {Lukin}}]{Andersen2021}%
  \BibitemOpen
  \bibfield  {author} {\bibinfo {author} {\bibfnamefont {T.~I.}\ \bibnamefont
  {Andersen}}, \bibinfo {author} {\bibfnamefont {G.}~\bibnamefont {Scuri}},
  \bibinfo {author} {\bibfnamefont {A.}~\bibnamefont {Sushko}}, \bibinfo
  {author} {\bibfnamefont {K.}~\bibnamefont {De~Greve}}, \bibinfo {author}
  {\bibfnamefont {J.}~\bibnamefont {Sung}}, \bibinfo {author} {\bibfnamefont
  {Y.}~\bibnamefont {Zhou}}, \bibinfo {author} {\bibfnamefont {D.~S.}\
  \bibnamefont {Wild}}, \bibinfo {author} {\bibfnamefont {R.~J.}\ \bibnamefont
  {Gelly}}, \bibinfo {author} {\bibfnamefont {H.}~\bibnamefont {Heo}}, \bibinfo
  {author} {\bibfnamefont {D.}~\bibnamefont {B{\'e}rub{\'e}}}, \bibinfo
  {author} {\bibfnamefont {A.~Y.}\ \bibnamefont {Joe}}, \bibinfo {author}
  {\bibfnamefont {L.~A.}\ \bibnamefont {Jauregui}}, \bibinfo {author}
  {\bibfnamefont {K.}~\bibnamefont {Watanabe}}, \bibinfo {author}
  {\bibfnamefont {T.}~\bibnamefont {Taniguchi}}, \bibinfo {author}
  {\bibfnamefont {P.}~\bibnamefont {Kim}}, \bibinfo {author} {\bibfnamefont
  {H.}~\bibnamefont {Park}},\ and\ \bibinfo {author} {\bibfnamefont {M.~D.}\
  \bibnamefont {Lukin}},\ }\bibfield  {title} {\bibinfo {title} {Excitons in a
  reconstructed moir{\'e} potential in twisted wse2/wse2 homobilayers},\ }\href
  {https://doi.org/10.1038/s41563-020-00873-5} {\bibfield  {journal} {\bibinfo
  {journal} {Nature Materials}\ }\textbf {\bibinfo {volume} {20}},\ \bibinfo
  {pages} {480} (\bibinfo {year} {2021})}\BibitemShut {NoStop}%
\bibitem [{\citenamefont {Campbell}\ \emph {et~al.}(2022)\citenamefont
  {Campbell}, \citenamefont {Brotons-Gisbert}, \citenamefont {Baek},
  \citenamefont {Vitale}, \citenamefont {Taniguchi}, \citenamefont {Watanabe},
  \citenamefont {Lischner},\ and\ \citenamefont {Gerardot}}]{Campbell2022}%
  \BibitemOpen
  \bibfield  {author} {\bibinfo {author} {\bibfnamefont {A.~J.}\ \bibnamefont
  {Campbell}}, \bibinfo {author} {\bibfnamefont {M.}~\bibnamefont
  {Brotons-Gisbert}}, \bibinfo {author} {\bibfnamefont {H.}~\bibnamefont
  {Baek}}, \bibinfo {author} {\bibfnamefont {V.}~\bibnamefont {Vitale}},
  \bibinfo {author} {\bibfnamefont {T.}~\bibnamefont {Taniguchi}}, \bibinfo
  {author} {\bibfnamefont {K.}~\bibnamefont {Watanabe}}, \bibinfo {author}
  {\bibfnamefont {J.}~\bibnamefont {Lischner}},\ and\ \bibinfo {author}
  {\bibfnamefont {B.~D.}\ \bibnamefont {Gerardot}},\ }\bibfield  {title}
  {\bibinfo {title} {Exciton-polarons in the presence of strongly correlated
  electronic states in a mose2/wse2 moir{\'e} superlattice},\ }\href
  {https://doi.org/10.1038/s41699-022-00358-w} {\bibfield  {journal} {\bibinfo
  {journal} {npj 2D Materials and Applications}\ }\textbf {\bibinfo {volume}
  {6}},\ \bibinfo {pages} {79} (\bibinfo {year} {2022})}\BibitemShut {NoStop}%
\bibitem [{\citenamefont {Feierabend}\ \emph {et~al.}(2017)\citenamefont
  {Feierabend}, \citenamefont {Morlet}, \citenamefont {Berghäuser},\ and\
  \citenamefont {Malic}}]{Feierabend:PRB96(2017)}%
  \BibitemOpen
  \bibfield  {author} {\bibinfo {author} {\bibfnamefont {M.}~\bibnamefont
  {Feierabend}}, \bibinfo {author} {\bibfnamefont {A.}~\bibnamefont {Morlet}},
  \bibinfo {author} {\bibfnamefont {G.}~\bibnamefont {Berghäuser}},\ and\
  \bibinfo {author} {\bibfnamefont {E.}~\bibnamefont {Malic}},\ }\bibfield
  {title} {\bibinfo {title} {Impact of strain on the optical fingerprint of
  monolayer transition-metal dichalcogenides},\ }\href
  {https://doi.org/10.1103/physrevb.96.045425} {\bibfield  {journal} {\bibinfo
  {journal} {Physical Review B}\ }\textbf {\bibinfo {volume} {96}},\ \bibinfo
  {pages} {045425} (\bibinfo {year} {2017})}\BibitemShut {NoStop}%
\bibitem [{\citenamefont {Weston}\ \emph {et~al.}(2020)\citenamefont {Weston},
  \citenamefont {Zou}, \citenamefont {Enaldiev}, \citenamefont {Summerfield},
  \citenamefont {Clark}, \citenamefont {Z{\'{o}}lyomi}, \citenamefont {Graham},
  \citenamefont {Yelgel}, \citenamefont {Magorrian}, \citenamefont {Zhou},
  \citenamefont {Zultak}, \citenamefont {Hopkinson}, \citenamefont {Barinov},
  \citenamefont {Bointon}, \citenamefont {Kretinin}, \citenamefont {Wilson},
  \citenamefont {Beton}, \citenamefont {Fal'ko}, \citenamefont {Haigh},\ and\
  \citenamefont {Gorbachev}}]{Weston:NatNano15(2020)}%
  \BibitemOpen
  \bibfield  {author} {\bibinfo {author} {\bibfnamefont {A.}~\bibnamefont
  {Weston}}, \bibinfo {author} {\bibfnamefont {Y.}~\bibnamefont {Zou}},
  \bibinfo {author} {\bibfnamefont {V.}~\bibnamefont {Enaldiev}}, \bibinfo
  {author} {\bibfnamefont {A.}~\bibnamefont {Summerfield}}, \bibinfo {author}
  {\bibfnamefont {N.}~\bibnamefont {Clark}}, \bibinfo {author} {\bibfnamefont
  {V.}~\bibnamefont {Z{\'{o}}lyomi}}, \bibinfo {author} {\bibfnamefont
  {A.}~\bibnamefont {Graham}}, \bibinfo {author} {\bibfnamefont
  {C.}~\bibnamefont {Yelgel}}, \bibinfo {author} {\bibfnamefont
  {S.}~\bibnamefont {Magorrian}}, \bibinfo {author} {\bibfnamefont
  {M.}~\bibnamefont {Zhou}}, \bibinfo {author} {\bibfnamefont {J.}~\bibnamefont
  {Zultak}}, \bibinfo {author} {\bibfnamefont {D.}~\bibnamefont {Hopkinson}},
  \bibinfo {author} {\bibfnamefont {A.}~\bibnamefont {Barinov}}, \bibinfo
  {author} {\bibfnamefont {T.~H.}\ \bibnamefont {Bointon}}, \bibinfo {author}
  {\bibfnamefont {A.}~\bibnamefont {Kretinin}}, \bibinfo {author}
  {\bibfnamefont {N.~R.}\ \bibnamefont {Wilson}}, \bibinfo {author}
  {\bibfnamefont {P.~H.}\ \bibnamefont {Beton}}, \bibinfo {author}
  {\bibfnamefont {V.~I.}\ \bibnamefont {Fal'ko}}, \bibinfo {author}
  {\bibfnamefont {S.~J.}\ \bibnamefont {Haigh}},\ and\ \bibinfo {author}
  {\bibfnamefont {R.}~\bibnamefont {Gorbachev}},\ }\bibfield  {title} {\bibinfo
  {title} {Atomic reconstruction in twisted bilayers of transition metal
  dichalcogenides},\ }\href {https://doi.org/10.1038/s41565-020-0682-9}
  {\bibfield  {journal} {\bibinfo  {journal} {Nature Nanotechnology}\ }\textbf
  {\bibinfo {volume} {15}},\ \bibinfo {pages} {592} (\bibinfo {year}
  {2020})}\BibitemShut {NoStop}%
\bibitem [{\citenamefont {Wang}\ \emph {et~al.}(2021)\citenamefont {Wang},
  \citenamefont {Zhu}, \citenamefont {Seyler}, \citenamefont {Rivera},
  \citenamefont {Zheng}, \citenamefont {Wang}, \citenamefont {He},
  \citenamefont {Taniguchi}, \citenamefont {Watanabe}, \citenamefont {Yan},
  \citenamefont {Mandrus}, \citenamefont {Gamelin}, \citenamefont {Yao},\ and\
  \citenamefont {Xu}}]{Wang:NatNano16(2021)}%
  \BibitemOpen
  \bibfield  {author} {\bibinfo {author} {\bibfnamefont {X.}~\bibnamefont
  {Wang}}, \bibinfo {author} {\bibfnamefont {J.}~\bibnamefont {Zhu}}, \bibinfo
  {author} {\bibfnamefont {K.~L.}\ \bibnamefont {Seyler}}, \bibinfo {author}
  {\bibfnamefont {P.}~\bibnamefont {Rivera}}, \bibinfo {author} {\bibfnamefont
  {H.}~\bibnamefont {Zheng}}, \bibinfo {author} {\bibfnamefont
  {Y.}~\bibnamefont {Wang}}, \bibinfo {author} {\bibfnamefont {M.}~\bibnamefont
  {He}}, \bibinfo {author} {\bibfnamefont {T.}~\bibnamefont {Taniguchi}},
  \bibinfo {author} {\bibfnamefont {K.}~\bibnamefont {Watanabe}}, \bibinfo
  {author} {\bibfnamefont {J.}~\bibnamefont {Yan}}, \bibinfo {author}
  {\bibfnamefont {D.~G.}\ \bibnamefont {Mandrus}}, \bibinfo {author}
  {\bibfnamefont {D.~R.}\ \bibnamefont {Gamelin}}, \bibinfo {author}
  {\bibfnamefont {W.}~\bibnamefont {Yao}},\ and\ \bibinfo {author}
  {\bibfnamefont {X.}~\bibnamefont {Xu}},\ }\bibfield  {title} {\bibinfo
  {title} {Moir{\'{e}} trions in {MoSe}2/{WSe}2 heterobilayers},\ }\href
  {https://doi.org/10.1038/s41565-021-00969-2} {\bibfield  {journal} {\bibinfo
  {journal} {Nature Nanotechnology}\ }\textbf {\bibinfo {volume} {16}},\
  \bibinfo {pages} {1208} (\bibinfo {year} {2021})}\BibitemShut {NoStop}%
\bibitem [{\citenamefont {Liu}\ \emph {et~al.}(2021)\citenamefont {Liu},
  \citenamefont {Barr{\'{e}}}, \citenamefont {van Baren}, \citenamefont
  {Wilson}, \citenamefont {Taniguchi}, \citenamefont {Watanabe}, \citenamefont
  {Cui}, \citenamefont {Gabor}, \citenamefont {Heinz}, \citenamefont {Chang},\
  and\ \citenamefont {Lui}}]{Liu:Nat594(2021)}%
  \BibitemOpen
  \bibfield  {author} {\bibinfo {author} {\bibfnamefont {E.}~\bibnamefont
  {Liu}}, \bibinfo {author} {\bibfnamefont {E.}~\bibnamefont {Barr{\'{e}}}},
  \bibinfo {author} {\bibfnamefont {J.}~\bibnamefont {van Baren}}, \bibinfo
  {author} {\bibfnamefont {M.}~\bibnamefont {Wilson}}, \bibinfo {author}
  {\bibfnamefont {T.}~\bibnamefont {Taniguchi}}, \bibinfo {author}
  {\bibfnamefont {K.}~\bibnamefont {Watanabe}}, \bibinfo {author}
  {\bibfnamefont {Y.-T.}\ \bibnamefont {Cui}}, \bibinfo {author} {\bibfnamefont
  {N.~M.}\ \bibnamefont {Gabor}}, \bibinfo {author} {\bibfnamefont {T.~F.}\
  \bibnamefont {Heinz}}, \bibinfo {author} {\bibfnamefont {Y.-C.}\ \bibnamefont
  {Chang}},\ and\ \bibinfo {author} {\bibfnamefont {C.~H.}\ \bibnamefont
  {Lui}},\ }\bibfield  {title} {\bibinfo {title} {Signatures of moir{\'{e}}
  trions in {WSe}2/{MoSe}2 heterobilayers},\ }\href
  {https://doi.org/10.1038/s41586-021-03541-z} {\bibfield  {journal} {\bibinfo
  {journal} {Nature}\ }\textbf {\bibinfo {volume} {594}},\ \bibinfo {pages}
  {46} (\bibinfo {year} {2021})}\BibitemShut {NoStop}%
\bibitem [{\citenamefont {Brotons-Gisbert}\ \emph {et~al.}(2021)\citenamefont
  {Brotons-Gisbert}, \citenamefont {Baek}, \citenamefont {Campbell},
  \citenamefont {Watanabe}, \citenamefont {Taniguchi},\ and\ \citenamefont
  {Gerardot}}]{Brotons-GisbertPRX11(2021)}%
  \BibitemOpen
  \bibfield  {author} {\bibinfo {author} {\bibfnamefont {M.}~\bibnamefont
  {Brotons-Gisbert}}, \bibinfo {author} {\bibfnamefont {H.}~\bibnamefont
  {Baek}}, \bibinfo {author} {\bibfnamefont {A.}~\bibnamefont {Campbell}},
  \bibinfo {author} {\bibfnamefont {K.}~\bibnamefont {Watanabe}}, \bibinfo
  {author} {\bibfnamefont {T.}~\bibnamefont {Taniguchi}},\ and\ \bibinfo
  {author} {\bibfnamefont {B.~D.}\ \bibnamefont {Gerardot}},\ }\bibfield
  {title} {\bibinfo {title} {Moir\'e-trapped interlayer trions in a
  charge-tunable ${\mathrm{wse}}_{2}/{\mathrm{mose}}_{2}$ heterobilayer},\
  }\href {https://doi.org/10.1103/PhysRevX.11.031033} {\bibfield  {journal}
  {\bibinfo  {journal} {Phys. Rev. X}\ }\textbf {\bibinfo {volume} {11}},\
  \bibinfo {pages} {031033} (\bibinfo {year} {2021})}\BibitemShut {NoStop}%
\bibitem [{\citenamefont {Song}\ \emph {et~al.}(2022)\citenamefont {Song},
  \citenamefont {Chiavazzo}, \citenamefont {Shelykh},\ and\ \citenamefont
  {Kyriienko}}]{Song2022}%
  \BibitemOpen
  \bibfield  {author} {\bibinfo {author} {\bibfnamefont {K.~W.}\ \bibnamefont
  {Song}}, \bibinfo {author} {\bibfnamefont {S.}~\bibnamefont {Chiavazzo}},
  \bibinfo {author} {\bibfnamefont {I.~A.}\ \bibnamefont {Shelykh}},\ and\
  \bibinfo {author} {\bibfnamefont {O.}~\bibnamefont {Kyriienko}},\ }\href
  {https://doi.org/10.48550/ARXIV.2204.00594} {\bibinfo {title} {Attractive
  trion-polariton nonlinearity due to coulomb scattering}} (\bibinfo {year}
  {2022}),\ \Eprint {https://arxiv.org/abs/22204.00594} {arXiv:22204.00594}
  \BibitemShut {NoStop}%
\end{thebibliography}%

\end{document}